\documentclass[papers,fleqn,usenatbib]{mnras}

\usepackage{newtxtext,newtxmath}

\usepackage[T1]{fontenc}
\usepackage{ae,aecompl}
\usepackage{url}

\usepackage{graphicx}	
\usepackage{amsmath}	
\usepackage{amssymb}	

\title[Protostellar accretion history with ArT\'eMiS]{The accretion history of high-mass stars: \\
An ArT\'eMiS pilot study of Infrared Dark Clouds}

\author[N. Peretto et al.]
{N.  Peretto$^{1}$\thanks{E-mail: nicolas.peretto@astro.cf.ac.uk},
A. Rigby$^{1}$,
Ph. Andr\'e$^{2}$,
V. K\"onyves$^{3}$,
G. Fuller$^{4}$,
A. Zavagno$^{5,6}$,
\newauthor F. Schuller$^{2,7}$,
D. Arzoumanian$^{8}$,
S. Bontemps$^{9}$,
T. Csengeri$^{9}$,
P. Didelon$^{2}$,
\newauthor  A. Duarte-Cabral$^{1}$,
P. Palmeirim$^{8}$,
S. Pezzuto$^{10}$,
V. Rev\'eret$^{2}$,
H. Roussel$^{11}$,
\newauthor Y. Shimajiri$^{12}$\\
$^{1}$Cardiff University, School of Physics \& Astronomy, Queen's buildings, The parade, Cardiff CF24 3AA, UK\\
$^{2}$ Laboratoire d'Astrophysique (AIM), CEA/DRF, CNRS, Universit\'e Paris-Saclay, Universit\'e Paris Diderot, Sorbonne Paris Cit\'e,\\ 
91191 Gif-sur-Yvette, France\\
$^{3}$ Jeremiah Horrocks Institute, University of Central Lancashire, Preston PR1 2HE, UK\\
$^{4}$ Jodrell Bank Centre for Astrophysics, Department of Physics and Astronomy, The University of Manchester, Oxford Road,\\
 Manchester M13 9PL, UK\\
$^{5}$ Aix Marseille Univ, CNRS, CNES, LAM, Marseille, France\\
$^{6}$ Institut Universitaire de France\\
$^{7}$ Leibniz-Institut f\"ur Astrophysik Potsdam (AIP), An der Sternwarte 16, 14482 Potsdam, Germany\\
$^{8}$ Instituto de Astrof\'isica e Ci{\^e}ncias do Espa\c{c}o, Universidade do Porto, CAUP, Rua das Estrelas, PT4150-762 Porto, Portugal\\
$^{9}$ Laboratoire d'Astrophysique de Bordeaux, Univ. Bordeaux, CNRS, B18N, allée Georoy Saint-Hilaire, 33615 Pessac, France\\
$^{10}$ INAF - IAPS Via fosso del Cavaliere 100, I-00133 Roma, Italy\\
$^{11}$ Institut d'Astrophysique de Paris, Sorbonne Universit\'e, CNRS (UMR7095), 75014 Paris, France\\
$^{12}$ National Astronomical Observatory of Japan, Osawa 2-21-1, Mitaka, Tokyo 181-8588, Japan\\
}

\date{Accepted XXX. Received YYY; in original form ZZZ}

\pubyear{2015}

\begin{document}
\label{firstpage}
\pagerange{\pageref{firstpage}--\pageref{lastpage}}
\maketitle

\begin{abstract}
The mass growth of protostars is a central element to the determination of fundamental stellar population properties such as the initial mass function. Constraining the accretion history of individual protostars is therefore an important aspect of star formation research. The goal of the study presented here is to determine whether high-mass (proto)stars gain their mass from a compact ($<0.1$~pc) fixed-mass reservoir of gas, often referred to as dense cores, in which they are embedded, or whether the mass growth of high-mass stars is governed by the dynamical evolution of the parsec-scale clump that typically surrounds them. To achieve this goal, we performed a 350$\mu$m continuum mapping of 11 infrared dark clouds, along side some of  their neighbouring clumps, with the ArT\'eMiS camera on APEX. By identifying about 200 compact ArT\'eMiS sources, and matching them with {\it Herschel} Hi-GAL 70$\mu$m sources, we have been able to  produce mass vs. temperature diagrams. We compare the nature (i.e. starless or protostellar) and location of the ArT\'eMiS sources in these diagrams with modelled evolutionary tracks of both core-fed and clump-fed accretion scenarios. We argue that the latter provide a better agreement with the observed distribution of high-mass star-forming cores. However, a robust and definitive conclusion on the question of the accretion history of high-mass stars requires larger number statistics.
\end{abstract}

\begin{keywords}
star formation - protostar - accretion
\end{keywords}



\begin{figure*}
	\includegraphics[width=17cm]{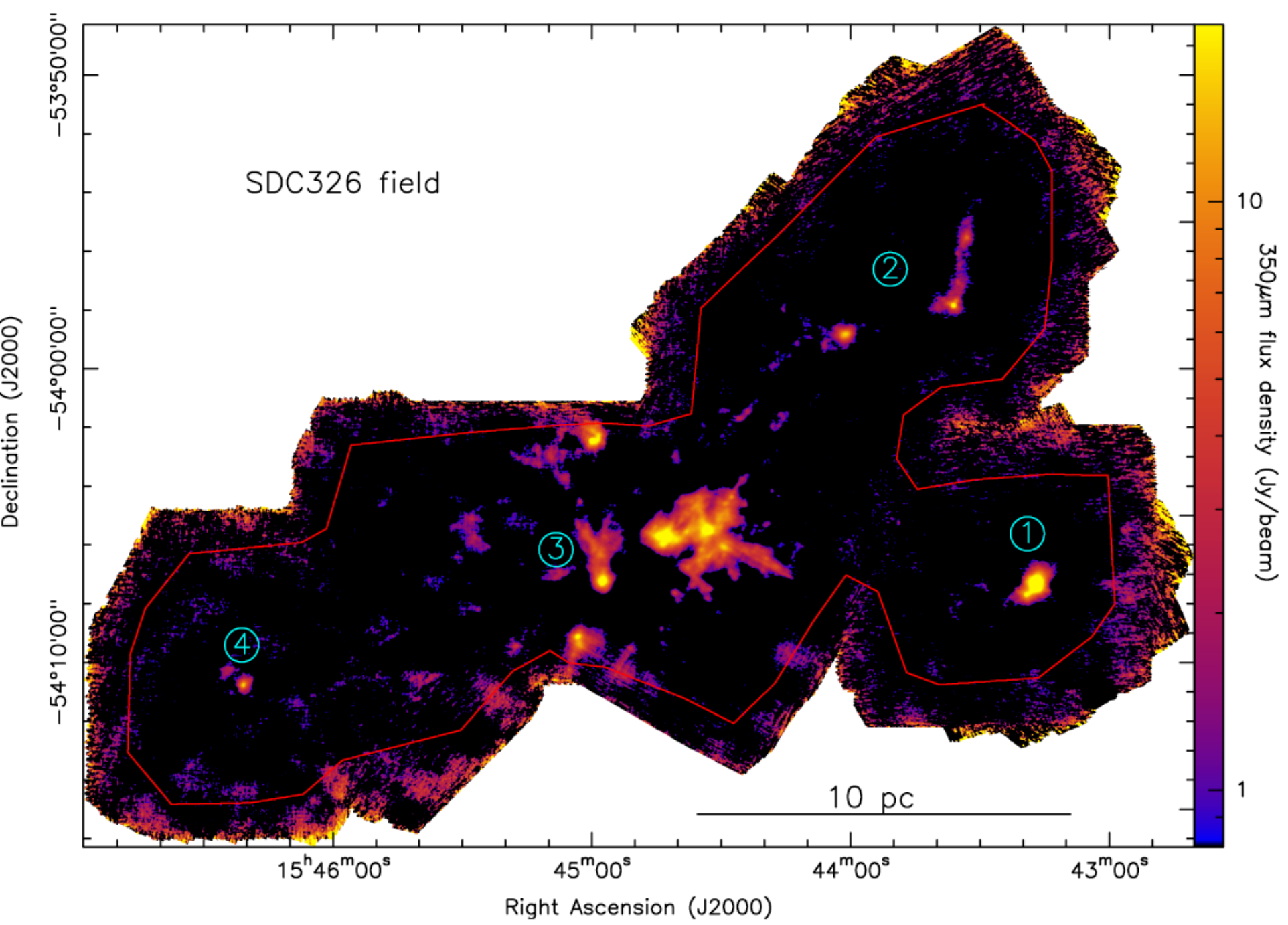}
    \caption{350$\mu m$ ArT\'eMiS image of the SDC326 field (in logarithmic scale), that includes 4 targeted IRDCs whose ID numbers (in cyan) correspond to those indicated in Table 1. The area within the red polygone roughly corresponds to the area of uniform noise. The 10~pc bar scale at the bottom of the image assumes a distance of 2.6~kpc.}
    \label{art_sdc326_nosource}
\end{figure*}

\section{Introduction}

Our knowledge about star formation has made tremendous progress in the past few years \citep{andre2014,motte2018a}. As a result of the science exploitation of {\it Herschel} data, our picture of how matter is condensed from diffuse clouds to stars has been significantly improved. In particular, one striking result has been the ubiquity of interstellar filaments in all types of molecular clouds, including low-mass star-forming \citep{andre2010}, massive-star forming \citep{molinari2010}, and even non-star-forming clouds \citep{menshchikov2010}. 
In nearby, and mostly low-mass, star-forming regions nearly all the dense gas ($n_{H_2} \ge 10^4$~cm$^{-3}$; $N_{H_2} \ge 10^{22}$ cm$^{-2}$) is concentrated within 0.1pc-width filaments \citep{arzoumanian2011,arzoumanian2019}. Moreover, the vast majority ($\sim$\,75\%) of 
prestellar cores are found to lie within these filaments \citep{konyves2015,konyves2020,ladjelate2020}. It has therefore been proposed that solar-type star-forming cores form as the result of Jeans-type gravitational instabilities developing along self-gravitating filaments \citep{inutsuka1997}, providing a compact and, at least to first order,  fixed mass reservoir for the protostars forming inside them. As a result, such protostars are said to be {\it core-fed}. This scenario is believed to explain the shape of the prestellar core mass function as observed in a number of low-mass star-forming clouds \citep[e.g.][]{motte1998,konyves2015},  and therefore the origin of the base of the initial mass function (IMF) from $\sim 0.1$ to 5~M$_{\odot}$  \citep[cf.][]{andre2014,andre2019,lee2017}.

\begin{table*}
	\centering
	\caption{Target properties. The V$_{\rm{lsr}}$ column corresponds to velocities that have been obtained using the MALT90 N$_2$H$^+$(1-0) line, with the exception of SDC340.928-1.042 (marked with an asterisk) for which we used the ThrUMMS $^{13}$CO(1-0) line (see text). IRDCs having several velocities/distances correspond to ArT\'eMiS maps where several clumps have been detected. Noise values correspond to the median rms noise in each of the 5 observed fields, for both the original ArT\'eMiS images and the Gaussian filtered ones.}
	\label{tab:IRDC_prop}
	\begin{tabular}{clccccccc} 
		\hline
		ID\#& IRDC name & Coordinates & V$_{\rm{lsr}}$ & Distance & Radius & Mass & Unfilt. noise & Filt. noise\\
				&   & (J2000) & (km/s) & (kpc) & (pc) & (M$_{\odot}$)&  (Jy/beam) &  (Jy/beam)\\
		\hline
		1 & SDC326.476+0.706  & 15:43:16.4 -54:07:13 & -40.5 & 2.7 & 1.08  & 3730 &0.48 & 0.26\\
		2 & SDC326.611+0.811  &  15:43:36.3 -53:57:45 & -37.0 & 2.5 & 1.29 & 3260 &0.48 & 0.26 \\
		3 & SDC326.672+0.585 &   15:44:57.3 -54:07:14 & -41.3 & 2.7 & 0.91 & 4120 & 0.48& 0.26 \\
		4 & SDC326.796+0.386 &   15:46:20.9  -54:10:44 & -20.4 & 1.6 & 0.42 & 240 &0.48& 0.26 \\
		5 & SDC328.199-0.588  &   15:57:59.6  -53:58:01 &-44.3 -38.7 & 2.9 2.6& 2.77 & 33220 & 0.63& 0.41 \\
		6 & SDC340.928-1.042 	& 16:55:01.4  -45:11:42  & -24.1$^*$ & 2.3& 0.73 & 640 &0.46& 0.31 \\
		7 & SDC340.969-1.020  & 16:54:57.1 -45:09:04 & -24.1 & 2.3 & 0.66 & 2630 &0.46& 0.31\\
		8 & SDC343.722-0.178  & 17:00:49.6 -42:26:05 & -28.0 -26.7 -25.6 & 2.8 2.7 2.6 & 1.42 & 5270 &0.51&0.38 \\
		9 & SDC343.735-0.110	& 17:00:32.6 -42:25:02 & -27.3 & 2.7 & 0.45 & 510 &0.51&0.38 \\
		10 & SDC343.781-0.236	& 17:01:13.0 -42:27:42 & -27.1 & 2.5 & 0.46 & 360 &0.51&0.38 \\
		11& SDC345.000-0.232  & 17:05:10.8 -41:29:08 & -27.8 & 2.9 & 2.14 & 16160 &0.43&0.30 \\
		\hline
	\end{tabular}
\end{table*}

At the high-mass end of the IMF, however, it is well known that thermal Jeans-type fragmentation cannot explain the formation of cores  more massive than a few solar masses \citep[e.g.][]{bontemps2010}. In fact, the search for compact high-mass prestellar cores with ALMA systematically reveals that reasonable candidates identified with single-dish telescopes are systematically sub-fragmented into low-mass cores \citep[e.g.][]{svoboda2019,sanhueza2019,louvet2019}. 
Therefore, the formation of massive stars requires additional physics.  One key difference between low and high-mass star formation is that self-gravitating mass reservoirs 
in massive star forming regions are larger (in mass and size) by several orders of magnitude \citep[e.g.][]{beuther2013}. 
These parsec-scale structures are often referred to as {\it clumps}.
Even though the question about how massive stars form is still very much debated, 
observations \citep[e.g.][]{peretto2006,peretto2013,schneider2010,duarte-cabral2013,urquhart2014,csengeri2017} 
and simulations \citep[e.g.][]{bonnell2004,smith2009,wang2010, vazquez-semadeni2019} converge toward a picture where massive stars 
form at the centre of globally collapsing clumps, quickly growing in mass as a result of large infall rates ($\dot{m}_{inf}\sim 10^{-3}$M$_{\odot}$/yr). 
In this picture, massive stars are said  to be {\it clump-fed}. Two questions naturally arise: 
i.) Is the clump-fed scenario the dominant mode of high-mass star formation? 
ii.) If yes, then around what core mass does the transition from core-fed to clump-fed star formation scenarios occur? 

Constraining the process through which stars gain mass is a major goal of star formation research. 
A number of studies have provided predictions regarding the mass and luminosity functions of protostars in the context of 
both the {\it core-fed} and {\it clump-fed} scenarios \citep{myers2009,myers2012,mckee2010,offner2011}. Such models have been compared to observations of nearby, {\it low-mass}, proto-clusters, 
but flux uncertainties, the limited protostellar mass range, and the low-number statistics have so far prevented model discrimination 
\citep[with the exception of the Single Isothermal Sphere model which is inconsistent with observations -][]{offner2011}. 
Mass-luminosity diagrams of protostellar cores have also been often used to constrain the time evolution of protostars. 
Theoretical evolutionary tracks have been computed, mostly assuming a fixed initial mass reservoir, i.e. {\it core-fed} \citep{bontemps1996,andre2000,andre2008b,molinari2008,duarte-cabral2013}. 

Recently, the most complete sample of clumps in the Galactic Plane has been identified using the {\it Herschel} Hi-GAL survey \citep{elia2017} 
and such tracks were used to constrain the time evolution of the Hi-GAL parsec-scale clumps. 
Even though the number statistics are impressive (with more than $10^5$ sources), 
the lack of angular resolution prevents us from probing the evolution of individual dense cores, the progenitors of single/small systems of stars. The ALMAGAL project (PIs: A. Molinari, P. Schilke, C. Battersby, P. Ho), i.e. the follow-up at high-angular resolution of $\sim1000$ Hi-GAL sources with ALMA, will provide in the near future the first large sample of individual protostellar cores. But even then, the selection bias towards the massive-star-forming clumps will likely prevent answering the question about the transition regime between high-mass and low-mass accretion scenarios.
\cite{duarte-cabral2013} have constrained the time evolution of a sample of massive-star forming cores observed 
at high angular resolution with the IRAM PdBI. In that study, the authors compute the $M_{\rm core}$ vs. $L_{\rm bol}$  and $F_{\rm{CO}}$ vs. $L_{\rm{bol}}$ evolution using simple core evolution models and use Monte Carlo simulations to populate the diagram. 
While very promising, the low-number statistics of 9 cores, and the focus on massive protostellar sources, limit the possibility of constraining core mass 
growth scenarios across all evolutionary stages and masses. 

In this paper, we present  new ArT\'eMiS/APEX 350$\mu m$ continuum observations of a sample of infrared dark clouds (IRDCs hereafter). The observed fields are large (the largest presented here is $\sim400$ \,arcmin$^2$ - see Fig.~\ref{art_sdc326_nosource}) allowing us to get a complete census of the source population within the targeted regions at relatively high angular resolution (i.e. $\sim8''$). This is the complementary approach to surveys such as Hi-GAL and ALMAGAL. The main goal of this study is to demonstrate the potential of ArT\'eMiS in determining the relative importance of {\it core-fed} and {\it clump-fed} scenarios in the context of high-mass star formation. This work will serve as a pilot study for CAFFEINE, an ArT\'eMiS large programme currently underway.

\begin{figure*}
	\includegraphics[width=17cm]{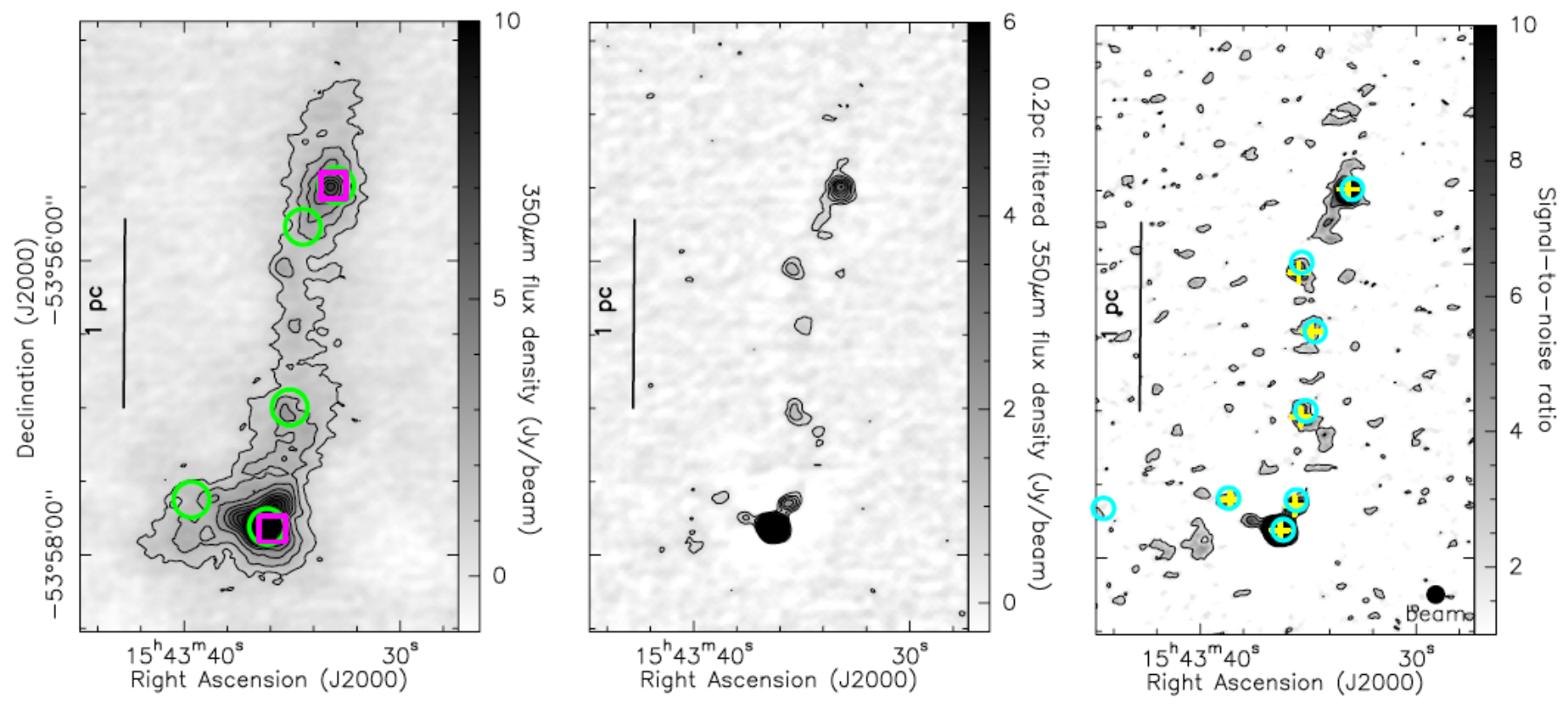}
    \caption{{\bf (left):} 350$\mu m$ ArT\'eMiS image of part of SDC326.611+0.811 field (linear scale). Contours start at 1~Jy/beam by steps of  1~Jy/beam. The green circles show the positions of {\it Herschel} clumps from \citet{elia2017}, while the magenta squares show the positions of ATLASGAL clumps from \citet{csengeri2014}. {\bf (middle):} Gaussian filtered image at 0.2~pc of the same source. Contours start at 0.5~Jy/beam in steps of  0.5~Jy/beam. {\bf (right):} Signal-to-noise ratio map of the same source. Contours start at 2 by steps of 3. The yellow crosses show the position of the ArT\'eMiS sources, while the cyan circles show the positions of 70$\mu$m sources from \citet{molinari2016}. }
    \label{sdc326p611}
\end{figure*}

\section{Targets and observations}

We targeted a total of 11 IRDCs from the \cite{peretto2009} catalogue, selected to have a H$_2$ column density peak above $10^{23}$~cm$^{-2}$ and located at very similar distances, i.e. $2.6\pm0.3$~kpc, with the exception of SDC326.796+0.386 (see Table 1).  Kinematic distances to all sources, but one, have been estimated using the MALT90 N$_2$H$^+$(1-0) data \citep{foster2013} and the \cite{reid2009,reid2014} galactic rotation model. Because SDC340.928-1.042 was not mapped by MALT90, we used the ThrUMMS $^{13}$CO(1-0) data instead \citep{barnes2015} to obtain its systemic velocity. These data show that it is part of the same molecular cloud as SDC340.969-1.020 and we therefore assigned the same velocity to both IRDCs. For all sources we adopted the near heliocentric distance as most IRDCs are located at the near distance \citep{ellsworth-bowers2013,giannetti2015}. When more than one ArT\'eMiS clumps are part of a single IRDC we checked that the velocity and corresponding distances of individual clumps are similar, which turned to be always the case. The typical uncertainty on these distances is 15\% \citep{reid2009}. Table 1 shows the main properties of the sources, including their effective radii and background-subtracted masses as estimated from the {\it Herschel} column density maps from \cite{peretto2016} within a H$_2$ column density contour level of $2\times10^{22}$ cm$^{-2}$. The 11 IRDCs have been mapped as part of 5 individual fields which in some cases (in particular for the largest of all, i.e the SDC326 field) include extra sources. For all of these extra sources, we ensured that their kinematic distances were similar to the average field distance by using the same method as described above.  

All targets were observed at 350$\mu$m with APEX and the ArT\'eMiS camera\footnote{Note that at the time of these observations the 450$\mu$m array was not available}  \citep{reveret2014,andre2016} between September 2013 and August 2014 (Onsala projects O-091.F-9301A and O-093.F-9317A). The angular resolution at 350$\mu$m with APEX is $\theta_{\rm{beam}}=8''$. Observations have been carried out with individual maps of $6'\times6'$, with a minimum of two coverages per field with different scanning angles. 
The scanning speed ranged from 20$''$/sec to $3''$/sec and the cross-scan step between consecutive scans from $6''$ to $12''$. The 350$\mu$m sky opacity (at zenith) and precipitable water vapour at the telescope were typically between 0.7 and 1.9  and between 0.35mm and 0.85mm, respectively. Absolute calibration was achieved by observing Mars as a primary calibrator, with a corresponding calibration uncertainty of $\sim$\,30\%.
Regular calibration and pointing checks were performed by taking short spiral scans toward the nearby secondary calibrators B13134, IRAS 16293, and G5.89 every $\sim$\,0.5-1.0\,h. 
The pointing accuracy was within $\sim$\,3$''$.  Data reduction was performed using the APIS pipeline running in IDL\footnote{ \url{http://www.apex-telescope.org/instruments/pi/artemis/data_reduction/}}. The ArT\'eMiS images can be seen in Fig.~\ref{art_sdc326_nosource} and in  Appendix A.

\section{Compact source identification}

In order to identify compact sources in all our fields, we first convolved all ArT\'eMIS images with a Gaussian of FWHM of 0.2~pc ($\sim$\,15$''$ at 2.6~kpc), and subtracted that convolved image from the original image. By doing so, we filter our ArT\'eMiS images from emission on spatial scales $\ge 0.2$~pc, and the comparison between sources becomes independent of their background properties. We then identify compact sources  using dendrograms  \citep[e.g.][]{rosolowsky2008,peretto2009} on signal-to-noise ratio  (SNR) maps (see Fig.~\ref{sdc326p611}). For that purpose we computed noise maps, $\sigma_{map}$, from the ArT\'eMiS  weight maps, $\omega_{map}$ (proportional to the integration time at every position in the map), and a noise calibration, $\sigma_{ref}$, estimated on an emission-free area of the filtered ArT\'eMiS maps following:  $\sigma_{map}=\sigma_{ref}\sqrt{(\omega_{ref}/\omega_{map})}$, where $\omega_{ref}$ is the average weight estimated in the same region as $\sigma_{ref}$. The calibration  $\sigma_{ref}$ is computed on the Gaussian filtered images (see Table 1 for median rms noise values).  Our dendrogram source identification uses a starting level of $2\sigma_{map}$, a step of $3\sigma_{map}$ (i.e. all sources must have a minimum SNR peak of 5), and a minimum source solid angle of 50\% of the beam solid angle which translates into a minimum effective diameter of $\sim$\,5.6$''$ ($\sim$\,0.07~pc at a distance of 2.6~kpc), i.e. 70\% of the beam FWHM. The {\it leaves} of the dendrogram (i.e. structures that exhibit no further fragmentation within the boundaries set by the input parameters of the extraction) are then used as masks in the filtered ArT\'eMiS images to measure the peak flux density of every source. In the context of the present study, this is the only parameter we are interested in (see Sec.~5.2). As it can be seen in, e.g., Fig.~\ref{art_sdc326_nosource}, the noise in the image is non-uniform, and increases towards the edge of the image. In order to reduce the potential bias in the source detection created by a non-uniform noise, we defined, by hand and for each field, a mask that cuts out the noisy edges. In the following we only consider the sources that fall within this mask. In total, across all fields, we detect 203 compact ArT\'eMiS sources. Table 2 provides information on individual sources, and individual cutout images of each source can be found in Appendix C.  Note that the source extraction parameters used in this paper are rather conservative and as a result faint sources might remain unidentified. However, the non-detection of such sources does not affect any of the results discussed here.

\begin{table*}
	\centering
	\caption{Properties of the first 10 ArT\'eMiS sources identified in the SDC326 field. 1$^{\rm{st}}$ col.: source ID number; 2$^{\rm{nd}}$ col.: galactic longitude; 3$^{\rm{rd}}$ col.: galactic latitude; 4$^{\rm{th}}$ col: Original ArT\'eMiS peak flux density (i.e. non-filtered) and associated uncertainties; 5$^{\rm{th}}$ col.: Filtered ArT\'eMiS peak flux density and associated uncertainties; 6$^{\rm{th}}$ col.: Dust temperature estimated on 0.1pc scale (see Sec.~6.1) and associated uncertainties; 7$^{\rm{th}}$ col.: Gas mass estimated on 0.1pc scale (see Sec.~6.1) and associated uncertainties; 8$^{\rm{th}}$: internal luminosity and associated uncertainties. If a value is given it means that the ArT\'eMiS source has a {\it Herschel} 70$\mu$m source from \citet{molinari2016} associated to it;  9$^{\rm{th}}$ col.: Is there a {\it Herschel} clump from \citet{elia2017} associated to it? 'y' for yes, 'n' for no; 10$^{\rm{th}}$ col.: Is there an ATLASGAL source from \citet{csengeri2014} associated to it? 'y' for yes, 'n' for no; 11$^{\rm{th}}$ col.: Can we visually identify a mid-infrared (70$\mu$m and/or 8$\mu$m peak) peak on the individual cutout images in Appendix C? 'y' for yes, 'n' for no.  The full table can be found online.}
	\label{tab:source_stats}
	\begin{tabular}{ccccccccccc} 
		\hline
		 ID $\#$  &  $l$ & $b$ &  $S_{\nu}^{\rm{pk}}$  & $S_{\nu}^{\rm{pk}}$[filt] &  T$_{\rm{dust}}$[0.1pc] & M$_{\rm{gas}}$[0.1pc] & L$_{\rm{int}}$ & H clump? &A clump? & mid-IR?\\
	                        & (degree) & (degree)  & (Jy/beam) & (Jy/beam) & (K) & (M$_{\odot}$) & ($\times10^3$L$_{\odot}$)&&&\\
				\hline
				&   &   &  & &SDC326 &  Field &  &  & &\\
				\hline
		  1 &  326.7951    & 0.3817  & $17.4\pm5.2$ & $9.8\pm2.9$ & $35.1\pm7.0$&$3.8^{+2.9}_{-1.6}$& $1.05^{+0.26}_{-0.32}$ & y& y &y \\
		  2  & 326.6328    & 0.5204  & $3.9\pm1.2$   & $1.4\pm0.4$  &  $30.3\pm6.1$& $2.0^{+1.6}_{-0.9}$&$0.33^{+0.10}_{-0.10}$  & y& n &y\\
		  3  & 326.6336    & 0.5288 & $2.6\pm0.8$  & $1.0\pm0.3$  &  $30.4\pm6.1$&$1.4^{+1.1}_{-0.6}$& --& y&y &n \\
		  4  & 326.6577   & 0.5104 & $6.7\pm2.0$ & $2.6\pm0.8$ & $30.7\pm6.1$&$3.6^{+2.8}_{-1.6}$&$0.63^{+0.19}_{-0.19}$& y& n &y\\
		  5  & 326.6622    & 0.5200 & $24.5\pm7.4$ & $11.6\pm3.5$  &$37.1\pm7.4$&$11.0^{+8.3}_{-4.6}$&$9.41^{+2.64}_{-2.80}$& y& y&y\\
		  6 &  326.6584    & 0.5169 & $13.8\pm4.1$ &$3.4\pm1.0$  & $25.1\pm5.0$&$6.5^{+5.4}_{-3.0}$&--& y &y  &n\\
		  7 &  326.6345 &  0.5328   & $3.3\pm1.0$ &$1.6\pm0.5$  &$34.6\pm6.9$&$1.7^{+1.4}_{-0.7}$&$0.87_{-0.26}^{+0.24}$& y&y &y \\
		  8 & 326.5636 & 0.5873  & $2.5\pm0.8$ &$0.9\pm0.3$  &$22.4\pm4.5$&$2.5^{+2.1}_{-1.2}$&--&n&n &y\\
		  9 & 326.6857 & 0.4950 & $3.9\pm1.2$ &$1.9\pm0.6$  & $28.2\pm5.6$&$3.0^{+2.8}_{-1.4}$&$0.21^{+0.07}_{-0.06}$&y &y &y\\
		  10 & 326.6272 & 0.5525  & $1.9\pm0.6$ &$1.2\pm0.4$  &$35.0\pm7.0$&$1.3^{+1.1}_{-0.6}$&$0.77^{+0.25}_{-0.23}$&y& n&y \\
		\hline
	\end{tabular}
\end{table*}

\begin{table*}
	\centering
	\caption{ArT\'eMiS source association statistics with Hi-GAL and ATLASGAL sources from \citet{molinari2016}, \citet{elia2017},  \citet{csengeri2014} catalogues. Provides the number of ArT\'eMiS sources in each field, and how many are associated with at least one Hi-GAL clump, one Hi-GAL 70$\mu$m source, and one ATLASGAL source (see text). The last column provides the number of sources with no association from any of these three catalogues. The bottom line gives the summary across all fields.}
	\label{tab:source_stats}
	\begin{tabular}{cccccc} 
		\hline
		Fields & $\#$ ArT\'eMiS &  $\#$ ArT\'eMiS & $\#$ ArT\'eMiS & $\#$ ArT\'eMiS & $\#$ ArT\'eMiS   \\
	                      & sources &  with Hi-GAL clumps & with Hi-GAL 70$\mu$m  & with ATLASGAL & no association \\
				\hline
		SDC326  &  129 &  104 & 52 & 42 & 19 \\
		SDC328  &  31  & 24    & 13 & 9   & 5  \\
		SDC340  &  11  & 8    & 4 & 6  & 2   \\
		SDC343  &  13  & 12   & 11 & 9 & 1 \\
		SDC345  &  19  & 18    & 6 & 8 & 1  \\
		\hline
		ALL	     & 203  & 166 & 86 &  74 & 28 \\
		\hline
	\end{tabular}
\end{table*}

\begin{table*}
	\centering
	\caption{Hi-GAL and ATLASGAL source association statistics (taken from \citet{molinari2016}, \citet{elia2017},  \citet{csengeri2014} catalogues) with ArT\'eMiS sources. Provides the number of Hi-GAL clumps, Hi-GAL 70$\mu$m sources, and ATLASGAL sources in each field,  and how many are associated with at least one ArT\'eMiS sources. The bottom line gives the summary across all fields.}
	\label{tab:source_stats}
	\begin{tabular}{ccccccc} 
		\hline
		Fields &  $\#$ Hi-GAL clump & $\#$ Hi-GAL clumps& $\#$ Hi-GAL 70$\mu$m & $\#$ Hi-GAL 70$\mu$m & $\#$ ATLASGAL & $\#$ ATLASGAL \\
	                     & in field &  with ArT\'eMiS  & in field &  with ArT\'eMiS & in field & with ArT\'eMiS \\
				\hline
		SDC326  & 87   & 43 & 275 & 52 & 37 & 22 \\
		SDC328  &  20 & 13    & 114 & 13   & 14 &  8  \\
		SDC340  &  12  & 5    & 38 & 4  & 8   & 5  \\
		SDC343  &  12  & 8 	  & 35 & 11 & 9 & 8  \\
		SDC345  &  19  & 12    & 43& 6 &10 &6  \\
		\hline
		ALL	      &  150  & 81  &  505 &   86 &  78 & 49 \\
		\hline
	\end{tabular}
\end{table*}

\section{Associations of ArT\'eMiS sources with Hi-GAL and ATLASGAL source catalogues}

In the past 10 years, far-IR and (sub-)millimetre continuum surveys of the Galactic plane have significantly contributed to improve our knowledge of massive star formation \citep{schuller2009,molinari2010,aguirre2011,moore2015}. However, even though these surveys have been, and still are, rich sources of information regarding massive star formation studies, one key issue is the lack of high-resolution, high-sensitivity observations of the cold dust on similar angular resolution as the Herschel 70$\mu$m band ($\sim7''$ resolution) which traces the protostars' luminosities. By filling in this gap, ArT\'eMiS observations allow us to unambiguously determine the envelope mass of young protostellar objects throughout the Galactic plane.  In order to demonstrate the advancement that sensitive ArT\'esMiS observations provide over existing surveys, we here compare sub-millimetre source detections with Hi-GAL \citep{elia2017}, and ATLASGAL \citep{csengeri2014}, along with performing a {\it Herschel} 70$\mu$m source association using the \cite{molinari2016} catalogue. 
  
Association between our ArT\'eMiS sources and sources in published catalogues is performed by searching sources whose published coordinates lie within one beam of the central coordinates of the ArT\'eMiS source. We therefore used an angular separation of $8''$ when performing the 70$\mu$m association, $19''$ when performing the association with ATLASGAL sources, and {\bf $36''$} when performing the Hi-GAL clump association. The statistics of the number of sources within each field and their respective association with ArT\'eMiS sources are given in Table~3 and Table~4.
These statistics show a number of important points. First, 14\% of the ArT\'eMiS sources are newly identified sources that do not belong to any of the three catalogues we searched for. Also, about 54\% of Hi-GAL clumps and 63\% of ATLASGAL sources have an ArT\'eMiS detection associated to them.  Finally,  about 42\% of the ArT\'eMiS sources have a published 70$\mu$m source associated to them, but when looking at the individual cutouts provided in Appendix C, one realises that an extra $\sim25\%$ of sources have locally peaked 70$\mu$m or 8$\mu$m emission towards them. This means that about 67\% of the ArT\'eMiS sources are protostellar, and about 33\% are starless \citep[down to the 70$\mu$m sensitivity of Hi-GAL of $\sim$\,0.1~Jy -][]{molinari2016}.  

\begin{figure}
	\includegraphics[width=8.5cm]{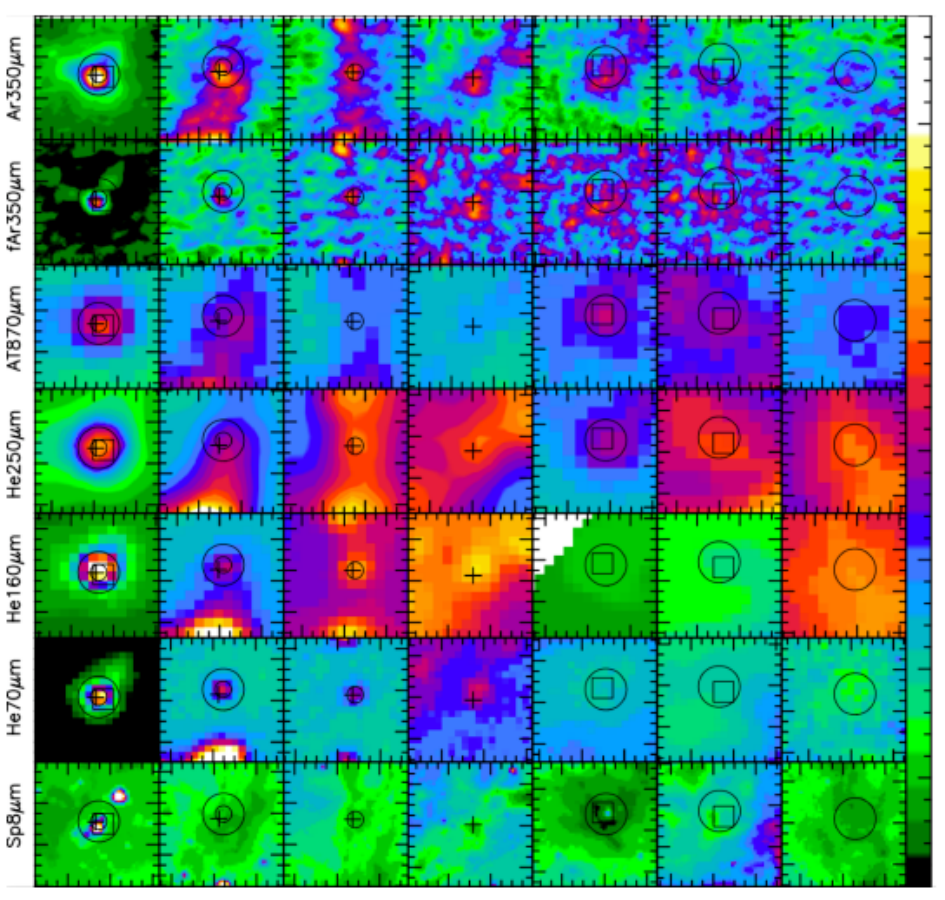}
    \caption{Examples of association types. Each column represent a different source, the first four being ArT\'eMiS detections (from left to right: SDC326 \#123, \#126, \#127, \#118), the last 3 being {\it Herschel} and/or ATLASGAL detections. Each row represents a given wavelength, from top to bottom: ArT\'eMiS 350$\mu$m; Filtered ArT\'eMiS 350$\mu$m; ATLASGAL 870$\mu$m; {\it Herschel} 250$\mu$m; {\it Herschel} 160$\mu$m; {\it Herschel} 70$\mu$m; {\it Spitzer} 8$\mu$m. The different symbols indicate if the source appear in a given catalogue: Crosses for ArT\'eMiS detections; Large circles for {\it Herschel} clumps; Small circles for {\it Herschel} 70$\mu$m detections; Squares for ATLASGAL detections.}
    \label{sourcesample}
\end{figure}

Figure~\ref{sourcesample} shows examples for each association type (see also Appendix B and C). In this figure we display 7 sources, 4 of which are detected with ArT\'eMiS, 3 which are not. We also show these 7 sources at different wavelengths in order to better understand the type of sources that we do, and do not, detect with ArT\'eMiS. On the same figure, the symbols indicate when a source has been identified in the three different source catalogues used. By looking at Figure~\ref{sourcesample}, it becomes clear that ArT\'eMiS is particularly good at identifying protostellar sources. In fact, even the source in the 4th column, which has not been identified in any of the three catalogue used, and which has therefore no {\it Herschel} 70$\mu$m entries in the \cite{molinari2016} catalogue, seems to be associated with a faint point-like 70$\mu$m emission (as mentioned above, $\sim$\,25\% of sources fall in this category of sources).  On the other hand, all three sources displayed in Fig.~\ref{sourcesample} that have not been detected with ArT\'eMiS have no 70$\mu$m emission associated to them.  The source in the 5th column is clearly seen in the ArT\'eMiS data, but falls just below our 5$\sigma$ threshold of detection. In a similar way as displayed in Fig.~\ref{sourcesample}, we looked at all individual ArT\'eMiS sources we identified to ensure the quality of the detection. Individual images of each ArT\'eMiS source can be found in Appendix C.

\section{Physical properties of ArT\'eMiS sources}

\subsection{Dust temperatures}

A key characteristic of the compact sources we identified within our ArT\'eMiS data is their dust temperature. Dust temperatures are needed to estimate the mass of these sources, but also can be used as an evolutionary tracer of the sources as dust tends to become warmer as star formation proceeds. We have here computed dust temperatures in two different ways. 

\subsubsection{Far-infrared colour temperature, T$_{\rm{col}}$}

In order to compute dust temperatures of interstellar structures one usually needs multi-wavelength observations to get a reasonable coverage of the spectral energy distribution. One problem we are facing is the lack of complementary far-IR sub-millimetre observations at similar angular resolution to our ArT\'eMiS data. {\it Herschel} observations represent the best dataset available regarding the characterisation of cold interstellar dust emission.  However, at 250$\micron$, the angular resolution of {\it Herschel} is $\sim2.5$ times worse than that of APEX at 350$\micron$. Another big difference between the two datasets is that  {\it Herschel} is sensitive to all spatial scales, and therefore recovers a lot more diffuse structures than within our ArT\'eMiS data. 
Here, we use the ratio between the 160$\mu$m and 250$\mu$m {\it Herschel} intensities at the location of each ArT\'eMiS source as a measure of the source dust temperature \citep{peretto2016}. In that respect, we first need to measure the local background intensities of each source. We do this by measuring the minimum 250$\mu$m intensity value within an annulus surrounding each of the ArT\'eMiS source, along with the corresponding 160$\mu$m intensity at the same position. The reason behind choosing the lowest 250$\mu$m intensity is that the local background around these sources can be complex, and made of other compact sources, filaments, etc... Therefore, taking, as it is often done, an average of the intensities within the annulus would result in an uncertain background intensity estimate. By focussing on the single faintest 250$\mu$m pixel, we are relatively confident to take the background at the lowest column density point within the annulus, which should provide a reasonable estimate of the local background of the compact sources we are interested in. 
 We finally subtract the local background measurements from the measured 250$\mu$m and 160$\mu$m peak intensities within the source mask. The resulting background-subtracted fluxes are used to compute the far-infrared colour dust temperatures of each  ArT\'eMiS sources \citep{peretto2016}.

\subsubsection{Internal temperature, T$_{\rm{int}}$}

For a spherical protostellar core, in the situation where dust emission is optically thin, and where the bulk of the source luminosity is in the far-infrared, one can show that flux conservation leads to the following temperature profile \citep{terebey1993}:

\begin{equation}
T_{\rm{int}} = T_0 \left(\frac{r}{r_0}\right)^{-2/(\beta+4)}\left(\frac{L_{\rm{int}}}{L_0}\right)^{1/(\beta+4)}
\end{equation}

\noindent where $\beta$ is the spectral index of the specific dust opacity law, and ($T_0,r_0, L_0$) are normalisation constants and, following   \cite{terebey1993}, are here set to (25\,K, 0.032\,pc, 520\,$L_{\odot}$), respectively.  By integrating over the volume of the core, and assuming  a given volume density profile, one can then obtain an expression for the mass-averaged temperature $\overline{T}_{\rm{int}}$. Here, we assume that $\rho\propto r^{-2}$, which leads to the following relation:

\begin{equation}
\overline{T}_{\rm{int}}=\left(\frac{\beta+4}{\beta+2}\right)T_{\rm{int}} 
\end{equation}

Given the luminosity of the source one can then compute the average dust temperature within a given radius $r$. In order to compute the bolometric luminosities of ArT\'eMiS sources  we exploit their tight relationship with 70$\mu$m fluxes \citep{dunham2008,ragan2012,elia2017}. Here, each ArT\'eMiS source has been checked against the \cite{molinari2016} 70$\mu$m source catalogue (see Sec.~4) and their corresponding 70$\mu$m fluxes come from the same catalogue. Then, we convert fluxes into luminosities using the following relation \citep{elia2017}:

\begin{equation}
L_{\rm{int}}=25.6 \,\left(\frac{F_{70\mu m}}{10\rm{Jy}}\right)\left( \frac{d}{1\rm{kpc}} \right)^2  L_{\odot}
\end{equation}

\noindent where $F_{70\mu m}$ is the 70$\mu$m flux of the ArT\'eMiS source in Jy. This relation is very similar to that obtained for low-mass protostellar objects by \cite{dunham2008}:

\begin{equation}
L_{\rm{int}}=20.0 \,\left(\frac{F_{70\mu m}}{10\rm{Jy}}\right)^{0.94}\left( \frac{d}{1\rm{kpc}} \right)^{1.88}  L_{\odot}
\end{equation}

\noindent  Since we are using the same {\it Herschel} datasets as in \cite{elia2017}, we here use the former relationship. Note that these authors have identified a third relation between $L_{\rm{int}}$ and $F_{70\mu m}$ for sources that do not have a known {\it Spitzer} or {\it WISE} mid-infrared at 24$\mu$m or 21$\mu$m respectively. However, for simplicity we here only use Eq.~(3), the dependence of $T_{\rm{int}}$ on $L_{\rm{int}}$ is in any case very shallow. Finally, by plugging in the corresponding luminosities in equation (1), and by setting $\beta=2$ (e.g. Hildebrand 1983), we can obtain $\overline{T}_{\rm{int}}$ for every ArT\'eMiS protostellar source.

\begin{figure}
	\includegraphics[width=8cm]{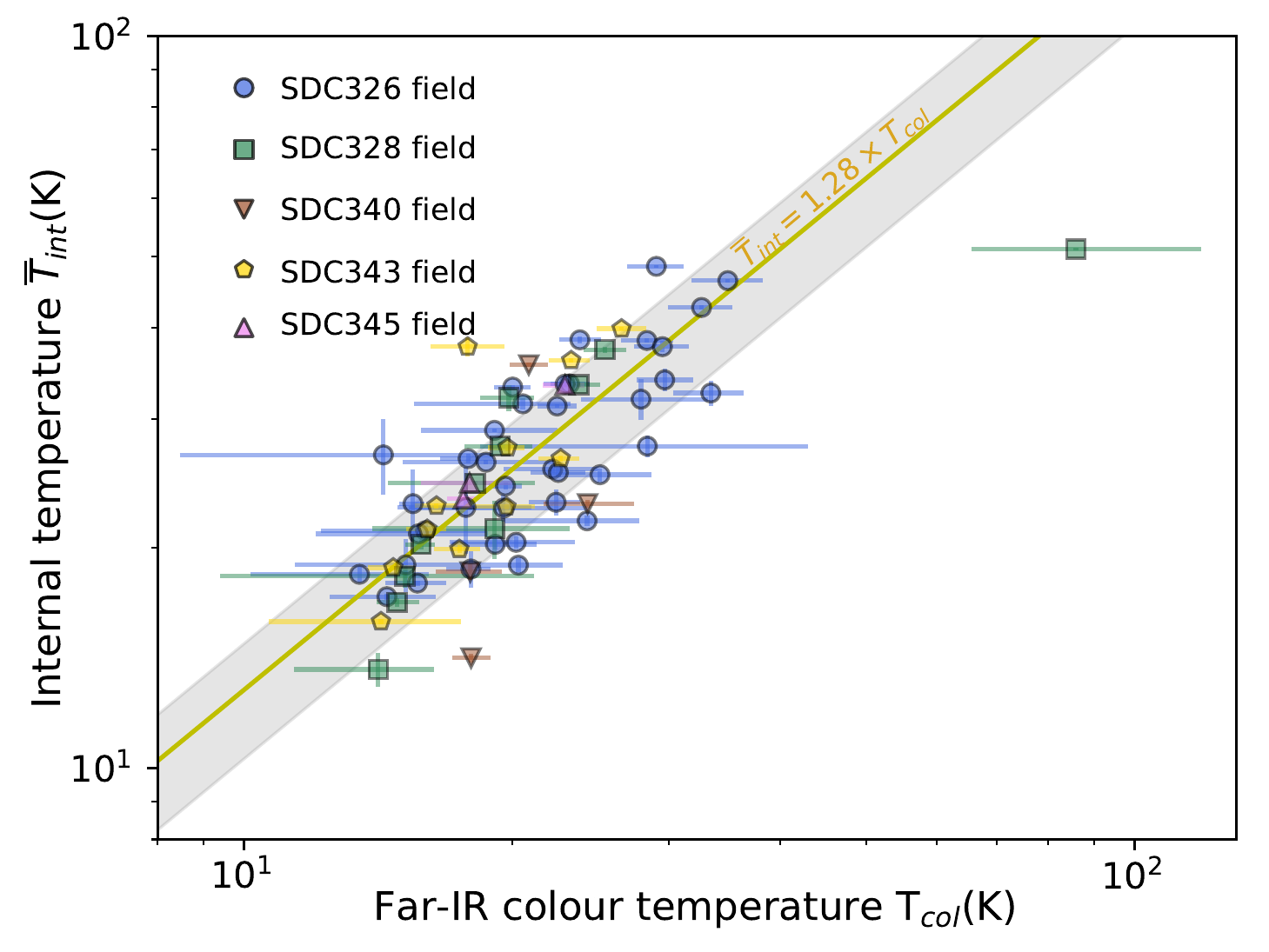}
    \caption{Far-infrared colour temperature obtained from the ratio of the {\it Herschel} 160$\mu$m to 250$\mu$m fluxes, versus the internal temperature obtained from the {\it Herschel} 70$\mu$m flux for each protostellar ArT\'eMiS source. The solid yellow line represents the median value of the $\overline{T}_{\rm{int}}/T_{col}$ ratio. The grey shaded area shows the 16$^{\rm{th}}$ to 84$^{\rm{th}}$ percentile range. }
    \label{tdustcomp}
\end{figure}

\subsubsection{Comparison between T$_{\rm{col}}$ and $\overline{T}_{\rm{int}}$}

Our estimates of  T$_{\rm{col}}$ and $\overline{T}_{\rm{int}}$ use independent {\it Herschel} data, and make use of different sets of assumptions  to compute the same quantity, i.e the dust temperature of ArT\'eMiS sources. In order to decide which of these two sets of temperatures is the most appropriate to use, we plotted them against each other (see Fig.~\ref{tdustcomp}). This can only be done for ArT\'eMiS sources with an associated 70$\mu$m source. For the purpose of making Fig.~\ref{tdustcomp},  $\overline{T}_{\rm{int}}$ has been here estimated within a radius equivalent to the {\it Herschel} 250$\mu$m beam (i.e. 0.23pc at 2.6\,kpc distance) so that the comparison remains valid. Uncertainties have been estimated by using Monte Carlo propagation. Uncertainties for $\overline{T}_{\rm{int}}$ are much lower as a result of its shallow dependency on $F_{70\mu m}$. One can see that the two sets of values are well correlated to each other, with a median ratio $\overline{T}_{\rm{int}} / T_{\rm{col}} =1.28^{+0.20}_{-0.25}$.
This  shows that, for most of the points in Fig.~\ref{tdustcomp}, the far-IR colour temperature is lower by $\sim28\%$ compared to its internal temperature counterpart. Interestingly, \cite{peretto2016} showed that far-IR colour temperature were also lower by $\sim20\%$ on average compared to dust temperatures estimated from a 4-point spectral energy distribution fit of the {\it Herschel} data. It is also worth noting that $\overline{T}_{\rm{int}}$ provides an upper limit to the temperature of compact sources as its calculation assumes optically thin emission and a spherically symmetric density profile that peaks at the location of the 70$\mu$m bright protostar. Deviations from these assumptions would lead to lower mass-averaged temperatures. As a consequence, in the remaining of the analysis, the quoted temperatures are computed using:
\begin{equation}
T_{\rm{dust}}= 1.2(\pm 0.2)\,T_{\rm{col}}
\end{equation}

 \noindent with the exception of the sources that have $T_{col} >  \overline{T}_{\rm{int}}$, for which we used $\overline{T}_{\rm{int}}$. Using Eq. (5) allows us to compute dust temperatures consistently for all ArT\'eMiS sources, something that the use of  $\overline{T}_{\rm{int}}$ would not allow us to do as it requires the detection of a 70\,$\mu$m source. Finally, note that these temperatures are estimated on the scale of the {\it Herschel} 250$\mu$m beam, i.e. 0.23\,pc at 2.6\,kpc distance, which is slightly more than twice larger than the ArT\'eMiS beam itself. According to Eq. (1), this can lead to a systematic underestimate of dust temperatures of $\sim$\,30\% for protostellar sources. The impact of this important systematic uncertainty on temperature is discussed in Section~6.

\subsection{Masses}

The mass of each ArT\'eMiS source is estimated assuming optically thin dust emission, uniform dust properties (temperature and dust emissivity) along the line of sight, and uniform dust-to-gas mass ratio. With these assumptions, the source mass is given by:

\begin{equation}
M_{\rm{gas}} = \frac{d^2 F_{\nu}}{R_{d2g}\kappa_{\nu} B_{\nu}(T_d)}
\end{equation}

\noindent where $d$ is the distance to the source, $F_{\nu}$ is the source flux, $R_{d2g}$ is the dust-to-gas mass ratio, $\kappa_{\nu}$ is the specific dust opacity at frequency $\nu$, and $B_{\nu}(T_d)$ is the Planck function at the same frequency and dust temperature $T_{d}$. Here, we used  $R_{d2g}=0.01$ and $\kappa_{\lambda}=4.44\left(\frac{\lambda}{450 \rm{\mu m}}\right)^{-2}$ cm$^2$/g \citep[e.g.][]{hildebrand1983,konyves2015}. Regarding distances, for each field we used the average distance of the individual clumps lying within them, with the exception of SDC326.796+0.386 which has been excluded from the rest of this study since it is much closer than all the other sources (see Table 1). Finally, regarding the dust temperature we use  $T_{\rm{dust}}$ as defined in Sec.~5.1.3. As far as uncertainties are concerned, we used 30\%, 15\%, and 20\% uncertainty for $F_{\nu}$, $d$, and $T_{\rm{dust}}$, respectively, that we propagated in Eq.~(6) using Monte Carlo uncertainty propagation.

The dendrogram analysis done here provides boundaries for every {\it leaf} identified in the ArT\'eMiS images. While we can use these to define the physical boundaries of compact sources, it is not clear if such an approach is the best. First, in some cases, especially for starless sources, these boundaries seem to encompass sub-structures that just fail to pass the detection criterion (i.e. local minimum to local maximum amplitude larger than 3$\sigma_{map}$). Also, nearly all high angular resolution ($\le 1''$) observations of similar sources show sub-fragmentation \citep[e.g.][]{svoboda2019,sanhueza2019,louvet2019} casting doubts on the true physical meaning of the identified ArT\'eMiS compact sources. Our approach here is more generic: we compute the mass within the ArT\'eMiS beam solid angle at the location of the peak flux density of every identified {\it leaf}. Because the sources analysed here  are all within a very narrow range of distances (see Table 1), the proposed approach provides a measure of compact source masses within a comparable physical diameter of $\sim$\,0.10$\pm0.01$~pc.

\section{The mass-temperature-luminosity diagram}

\subsection{The ArT\'eMiS view}
\begin{figure}
	\includegraphics[width=9cm]{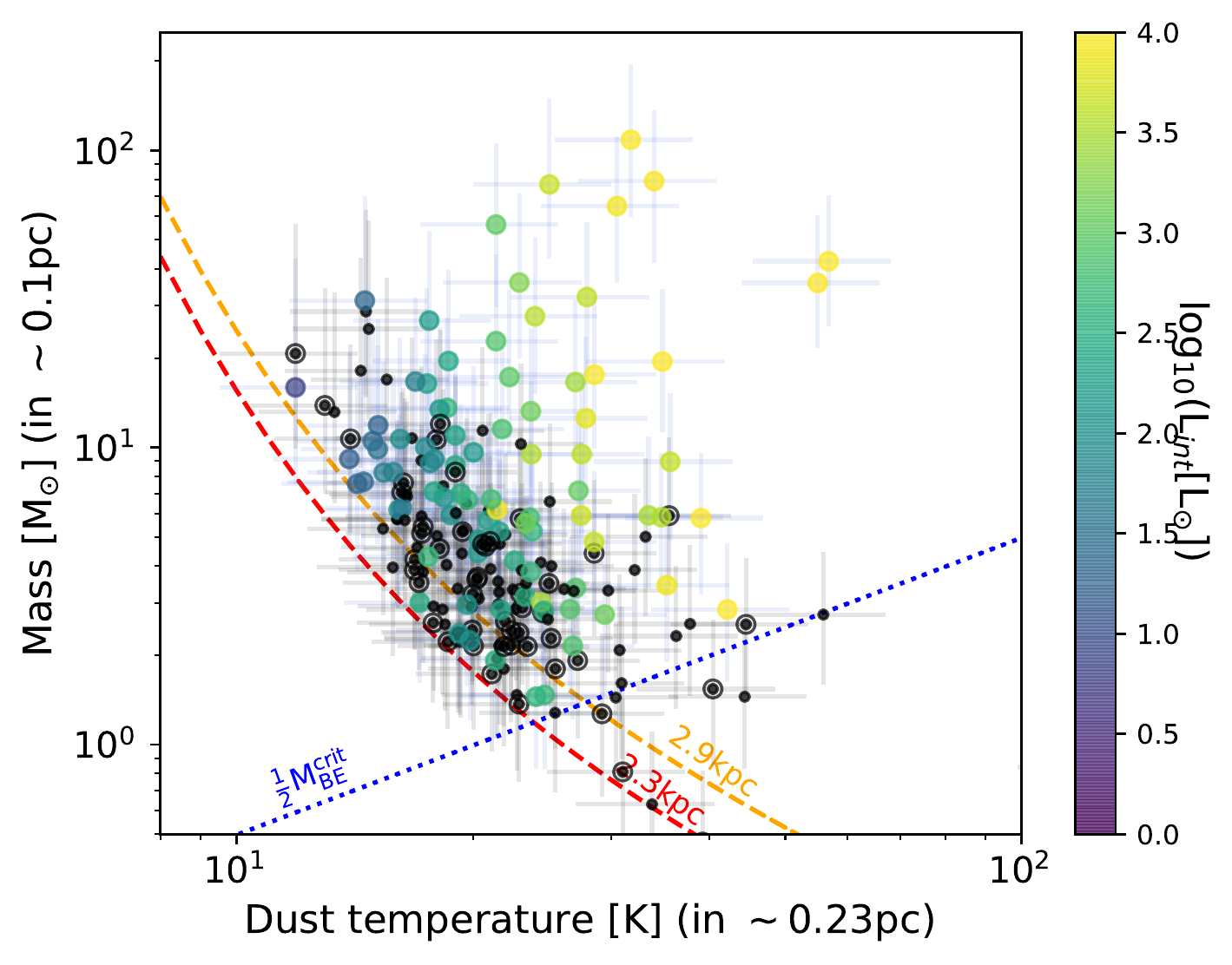}
    \caption{Mass versus dust temperature for all ArT\'eMiS sources. The temperatures are here estimated on the {\it Herschel} 250$\mu$m beam size (i.e. 0.23pc), while the ArT\'eMiS flux used to compute the mass is estimated on the ArT\'eMiS beam size (i.e. 0.1pc). The symbols are coded by the sources' internal luminosities. Sources with no 70$\mu$m association from the \citet{molinari2016} catalogue are represented as black symbols. Amongst these, those that do not display any local 70$\mu$m or 8$\mu$m peaks on the individual images presented in Appendix C are represented as filled circles. Those that do present a visually-identified mid-infrared peak  in these individual images have in addition a larger empty circular symbol. The dashed lines show the mass sensitivity limits at the two extreme distances of the sources in our sample. The blue dotted line gives half the thermal critical Bonnor Ebert mass for a core radius of 0.05~pc as a function of temperature.}
    \label{mass_temp}
\end{figure}

As protostars evolve with time, the temperature, luminosity, and mass of their envelopes change. The accretion history of these protostellar envelopes will define what their tracks will be on a mass vs. dust temperature diagram. Large statistical samples of protostellar sources within star-forming regions can therefore help constraining the accretion histories of these objects.  In Figure~\ref{mass_temp} we show the mass vs. dust temperature diagram for all identified ArT\'eMiS sources with masses estimated using the temperatures given by Eq.~(5) and the ArT\'eMiS peak flux density. On the same figure we have added the mass sensitivity limits for the minimum and maximum distances of our sample. One advantage of a  mass vs. dust temperature diagrams over a more standard mass vs. luminosity one is that all sources, starless and protostellar, can easily be represented on it.

Figure ~\ref{mass_temp} displays a couple of important features. First, we notice the presence of warm ($T_{\rm{dust}} > 30$~K) starless sources, which might seem surprising at first. However, these sources are all located in very specific environments, that is in the direct vicinity of some of the more luminous young stellar objects we have mapped. For instance, starless sources \#14, 17, 18, and 20 in the SDC328 field have all dust temperatures larger than 30K  (including the two warmest ones displayed on Fig.~\ref{mass_temp} at 44K and 55K) and are all located within a radius of 0.6\,pc of sources \#13 and \#19. These two sources have internal luminosities of  $\sim8,500$~L$_{\odot}$ and $\sim55,000 $~L$_{\odot}$, respectively. According to Eq.(1), sources with such luminosities can warm up dust up to 30K within a radius of 0.3\,pc and 0.6\,pc. It is therefore unsurprising to find starless sources with temperatures in excess of 30K. However, it is unclear if such sources are gravitationally bound and will form stars in the future. As a reference, a Bonnor-Ebert sphere of 0.05~pc radius and 40~K gas temperature has a critical mass of $\sim4$~M$_{\odot}$ \citep{bonnor1956,ebert1955}. In Fig.~\ref{mass_temp} we added, as a blue-dotted line, the critical half-mass Bonnor-Ebert relationship for a core radius of 0.05~pc, $\frac{1}{2}M_{\rm{BE}}^{\rm{crit}}=1 \times \left(\frac{T}{20K}\right)$~M$_{\odot}$. Starless sources below that line are very likely to be unbound structures. 

\begin{figure}
	\includegraphics[width=9cm]{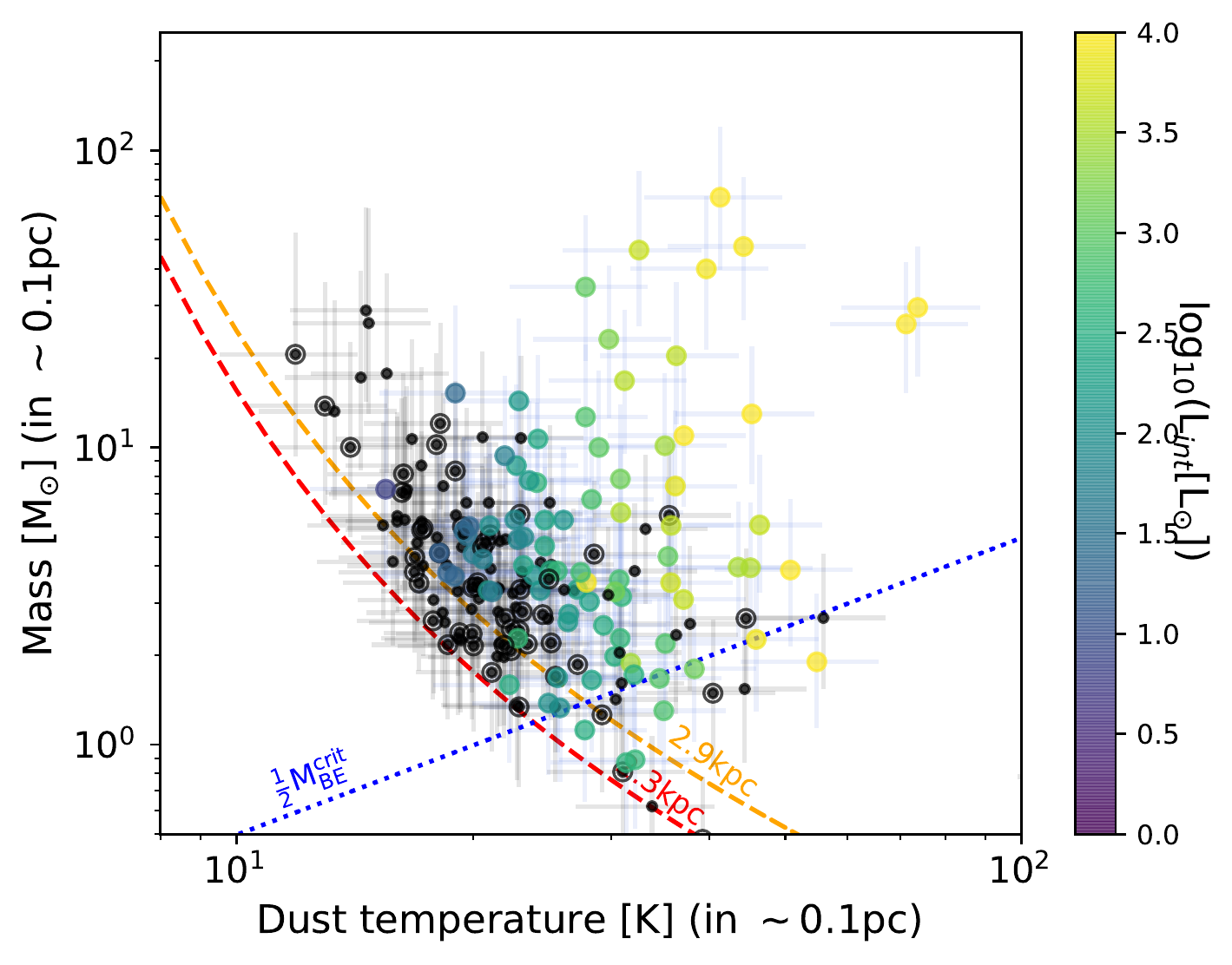}
    \caption{Same as Fig.~\ref{mass_temp} but temperatures and masses have been rescaled to a common spatial scale of 0.1pc.}
    \label{mass_temp_0p05pc}
\end{figure}

An even more important feature of Fig.~\ref{mass_temp} is the presence of massive protostellar sources with masses beyond 30 M$_{\odot}$ and the absence of equally massive starless counterparts. This is in line with the early result by \citet{motte2007} on the lack of massive pre-stellar cores in Cygnus. We also note that a luminosity gradient seems to run from the low-mass low-temperature corner to the high-mass high-temperature one. These trends, however, are very much subject to the relative temperature difference between starless and protostellar sources. As noted in Sec.~5.1.3, the flux and temperature measurements used to build Fig.~\ref{mass_temp} are inconsistent with each other since they are estimated on different spatial scales, i.e. 0.1\,pc and 0.23\,pc, respectively. Because the temperature profiles of starless and protostellar sources scales are different, this inconsistency could create artificial trends in a diagram such as that of Fig.~\ref{mass_temp}. We attempt to correct for it using Eq.~(1) for protostellar sources, and assuming that starless sources are isothermal. The mass-averaged temperature correction factor for protostellar sources is given by $(0.23/0.1)^{1/3}=1.32$ (from Eq.~(1), with $\beta=2$). The temperature of starless sources are left unchanged. The resulting corrected temperature vs mass diagram is shown in Fig.~\ref{mass_temp_0p05pc}.  On this figure, we see that the trends  observed in Fig.~\ref{mass_temp} (i.e. protostellar sources being more massive than starless ones, and the presence of a diagonal luminosity gradient) are mostly still present, albeit with slightly decreased significance. All data (temperature, mass, and luminosity) used to produce that figure is provided in Table 1.

The correction we made on the source temperatures relies on the fact that our starless/protostellar classification is robust. However, as mentioned in Sec.~4, $\sim43\%$ of the ArT\'eMiS sources that do not have a 70$\mu$m association from the \citet{molinari2016} catalogue seem to have a 70$\mu$m and/or 8$\mu$m emission peak when looking at the individual source images provided in Appendix C (in Figs.~\ref{mass_temp} and \ref{mass_temp_0p05pc} these sources are marked as black empty circular symbols with a smaller black filled symbol in them). 
Also, when observed with ALMA at high angular resolution, single-dish starless sources observed in high-mass star-forming regions systematically fragment into a set of low-mass protostellar cores  \citep[e.g.][]{svoboda2019}. This shows that the classification of sources as starless based 
on single-dish continuum observations (e.g. with {\it Herschel})  should be viewed with caution in these regions. The net impact of wrongly  classifying a protostellar source as starless would be to underestimate its temperature and therefore overestimate its mass. In other words, the trends mentioned earlier can only be strengthened by correcting for such misclassifications. This is particularly true if the handful of cold massive sources above 10\,$M_{\odot}$ would turn out to be protostellar (as Fig.~\ref{mass_temp_0p05pc} shows it is likely to be the case for at least three of these sources).

Finally, we also note that the relationship provided by Eq.~(5), even though established on protostellar sources only, has been applied to all sources, including starless ones. This seems to be the most appropriate approach since the ratio $\overline{T}_{\rm{int}}$ over $T_{\rm{col}}$ does not appear to be a function of the internal luminosity (see Fig.~\ref{tdustcomp}). However, for completeness, we do show in Appendix D a version of the mass versus temperature diagram in which we used $T_{\rm{dust}}=T_{\rm{col}}$ for the starless sources while applying the same correction factors for the protosellar sources.

Given the relatively low number of sources in our sample, the trends mentioned above are rather speculative. Nevertheless, it remains interesting to determine whether or not one can recover these trends with simple models that mimic both core-fed and clump-fed accretion scenarios.

\subsection{Accretion models}

Following \cite{bontemps1996}, \citet{andre2008b}, and \cite{duarte-cabral2013}, we built a simple accretion model that is aimed at reproducing the evolution of a protostellar core as the central protostar grows in mass. The set of equations that describes the mass growth of a protostar, and the parallel mass evolution of the core, is:

\begin{equation}
\frac{dm_*}{dt}=\dot{m}_*
\end{equation}

\begin{equation}
\frac{dm_{\rm{core}}}{dt}=-\dot{m}_*+\dot{m}_{\rm{clump}}
\end{equation}

\begin{equation}
\dot{m}_*=\epsilon_{cs}\frac{m_{\rm{core}}}{\tau_{\rm{core}}}
\end{equation}

\begin{equation}
\dot{m}_{\rm{clump}}=\epsilon_{cc}\frac{m_{\rm{clump}}}{\tau_{\rm{clump}}}
\end{equation}

\begin{figure}
	\includegraphics[width=9cm]{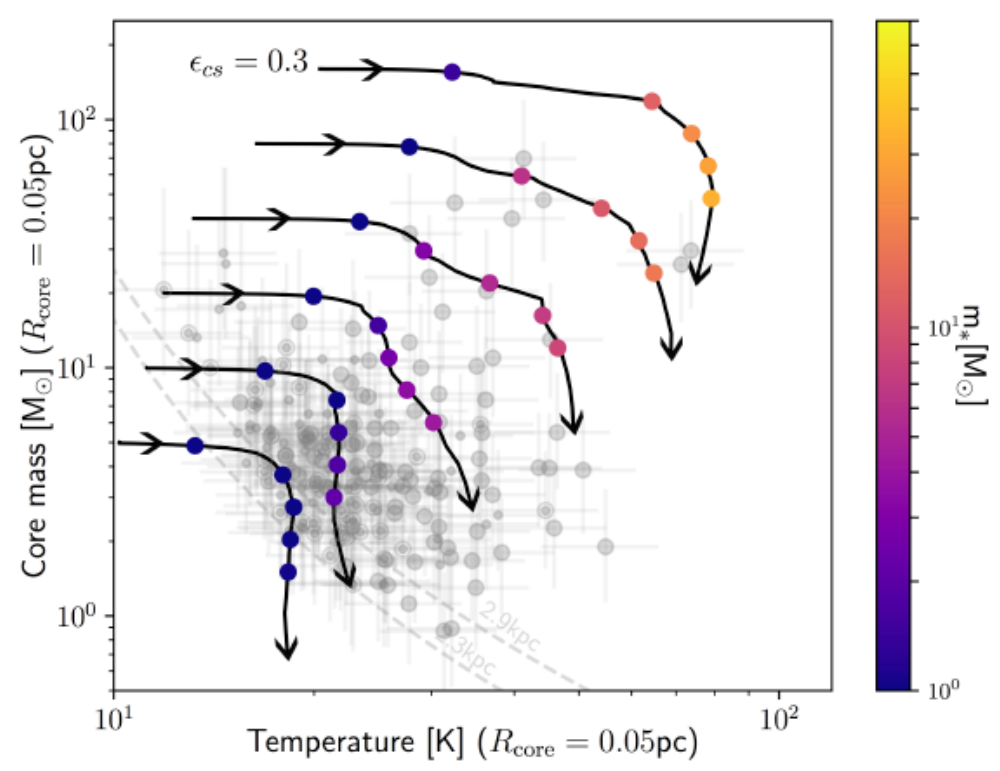}
    \caption{Core-fed models. Each track has been computed for a different initial core mass, from bottom to top ${m}_{\rm{core}}(t=0)= [5,10,20,40,80,160]$~M$_{\odot}$. The coloured symbols represent the position of the cores at times $t=[9\times10^3,9\times10^4,1.8\times10^5,2.7\times10^5,3.6\times10^5]$~yr. The colour codes the stellar mass at these times as displayed by the colour bar. The background grey symbols are those presented in Fig.~\ref{mass_temp_0p05pc}. Note that sources with $M_{\rm{gas}} < \frac{1}{2}M_{\rm{BE}}^{\rm{crit}}$ have been removed. }
    \label{corefed}
\end{figure}

\noindent where $m_*$ is the mass of the protostar, $m_{\rm{core}}$ is the mass of the core, $\dot{m}_*$ is the mass accretion rate of the protostar, $\dot{m}_{\rm{clump}}$ is the mass accretion rate of the core from the clump, $\tau_{\rm{core}}$ is the characteristic star formation timescale on core scale, $\tau_{\rm{clump}}$ is the characteristic star formation timescale on clump scale, $\epsilon_{cs}$ is the star formation efficiency from core to star (the fraction of the core mass that is being accreted onto the protostar), and finally $\epsilon_{cc}$ is the core formation efficiency from clump to core (the fraction of the clump mass that ends up in a core). 
In the context of this set of equations, core-fed scenarios differentiate themselves from clump-fed ones by having $\dot{m}_{\rm{clump}}=0$. This is the framework \cite{duarte-cabral2013} worked in. The clump-fed models, on the other hand, are presented here for the first time. In the following, we explore both type of scenarios.

\begin{figure}
	\includegraphics[width=9.cm]{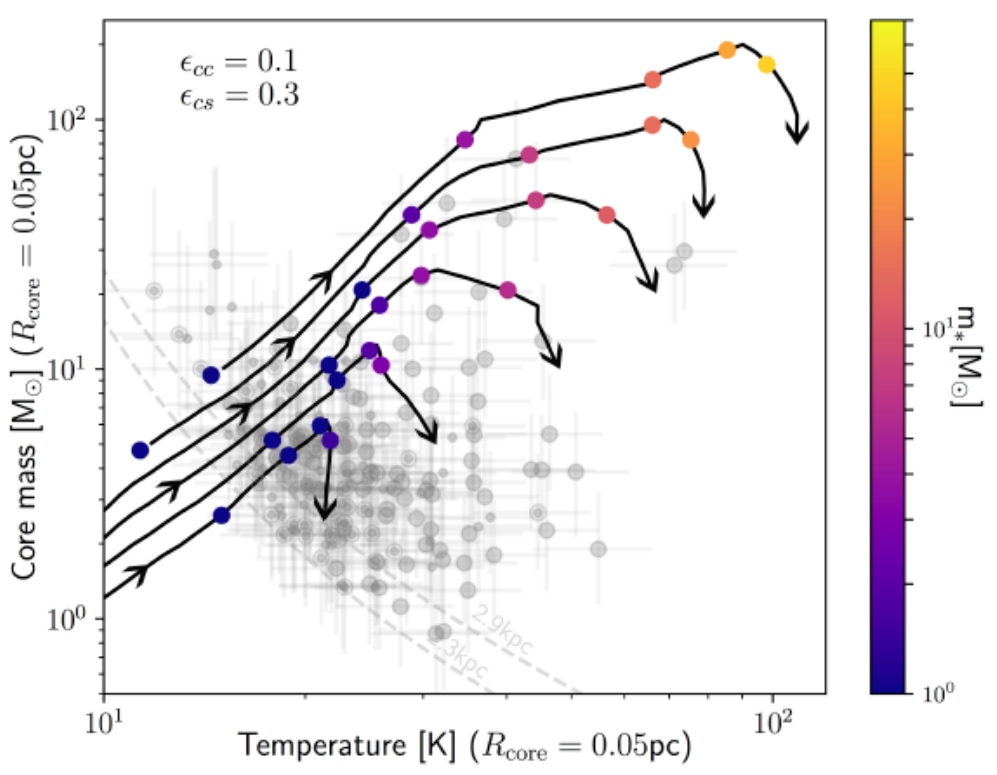}
    \caption{Clump-fed models. Each track has been computed for a different clump mass, from bottom to top $m_{\rm{clump}}=[100, 200, 400, 800, 1600, 3200]$~M$_{\odot}$.The coloured symbols represent the position of the cores at times $t=[3\times10^4,3\times10^5,6\times10^5,9\times10^5,1.2\times10^6]$~yr. The colour codes the stellar mass at these times as displayed by the colour bar. The background grey symbols are those presented in Fig.~\ref{mass_temp_0p05pc}. Note that sources with $M_{\rm{gas}} < \frac{1}{2}M_{\rm{BE}}^{\rm{crit}}$ have been removed.}
    \label{clumpfed}
\end{figure}

Equations 7 to 10 provide a description of the mass evolution of both the protostar and the surrounding core. However, in order to produce a mass vs temperature diagram one needs to compute, in parallel to the mass evolution, the evolution of the luminosity of the system.  To do this, we used the protostellar evolutionary tracks from \cite{hosokawa2009}. These are well adapted to the formation of massive stars. These tracks provide, for a given mass accretion rate and given protostar mass, the total luminosity of the system that includes both accretion luminosity and stellar luminosity. At each time step of our numerical integration of Eq. (7) to (10), we linearly interpolate the luminosity between the closest tracks. Finally, using Equations (1), (2) and (5) one can then compute the theoretical equivalent of Fig.~\ref{mass_temp_0p05pc}.

In the context of core-fed scenarios, cores refer to the fixed-mass reservoir of individual protostars. In nearby low-mass star-forming regions these cores have typical sizes ranging from 0.01pc to 0.1pc \citep[e.g.][]{konyves2015,konyves2020}. These can be understood as the typical sizes of the gravitational potential well's local minima, decoupled from their larger-scale surroundings. In the context of clump-fed scenarios, these cores are located within a larger-scale minimum defined by the presence of a surrounding parsec-scale clump that continuously feeds the cores with more mass. While Eqs (7) to (10) do not explicitly refer to any size-scale, the calculation of the mass-averaged temperature, Eq. (1), does require setting a characteristic core scale.
Here, we are limited by the spatial resolution of the ArT\'eMiS observations, i.e. $\sim0.1$\,pc at the distance of the observed regions. Hence, in the following models, we use $R_{\rm{core}}=0.05$~pc.

Figure~\ref{corefed} shows a set of models with $\dot{m}_{\rm{clump}}=0$ (effectively core-fed models), and for 6 different initial (prestellar) core masses, ${m}_{\rm{core}}(t=0)= [5,10,20,40,80,160]$~M$_{\odot}$, all with a radius $R_{\rm{core}}=0.05$~pc. As suggested by  \cite{duarte-cabral2013}  we set $\tau_{\rm{core}}=3\times10^5$~yr for all sources. Note that the exact value used for this timescale does not change the shape of the modelled tracks, a shorter timescale would only make the evolution faster. We also set $\epsilon_{\rm{cs}}=0.3$, lower than the value of 0.5 used in \cite{duarte-cabral2013} to represent the fact that the modelled cores are larger. In essence, the tracks presented in Fig.~\ref{corefed} are identical to those presented in Fig.~5 of  \cite{duarte-cabral2013} (albeit the slightly different set of parameter values).  While these models cover a similar range of mass and temperature as the ArT\'eMiS sources, they require the existence of massive prestellar cores that should reside in the top left corner of the plot. For the tracks describing the evolution of the most massive stars (${m}_{\rm{core}}(t=0)= [80,160]$~M$_{\odot}$), such starless sources are not present in our ArT\'eMiS sample. But one could argue though that such core-fed models provide a good description of the data for initial core masses $m_{\rm{core}} \le 30$~M$_{\odot}$ which, according to the models, would form stars with  $m_* \le 9$~$M_{\odot}$. These same intermediate-mass tracks also explain the presence of luminous objects (i.e. $L_{\rm{int}} \ge 10^3$\,L$_{\odot}$) with low associated core masses as sources that arrive at the end of their accretion phase.

Figure~\ref{clumpfed} shows a set of tracks with $\dot{m}_{\rm{clump}}\ne0$ (effectively clump-fed models) and ${m}_{\rm{clump}}$=[100, 200, 400, 800, 1600, 3200] M$_{\odot}$. They all start with the same initial  core mass $m_{\rm{core}}(t=0)$=1~M$_{\odot}$, the typical Jeans mass in dense molecular clumps. We also set $\epsilon_{cc}=0.1$, $\epsilon_{\rm{cs}}=0.3$, and $\tau_{\rm{clump}}=\tau_{\rm{core}}=1\times10^6$~yr, i.e. the clump crossing time controls the infall. This assumption remains valid as long as the time to regenerate the mass of the core, i.e. $\frac{m_{\rm{core}}}{m_{\rm{clump}}}\tau_{\rm{clump}}$, is shorter than the core freefall time. This is verified at all times in the models. We set a longer timescale for clump-fed models than for core-fed models since the gas density of clumps is necessarily lower than that of the cores embedded within them. However, as for the core-fed models, the exact value used for the timescale in the clump-fed models does not change the shape of the tracks.   Finally, the core accretion phase is stopped once $t>\tau_{\rm{clump}}$. Note that the point of this paper is not to proceed to a thorough examination of the parameter space of the proposed model but rather to evaluate if such models could generate a reasonable agreement with the observations. As we can see in Fig.~\ref{clumpfed}, these models do also cover a similar range in mass and temperature as the observations, and are able to explain the formation of the most massive stars without the need for massive starless sources. In addition, the modelled tracks evolved along the evolutionary gradients that we tentatively see in the observations. These models are therefore rather promising in the context of trying to pinpoint the physical mechanisms lying behind the mass accretion history of the most massive stars.

One could argue that the spatial resolution of the ArT\'eMiS data presented here (i.e. 0.1pc) is not enough to probe individual pre/protostellar cores, and that the ArT\'eMiS sources are  therefore likely to be sub-fragmented. While this might be true, it is also likely that the measured ArT\'eMiS flux  of each source is dominated by the brightest unresolved core lying within the ArT\'eMiS beam. In fact, there is evidence that this is indeed the case as \citet{csengeri2017} observed 8 of the most massive ArT\'eMiS sources presented here with ALMA at $\sim3''$ resolution, nearly 3 times better resolution than ArT\'eMiS and corresponding to a size scale of $\sim 8000$~AU. On that scale, the fraction of the ALMA flux locked in the brightest ALMA compact source is between 50\% to 90\% of the total flux. Also, \citet{csengeri2018} presented ALMA observations of source SDC328\#19 (one of the two warmest sources presented in, e.g., Fig.~\ref{mass_temp}, at an angular resolution of 0.17$''$ (i.e. $\sim500$AU at 2.75~kpc). There, no sub-fragmentation is observed. 

A comparison between our ArT\'eMiS observations and models on scales smaller than the ArT\'eMiS beam requires a set of extra assumptions and is therefore most uncertain. Such comparison is provided in Appendix E.

\section{Conclusions}

The key observational constraint regarding core-fed star formation is the existence of massive prestellar cores. The most massive starless sources identified here have masses of $\sim30$~M$_{\odot}$ in a 0.1pc source size, which is  3 to 4 times less massive that the most massive protostellar sources identified in the observed fields (within the same size). Taken at face value, this would suggest that the most massive ArT\'eMiS  sources we identified keep growing in mass while simultaneously feeding massive protostar(s) at their centre, and that clump-fed models describe best the formation of massive stars. Our data though does not exclude the possibility of core-fed star formation for intermediate-mass stars. Therefore, a transition regime could exist between core-fed and clump-fed star formation scenarios around $m_*=8$~M$_{\odot}$.

Most of the ArT\'eMiS sources studied here are likely to be sub-fragmented into a number of unresolved individual cores. A larger fragmentation level in our ArT\'eMiS {\it protostellar} sources, compared to the starless ones, could invalidate our former conclusion and instead favour core-fed scenarios. High-angular resolution observations on 1000~AU scale of massive 0.1pc-size sources, both starless and protostellar, have indeed revealed  sub-fragmentation \cite[e.g.][]{bontemps2010,palau2013,svoboda2019,sanhueza2019}. There is however no evidence that starless sources are less fragmented than protostellar ones, and if anything, these studies show the opposite. We already know that for 8 of the most massive sources from our sample, ALMA observations at $\sim8000$~AU resolution reveal that most of the ALMA flux comes from the brightest core \citep{csengeri2017}, and for the one source observed at $\sim500$~AU resolution, a single core is identified\citep{csengeri2018}. It is therefore likely that our conclusions remain valid even on small scales (see also Appendix E).

Another argument that seems to favour the clump-fed scenario is the the shape of the upper envelope of the  data point distribution in Fig.~\ref{mass_temp}. As it can be seen in Fig.~\ref{clumpfed}, this envelope is naturally reproduced by clump-fed tracks. Ideally, we would like to generate modelled density plots of such diagrams and compare to its observed equivalent. However, the number of sources at our disposition is currently too small to perform such an analysis. Larger number statistics would also allow us to set stronger constraints on the existence of starless sources with masses above 30\,M$_{\odot}$ and their statistical lifetimes. By mapping all observable massive star-forming regions within a 3~kpc distance radius from the Sun,  the CAFFEINE large programme on APEX with ArT\'eMiS aims at providing enough source statistics to build temperature vs mass density plots, allowing us to definitely conclude on the dominant scenario regulating the formation of massive stars and on the existence of a transition regime between core-fed and clump-fed star formation.

\section*{Acknowledgements}

We would like to thank the referee for their report that contributed to improve the quality of this paper. NP acknowledges the support of STFC consolidated grant number ST/N000706/1. DA and PP acknowledge support from FCT through the research grants UIDB/04434/2020 and UIDP/04434/2020. PP receives support from fellowship SFRH/BPD/110176/2015 funded by FCT (Portugal) and POPH/FSE (EC). ADC acknowledges the support from the Royal Society University Research Fellowship (URF/R1/191609). SB acknowledges support by the french ANR  through the project "GENESIS" (ANR-16-CE92-0035-01). Part of this work was also supported by the European Research Council under the  European Union's Seventh Framework Programme (ERC Advanced Grant Agreement  No. 291294 - "ORISTARS"). We also acknowledge the financial support of the French national  programs on stellar and ISM physics (PNPS and PCMI). 
This work is based on observations with the Atacama Pathfinder EXperiment (APEX) telescope. APEX is a collaboration between the Max Planck Institute for Radio Astronomy, the European Southern Observatory, and the Onsala Space Observatory. Swedish observations on APEX are supported through Swedish Research Council grant No 2017-00648.




\bibliographystyle{mnras}
\bibliography{references} 

\begin{thebibliography}{}
\makeatletter
\relax
\def\mn@urlcharsother{\let\do\@makeother \do\$\do\&\do\#\do\^\do\_\do\%\do\~}
\def\mn@doi{\begingroup\mn@urlcharsother \@ifnextchar [ {\mn@doi@}
  {\mn@doi@[]}}
\def\mn@doi@[#1]#2{\def\@tempa{#1}\ifx\@tempa\@empty \href
  {http://dx.doi.org/#2} {doi:#2}\else \href {http://dx.doi.org/#2} {#1}\fi
  \endgroup}
\def\mn@eprint#1#2{\mn@eprint@#1:#2::\@nil}
\def\mn@eprint@arXiv#1{\href {http://arxiv.org/abs/#1} {{\tt arXiv:#1}}}
\def\mn@eprint@dblp#1{\href {http://dblp.uni-trier.de/rec/bibtex/#1.xml}
  {dblp:#1}}
\def\mn@eprint@#1:#2:#3:#4\@nil{\def\@tempa {#1}\def\@tempb {#2}\def\@tempc
  {#3}\ifx \@tempc \@empty \let \@tempc \@tempb \let \@tempb \@tempa \fi \ifx
  \@tempb \@empty \def\@tempb {arXiv}\fi \@ifundefined
  {mn@eprint@\@tempb}{\@tempb:\@tempc}{\expandafter \expandafter \csname
  mn@eprint@\@tempb\endcsname \expandafter{\@tempc}}}

\bibitem[\protect\citeauthoryear{{Aguirre} et~al.,}{{Aguirre}
  et~al.}{2011}]{aguirre2011}
{Aguirre} J.~E.,  et~al., 2011, \mn@doi [\apjs] {10.1088/0067-0049/192/1/4},
  \href {https://ui.adsabs.harvard.edu/abs/2011ApJS..192....4A} {192, 4}

\bibitem[\protect\citeauthoryear{{Andre}, {Ward-Thompson}  \&
  {Barsony}}{{Andre} et~al.}{2000}]{andre2000}
{Andre} P.,  {Ward-Thompson} D.,   {Barsony} M.,  2000, in {Mannings} V.,
  {Boss} A.~P.,   {Russell} S.~S.,  eds, Protostars and Planets IV. p.~59
  (\mn@eprint {arXiv} {astro-ph/9903284})

\bibitem[\protect\citeauthoryear{{Andr{\'e}} et~al.,}{{Andr{\'e}}
  et~al.}{2008}]{andre2008b}
{Andr{\'e}} P.,  et~al., 2008, \mn@doi [\aap] {10.1051/0004-6361:200810957},
  \href {https://ui.adsabs.harvard.edu/abs/2008A&A...490L..27A} {490, L27}

\bibitem[\protect\citeauthoryear{{Andr{\'e}} et~al.,}{{Andr{\'e}}
  et~al.}{2010}]{andre2010}
{Andr{\'e}} P.,  et~al., 2010, \mn@doi [\aap] {10.1051/0004-6361/201014666},
  \href {http://adsabs.harvard.edu/abs/2010A%26A...518L.102A} {518, L102+}

\bibitem[\protect\citeauthoryear{{Andr{\'e}}, {Di Francesco}, {Ward-Thompson},
  {Inutsuka}, {Pudritz}  \& {Pineda}}{{Andr{\'e}} et~al.}{2014}]{andre2014}
{Andr{\'e}} P.,  {Di Francesco} J.,  {Ward-Thompson} D.,  {Inutsuka} S.~I.,
  {Pudritz} R.~E.,   {Pineda} J.~E.,  2014, in {Beuther} H.,  {Klessen} R.~S.,
  {Dullemond} C.~P.,   {Henning} T.,  eds, Protostars and Planets VI. p.~27
  (\mn@eprint {arXiv} {1312.6232}),
  \mn@doi{10.2458/azu_uapress_9780816531240-ch002}

\bibitem[\protect\citeauthoryear{{Andr{\'e}} et~al.,}{{Andr{\'e}}
  et~al.}{2016}]{andre2016}
{Andr{\'e}} P.,  et~al., 2016, \mn@doi [\aap] {10.1051/0004-6361/201628378},
  \href {https://ui.adsabs.harvard.edu/abs/2016A&A...592A..54A} {592, A54}

\bibitem[\protect\citeauthoryear{{Andr{\'e}}, {Arzoumanian}, {K{\"o}nyves},
  {Shimajiri}  \& {Palmeirim}}{{Andr{\'e}} et~al.}{2019}]{andre2019}
{Andr{\'e}} P.,  {Arzoumanian} D.,  {K{\"o}nyves} V.,  {Shimajiri} Y.,
  {Palmeirim} P.,  2019, \mn@doi [\aap] {10.1051/0004-6361/201935915}, \href
  {https://ui.adsabs.harvard.edu/abs/2019A&A...629L...4A} {629, L4}

\bibitem[\protect\citeauthoryear{{Arzoumanian} et~al.,}{{Arzoumanian}
  et~al.}{2011}]{arzoumanian2011}
{Arzoumanian} D.,  et~al., 2011, \mn@doi [\aap] {10.1051/0004-6361/201116596},
  \href {http://adsabs.harvard.edu/abs/2011A%26A...529L...6A} {529, L6+}

\bibitem[\protect\citeauthoryear{{Arzoumanian} et~al.,}{{Arzoumanian}
  et~al.}{2019}]{arzoumanian2019}
{Arzoumanian} D.,  et~al., 2019, \mn@doi [\aap] {10.1051/0004-6361/201832725},
  \href {https://ui.adsabs.harvard.edu/abs/2019A&A...621A..42A} {621, A42}

\bibitem[\protect\citeauthoryear{{Barnes}, {Muller}, {Indermuehle},
  {O'Dougherty}, {Lowe}, {Cunningham}, {Hernandez}  \& {Fuller}}{{Barnes}
  et~al.}{2015}]{barnes2015}
{Barnes} P.~J.,  {Muller} E.,  {Indermuehle} B.,  {O'Dougherty} S.~N.,  {Lowe}
  V.,  {Cunningham} M.,  {Hernandez} A.~K.,   {Fuller} G.~A.,  2015, \mn@doi
  [\apj] {10.1088/0004-637X/812/1/6}, \href
  {https://ui.adsabs.harvard.edu/abs/2015ApJ...812....6B} {812, 6}

\bibitem[\protect\citeauthoryear{{Beuther} et~al.,}{{Beuther}
  et~al.}{2013}]{beuther2013}
{Beuther} H.,  et~al., 2013, \mn@doi [\aap] {10.1051/0004-6361/201220475},
  \href {https://ui.adsabs.harvard.edu/abs/2013A&A...553A.115B} {553, A115}

\bibitem[\protect\citeauthoryear{{Beuther} et~al.,}{{Beuther}
  et~al.}{2018}]{beuther2018}
{Beuther} H.,  et~al., 2018, \mn@doi [\aap] {10.1051/0004-6361/201833021},
  \href {https://ui.adsabs.harvard.edu/abs/2018A&A...617A.100B} {617, A100}

\bibitem[\protect\citeauthoryear{{Bonnell}, {Vine}  \& {Bate}}{{Bonnell}
  et~al.}{2004}]{bonnell2004}
{Bonnell} I.~A.,  {Vine} S.~G.,   {Bate} M.~R.,  2004, \mn@doi [\mnras]
  {10.1111/j.1365-2966.2004.07543.x}, \href
  {http://adsabs.harvard.edu/abs/2004MNRAS.349..735B} {349, 735}

\bibitem[\protect\citeauthoryear{{Bonnor}}{{Bonnor}}{1956}]{bonnor1956}
{Bonnor} W.~B.,  1956, \mn@doi [\mnras] {10.1093/mnras/116.3.351}, \href
  {https://ui.adsabs.harvard.edu/abs/1956MNRAS.116..351B} {116, 351}

\bibitem[\protect\citeauthoryear{{Bontemps}, {Andre}, {Terebey}  \&
  {Cabrit}}{{Bontemps} et~al.}{1996}]{bontemps1996}
{Bontemps} S.,  {Andre} P.,  {Terebey} S.,   {Cabrit} S.,  1996, \aap, \href
  {https://ui.adsabs.harvard.edu/abs/1996A&A...311..858B} {311, 858}

\bibitem[\protect\citeauthoryear{{Bontemps}, {Motte}, {Csengeri}  \&
  {Schneider}}{{Bontemps} et~al.}{2010}]{bontemps2010}
{Bontemps} S.,  {Motte} F.,  {Csengeri} T.,   {Schneider} N.,  2010, \mn@doi
  [\aap] {10.1051/0004-6361/200913286}, \href
  {http://adsabs.harvard.edu/abs/2010A%26A...524A..18B} {524, A18}

\bibitem[\protect\citeauthoryear{{Csengeri} et~al.,}{{Csengeri}
  et~al.}{2014}]{csengeri2014}
{Csengeri} T.,  et~al., 2014, \mn@doi [\aap] {10.1051/0004-6361/201322434},
  \href {https://ui.adsabs.harvard.edu/abs/2014A&A...565A..75C} {565, A75}

\bibitem[\protect\citeauthoryear{{Csengeri} et~al.,}{{Csengeri}
  et~al.}{2017}]{csengeri2017}
{Csengeri} T.,  et~al., 2017, \mn@doi [\aap] {10.1051/0004-6361/201629754},
  \href {https://ui.adsabs.harvard.edu/abs/2017A&A...600L..10C} {600, L10}

\bibitem[\protect\citeauthoryear{{Csengeri} et~al.,}{{Csengeri}
  et~al.}{2018}]{csengeri2018}
{Csengeri} T.,  et~al., 2018, \mn@doi [\aap] {10.1051/0004-6361/201832753},
  \href {https://ui.adsabs.harvard.edu/abs/2018A&A...617A..89C} {617, A89}

\bibitem[\protect\citeauthoryear{{Duarte-Cabral}, {Bontemps}, {Motte},
  {Hennemann}, {Schneider}  \& {Andr{\'e}}}{{Duarte-Cabral}
  et~al.}{2013}]{duarte-cabral2013}
{Duarte-Cabral} A.,  {Bontemps} S.,  {Motte} F.,  {Hennemann} M.,  {Schneider}
  N.,   {Andr{\'e}} P.,  2013, \mn@doi [\aap] {10.1051/0004-6361/201321393},
  \href {https://ui.adsabs.harvard.edu/abs/2013A&A...558A.125D} {558, A125}

\bibitem[\protect\citeauthoryear{{Dunham}, {Crapsi}, {Evans}, {Bourke},
  {Huard}, {Myers}  \& {Kauffmann}}{{Dunham} et~al.}{2008}]{dunham2008}
{Dunham} M.~M.,  {Crapsi} A.,  {Evans} II N.~J.,  {Bourke} T.~L.,  {Huard}
  T.~L.,  {Myers} P.~C.,   {Kauffmann} J.,  2008, \mn@doi [\apjs]
  {10.1086/591085}, \href {http://adsabs.harvard.edu/abs/2008ApJS..179..249D}
  {179, 249}

\bibitem[\protect\citeauthoryear{{Ebert}}{{Ebert}}{1955}]{ebert1955}
{Ebert} R.,  1955, \zap, \href
  {https://ui.adsabs.harvard.edu/abs/1955ZA.....37..217E} {37, 217}

\bibitem[\protect\citeauthoryear{{Elia} et~al.,}{{Elia}
  et~al.}{2017}]{elia2017}
{Elia} D.,  et~al., 2017, \mn@doi [\mnras] {10.1093/mnras/stx1357}, \href
  {https://ui.adsabs.harvard.edu/abs/2017MNRAS.471..100E} {471, 100}

\bibitem[\protect\citeauthoryear{{Ellsworth-Bowers} et~al.,}{{Ellsworth-Bowers}
  et~al.}{2013}]{ellsworth-bowers2013}
{Ellsworth-Bowers} T.~P.,  et~al., 2013, \mn@doi [\apj]
  {10.1088/0004-637X/770/1/39}, \href
  {https://ui.adsabs.harvard.edu/abs/2013ApJ...770...39E} {770, 39}

\bibitem[\protect\citeauthoryear{{Foster} et~al.,}{{Foster}
  et~al.}{2013}]{foster2013}
{Foster} J.~B.,  et~al., 2013, \mn@doi [\pasa] {10.1017/pasa.2013.18}, \href
  {https://ui.adsabs.harvard.edu/abs/2013PASA...30...38F} {30, e038}

\bibitem[\protect\citeauthoryear{{Giannetti}, {Wyrowski}, {Leurini},
  {Urquhart}, {Csengeri}, {Menten}, {Bronfman}  \& {van der Tak}}{{Giannetti}
  et~al.}{2015}]{giannetti2015}
{Giannetti} A.,  {Wyrowski} F.,  {Leurini} S.,  {Urquhart} J.,  {Csengeri} T.,
  {Menten} K.~M.,  {Bronfman} L.,   {van der Tak} F.~F.~S.,  2015, \mn@doi
  [\aap] {10.1051/0004-6361/201526474}, \href
  {https://ui.adsabs.harvard.edu/abs/2015A&A...580L...7G} {580, L7}

\bibitem[\protect\citeauthoryear{{Hildebrand}}{{Hildebrand}}{1983}]{hildebrand1983}
{Hildebrand} R.~H.,  1983, \qjras, \href
  {http://adsabs.harvard.edu/abs/1983QJRAS..24..267H} {24, 267}

\bibitem[\protect\citeauthoryear{{Hosokawa} \& {Omukai}}{{Hosokawa} \&
  {Omukai}}{2009}]{hosokawa2009}
{Hosokawa} T.,  {Omukai} K.,  2009, \mn@doi [\apj]
  {10.1088/0004-637X/691/1/823}, \href
  {https://ui.adsabs.harvard.edu/abs/2009ApJ...691..823H} {691, 823}

\bibitem[\protect\citeauthoryear{{Inutsuka} \& {Miyama}}{{Inutsuka} \&
  {Miyama}}{1997}]{inutsuka1997}
{Inutsuka} S.-i.,  {Miyama} S.~M.,  1997, \mn@doi [\apj] {10.1086/303982},
  \href {https://ui.adsabs.harvard.edu/abs/1997ApJ...480..681I} {480, 681}

\bibitem[\protect\citeauthoryear{{K{\"o}nyves} et~al.,}{{K{\"o}nyves}
  et~al.}{2015}]{konyves2015}
{K{\"o}nyves} V.,  et~al., 2015, \mn@doi [\aap] {10.1051/0004-6361/201525861},
  \href {https://ui.adsabs.harvard.edu/abs/2015A&A...584A..91K} {584, A91}

\bibitem[\protect\citeauthoryear{{K{\"o}nyves} et~al.,}{{K{\"o}nyves}
  et~al.}{2020}]{konyves2020}
{K{\"o}nyves} V.,  et~al., 2020, \mn@doi [\aap] {10.1051/0004-6361/201834753},
  \href {https://ui.adsabs.harvard.edu/abs/2020A&A...635A..34K} {635, A34}

\bibitem[\protect\citeauthoryear{{Ladjelate} et~al.,}{{Ladjelate}
  et~al.}{2020}]{ladjelate2020}
{Ladjelate} B.,  et~al., 2020, arXiv e-prints, \href
  {https://ui.adsabs.harvard.edu/abs/2020arXiv200111036L} {p. arXiv:2001.11036}

\bibitem[\protect\citeauthoryear{{Lee}, {Hennebelle}  \& {Chabrier}}{{Lee}
  et~al.}{2017}]{lee2017}
{Lee} Y.-N.,  {Hennebelle} P.,   {Chabrier} G.,  2017, \mn@doi [\apj]
  {10.3847/1538-4357/aa898f}, \href
  {https://ui.adsabs.harvard.edu/abs/2017ApJ...847..114L} {847, 114}

\bibitem[\protect\citeauthoryear{{Louvet} et~al.,}{{Louvet}
  et~al.}{2019}]{louvet2019}
{Louvet} F.,  et~al., 2019, \mn@doi [\aap] {10.1051/0004-6361/201732282}, \href
  {https://ui.adsabs.harvard.edu/abs/2019A&A...622A..99L} {622, A99}

\bibitem[\protect\citeauthoryear{{McKee} \& {Offner}}{{McKee} \&
  {Offner}}{2010}]{mckee2010}
{McKee} C.~F.,  {Offner} S. S.~R.,  2010, \mn@doi [\apj]
  {10.1088/0004-637X/716/1/167}, \href
  {https://ui.adsabs.harvard.edu/abs/2010ApJ...716..167M} {716, 167}

\bibitem[\protect\citeauthoryear{{Men'shchikov} et~al.,}{{Men'shchikov}
  et~al.}{2010}]{menshchikov2010}
{Men'shchikov} A.,  et~al., 2010, \mn@doi [\aap] {10.1051/0004-6361/201014668},
  \href {https://ui.adsabs.harvard.edu/abs/2010A&A...518L.103M} {518, L103}

\bibitem[\protect\citeauthoryear{{Molinari}, {Pezzuto}, {Cesaroni}, {Brand},
  {Faustini}  \& {Testi}}{{Molinari} et~al.}{2008}]{molinari2008}
{Molinari} S.,  {Pezzuto} S.,  {Cesaroni} R.,  {Brand} J.,  {Faustini} F.,
  {Testi} L.,  2008, \mn@doi [\aap] {10.1051/0004-6361:20078661}, \href
  {https://ui.adsabs.harvard.edu/abs/2008A&A...481..345M} {481, 345}

\bibitem[\protect\citeauthoryear{{Molinari} et~al.,}{{Molinari}
  et~al.}{2010}]{molinari2010}
{Molinari} S.,  et~al., 2010, \mn@doi [\aap] {10.1051/0004-6361/201014659},
  \href {http://adsabs.harvard.edu/abs/2010A%26A...518L.100M} {518, L100+}

\bibitem[\protect\citeauthoryear{{Molinari} et~al.,}{{Molinari}
  et~al.}{2016}]{molinari2016}
{Molinari} S.,  et~al., 2016, \mn@doi [\aap] {10.1051/0004-6361/201526380},
  \href {https://ui.adsabs.harvard.edu/abs/2016A&A...591A.149M} {591, A149}

\bibitem[\protect\citeauthoryear{{Moore} et~al.,}{{Moore}
  et~al.}{2015}]{moore2015}
{Moore} T.~J.~T.,  et~al., 2015, \mn@doi [\mnras] {10.1093/mnras/stv1833},
  \href {https://ui.adsabs.harvard.edu/abs/2015MNRAS.453.4264M} {453, 4264}

\bibitem[\protect\citeauthoryear{{Motte}, {Andre}  \& {Neri}}{{Motte}
  et~al.}{1998}]{motte1998}
{Motte} F.,  {Andre} P.,   {Neri} R.,  1998, \aap, \href
  {http://ads.astro.puc.cl/abs/1998A%26A...336..150M} {336, 150}

\bibitem[\protect\citeauthoryear{{Motte}, {Bontemps}, {Schilke}, {}, {Menten}
  \& {Brogui{\`e}re}}{{Motte} et~al.}{2007}]{motte2007}
{Motte} F.,  {Bontemps} S.,  {Schilke} P.,  {} N.,  {Menten} K.~M.,
  {Brogui{\`e}re} D.,  2007, \mn@doi [\aap] {10.1051/0004-6361:20077843}, \href
  {http://ads.astro.puc.cl/abs/2007A%26A...476.1243M} {476, 1243}

\bibitem[\protect\citeauthoryear{{Motte} et~al.,}{{Motte}
  et~al.}{2018a}]{motte2018b}
{Motte} F.,  et~al., 2018a, \mn@doi [Nature Astronomy]
  {10.1038/s41550-018-0452-x}, \href
  {https://ui.adsabs.harvard.edu/abs/2018NatAs...2..478M} {2, 478}

\bibitem[\protect\citeauthoryear{{Motte}, {Bontemps}  \& {Louvet}}{{Motte}
  et~al.}{2018b}]{motte2018a}
{Motte} F.,  {Bontemps} S.,   {Louvet} F.,  2018b, \mn@doi [\araa]
  {10.1146/annurev-astro-091916-055235}, \href
  {https://ui.adsabs.harvard.edu/abs/2018ARA&A..56...41M} {56, 41}

\bibitem[\protect\citeauthoryear{{Myers}}{{Myers}}{2009}]{myers2009}
{Myers} P.~C.,  2009, \mn@doi [\apj] {10.1088/0004-637X/700/2/1609}, \href
  {http://adsabs.harvard.edu/abs/2009ApJ...700.1609M} {700, 1609}

\bibitem[\protect\citeauthoryear{{Myers}}{{Myers}}{2012}]{myers2012}
{Myers} P.~C.,  2012, \mn@doi [\apj] {10.1088/0004-637X/752/1/9}, \href
  {https://ui.adsabs.harvard.edu/abs/2012ApJ...752....9M} {752, 9}

\bibitem[\protect\citeauthoryear{{Offner} \& {McKee}}{{Offner} \&
  {McKee}}{2011}]{offner2011}
{Offner} S. S.~R.,  {McKee} C.~F.,  2011, \mn@doi [\apj]
  {10.1088/0004-637X/736/1/53}, \href
  {https://ui.adsabs.harvard.edu/abs/2011ApJ...736...53O} {736, 53}

\bibitem[\protect\citeauthoryear{{Palau} et~al.,}{{Palau}
  et~al.}{2013}]{palau2013}
{Palau} A.,  et~al., 2013, \mn@doi [\apj] {10.1088/0004-637X/762/2/120}, \href
  {https://ui.adsabs.harvard.edu/abs/2013ApJ...762..120P} {762, 120}

\bibitem[\protect\citeauthoryear{{Peretto} \& {Fuller}}{{Peretto} \&
  {Fuller}}{2009}]{peretto2009}
{Peretto} N.,  {Fuller} G.~A.,  2009, \mn@doi [\aap]
  {10.1051/0004-6361/200912127}, \href
  {http://adsabs.harvard.edu/abs/2009A%26A...505..405P} {505, 405}

\bibitem[\protect\citeauthoryear{{Peretto}, {Andr{\'e}}  \&
  {Belloche}}{{Peretto} et~al.}{2006}]{peretto2006}
{Peretto} N.,  {Andr{\'e}} P.,   {Belloche} A.,  2006, \mn@doi [\aap]
  {10.1051/0004-6361:20053324}, \href
  {http://ads.astro.puc.cl/abs/2006A%26A...445..979P} {445, 979}

\bibitem[\protect\citeauthoryear{{Peretto} et~al.,}{{Peretto}
  et~al.}{2013}]{peretto2013}
{Peretto} N.,  et~al., 2013, \mn@doi [\aap] {10.1051/0004-6361/201321318},
  \href {http://adsabs.harvard.edu/abs/2013A%26A...555A.112P} {555, A112}

\bibitem[\protect\citeauthoryear{{Peretto}, {Lenfestey}, {Fuller},
  {Traficante}, {Molinari}, {Thompson}  \& {Ward-Thompson}}{{Peretto}
  et~al.}{2016}]{peretto2016}
{Peretto} N.,  {Lenfestey} C.,  {Fuller} G.~A.,  {Traficante} A.,  {Molinari}
  S.,  {Thompson} M.~A.,   {Ward-Thompson} D.,  2016, \mn@doi [\aap]
  {10.1051/0004-6361/201527064}, \href
  {https://ui.adsabs.harvard.edu/abs/2016A&A...590A..72P} {590, A72}

\bibitem[\protect\citeauthoryear{{Ragan}, {Heitsch}, {Bergin}  \&
  {Wilner}}{{Ragan} et~al.}{2012}]{ragan2012}
{Ragan} S.~E.,  {Heitsch} F.,  {Bergin} E.~A.,   {Wilner} D.,  2012, \mn@doi
  [\apj] {10.1088/0004-637X/746/2/174}, \href
  {http://adsabs.harvard.edu/abs/2012ApJ...746..174R} {746, 174}

\bibitem[\protect\citeauthoryear{{Reid} et~al.,}{{Reid}
  et~al.}{2009}]{reid2009}
{Reid} M.~J.,  et~al., 2009, \mn@doi [\apj] {10.1088/0004-637X/700/1/137},
  \href {http://adsabs.harvard.edu/abs/2009ApJ...700..137R} {700, 137}

\bibitem[\protect\citeauthoryear{{Reid} et~al.,}{{Reid}
  et~al.}{2014}]{reid2014}
{Reid} M.~J.,  et~al., 2014, \mn@doi [\apj] {10.1088/0004-637X/783/2/130},
  \href {https://ui.adsabs.harvard.edu/abs/2014ApJ...783..130R} {783, 130}

\bibitem[\protect\citeauthoryear{{Rev{\'e}ret} et~al.,}{{Rev{\'e}ret}
  et~al.}{2014}]{reveret2014}
{Rev{\'e}ret} V.,  et~al., 2014, {The ArT{\'e}MiS wide-field sub-millimeter
  camera: preliminary on-sky performance at 350 microns}.
p. 915305, \mn@doi{10.1117/12.2055985}

\bibitem[\protect\citeauthoryear{{Rosolowsky}, {Pineda}, {Kauffmann}  \&
  {Goodman}}{{Rosolowsky} et~al.}{2008}]{rosolowsky2008}
{Rosolowsky} E.~W.,  {Pineda} J.~E.,  {Kauffmann} J.,   {Goodman} A.~A.,  2008,
  \mn@doi [\apj] {10.1086/587685}, \href
  {http://adsabs.harvard.edu/abs/2008ApJ...679.1338R} {679, 1338}

\bibitem[\protect\citeauthoryear{{Sanhueza} et~al.,}{{Sanhueza}
  et~al.}{2019}]{sanhueza2019}
{Sanhueza} P.,  et~al., 2019, \mn@doi [\apj] {10.3847/1538-4357/ab45e9}, \href
  {https://ui.adsabs.harvard.edu/abs/2019ApJ...886..102S} {886, 102}

\bibitem[\protect\citeauthoryear{{Schneider}, {Csengeri}, {Bontemps}, {Motte},
  {Simon}, {Hennebelle}, {Federrath}  \& {Klessen}}{{Schneider}
  et~al.}{2010}]{schneider2010}
{Schneider} N.,  {Csengeri} T.,  {Bontemps} S.,  {Motte} F.,  {Simon} R.,
  {Hennebelle} P.,  {Federrath} C.,   {Klessen} R.,  2010, \mn@doi [\aap]
  {10.1051/0004-6361/201014481}, \href
  {http://adsabs.harvard.edu/abs/2010A%26A...520A..49S} {520, A49}

\bibitem[\protect\citeauthoryear{{Schuller} et~al.,}{{Schuller}
  et~al.}{2009}]{schuller2009}
{Schuller} F.,  et~al., 2009, \mn@doi [\aap] {10.1051/0004-6361/200811568},
  \href {https://ui.adsabs.harvard.edu/abs/2009A&A...504..415S} {504, 415}

\bibitem[\protect\citeauthoryear{{Smith}, {Longmore}  \& {Bonnell}}{{Smith}
  et~al.}{2009}]{smith2009}
{Smith} R.~J.,  {Longmore} S.,   {Bonnell} I.,  2009, \mn@doi [\mnras]
  {10.1111/j.1365-2966.2009.15621.x}, \href
  {http://adsabs.harvard.edu/abs/2009MNRAS.400.1775S} {400, 1775}

\bibitem[\protect\citeauthoryear{{Svoboda} et~al.,}{{Svoboda}
  et~al.}{2019}]{svoboda2019}
{Svoboda} B.~E.,  et~al., 2019, \mn@doi [\apj] {10.3847/1538-4357/ab40ca},
  \href {https://ui.adsabs.harvard.edu/abs/2019ApJ...886...36S} {886, 36}

\bibitem[\protect\citeauthoryear{{Terebey}, {Chandler}  \& {Andre}}{{Terebey}
  et~al.}{1993}]{terebey1993}
{Terebey} S.,  {Chandler} C.~J.,   {Andre} P.,  1993, \mn@doi [\apj]
  {10.1086/173121}, \href
  {https://ui.adsabs.harvard.edu/abs/1993ApJ...414..759T} {414, 759}

\bibitem[\protect\citeauthoryear{{Urquhart} et~al.,}{{Urquhart}
  et~al.}{2014}]{urquhart2014}
{Urquhart} J.~S.,  et~al., 2014, \mn@doi [\mnras] {10.1093/mnras/stu1207},
  \href {https://ui.adsabs.harvard.edu/abs/2014MNRAS.443.1555U} {443, 1555}

\bibitem[\protect\citeauthoryear{{V{\'a}zquez-Semadeni}, {Palau},
  {Ballesteros-Paredes}, {G{\'o}mez}  \&
  {Zamora-Avil{\'e}s}}{{V{\'a}zquez-Semadeni}
  et~al.}{2019}]{vazquez-semadeni2019}
{V{\'a}zquez-Semadeni} E.,  {Palau} A.,  {Ballesteros-Paredes} J.,  {G{\'o}mez}
  G.~C.,   {Zamora-Avil{\'e}s} M.,  2019, \mn@doi [\mnras]
  {10.1093/mnras/stz2736}, \href
  {https://ui.adsabs.harvard.edu/abs/2019MNRAS.490.3061V} {490, 3061}

\bibitem[\protect\citeauthoryear{{Wang}, {Li}, {Abel}  \& {Nakamura}}{{Wang}
  et~al.}{2010}]{wang2010}
{Wang} P.,  {Li} Z.-Y.,  {Abel} T.,   {Nakamura} F.,  2010, \mn@doi [\apj]
  {10.1088/0004-637X/709/1/27}, \href
  {https://ui.adsabs.harvard.edu/abs/2010ApJ...709...27W} {709, 27}

\makeatother
\end{thebibliography}



\appendix

\clearpage

\section{ArT\'eMiS images}

In this Appendix we present the ArT\'eMiS images for the SDC328, SDC340, SDC343, and SDC345 fields.


\begin{figure}
	\includegraphics[width=9cm]{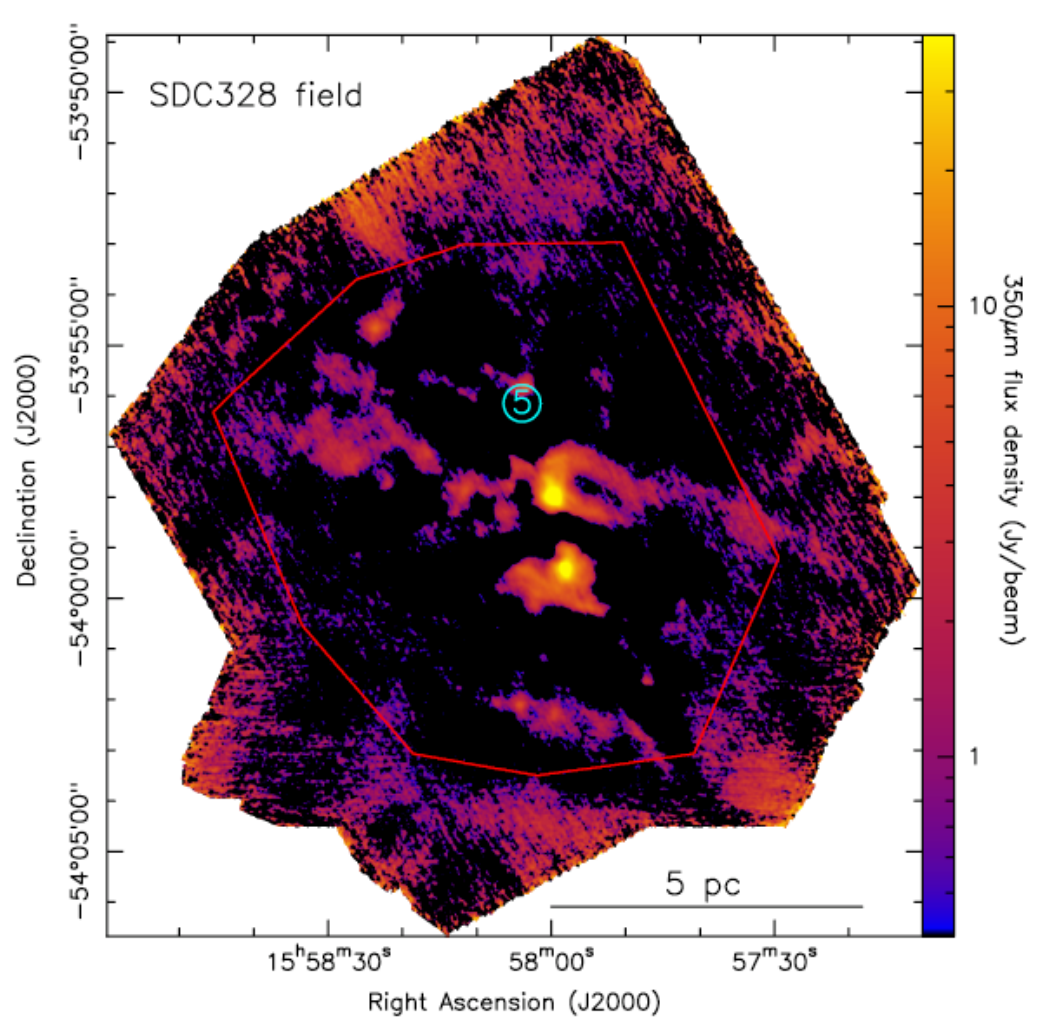}
    \caption{Same as Fig.~\ref{art_sdc326_nosource} for the SDC328 field}
    \label{fig:example_figure}
\end{figure}

\begin{figure}
	\includegraphics[width=9cm]{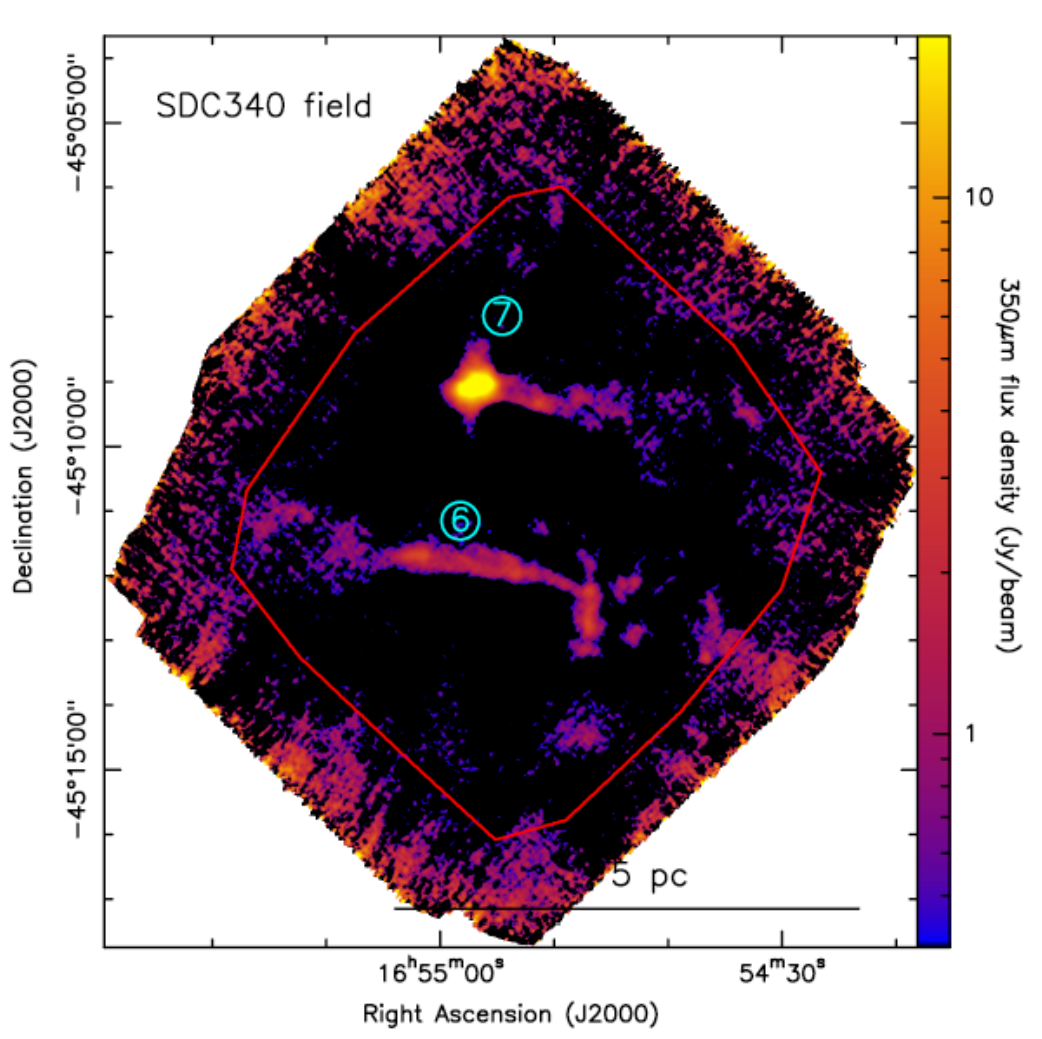}
    \caption{Same as Fig.~\ref{art_sdc326_nosource} for the SDC340 field}
    \label{fig:example_figure}
\end{figure}

\begin{figure}
	\includegraphics[width=9cm]{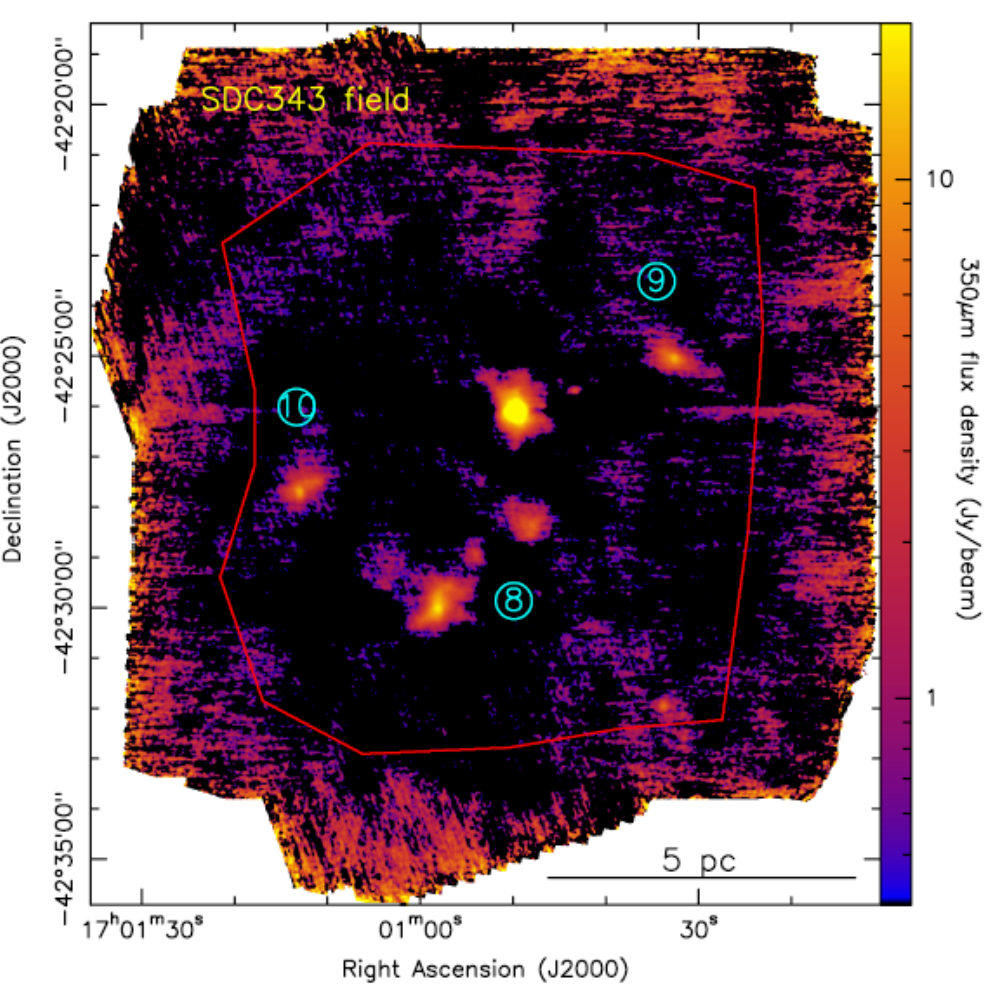}
    \caption{Same as Fig.~\ref{art_sdc326_nosource} for the SDC343 field}
    \label{fig:example_figure}
\end{figure}

\begin{figure}
	\includegraphics[width=9cm]{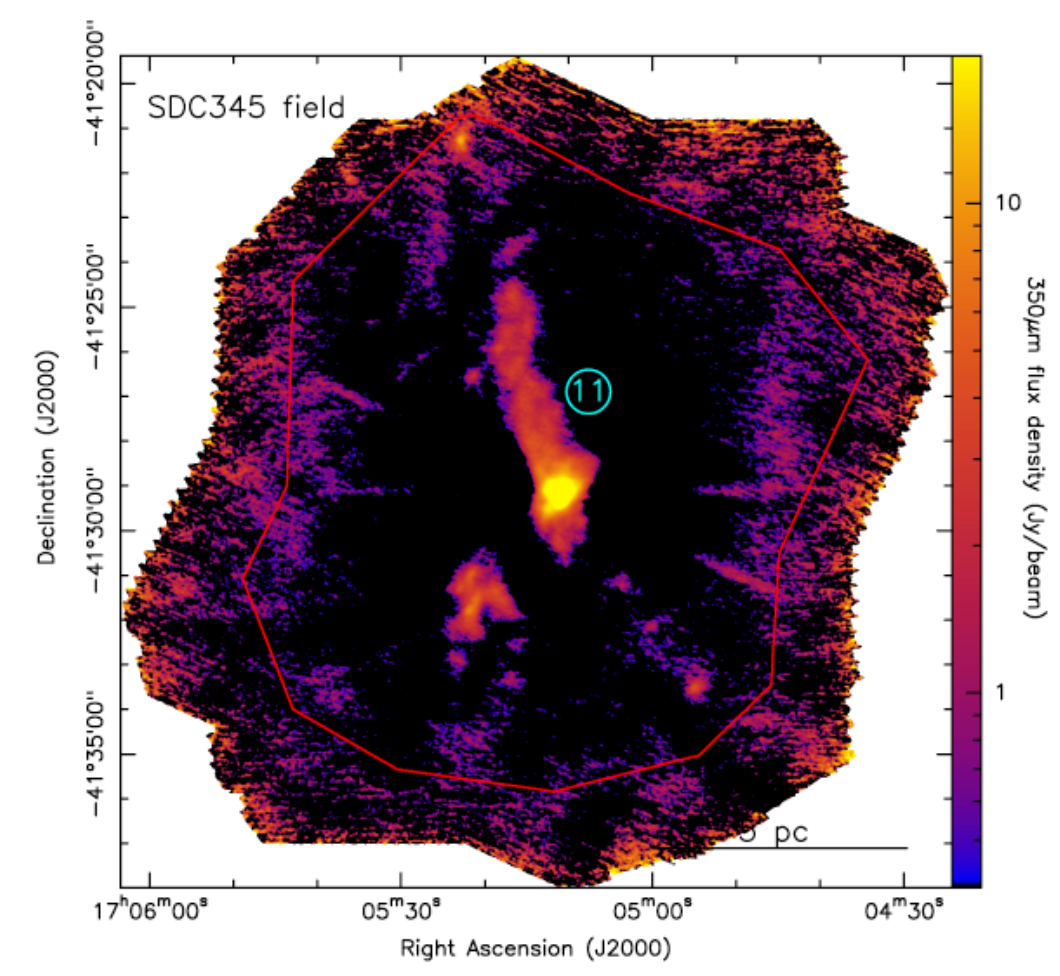}
    \caption{Same as Fig.~\ref{art_sdc326_nosource} for the SDC345 field}
    \label{fig:example_figure}
\end{figure}

\clearpage

\section{Images of ArT\'eMiS sources associations}

In this Appendix we present the ArT\'eMiS images with the locations of the {\it Herschel} 70$\mu$m sources \citep{molinari2016}, {\it Herschel} clumps \citep{elia2017}, and ATLASGAL clumps \citep{csengeri2014} for the SDC326, SDC328, SDC340, SDC343, and SDC345 fields.

\begin{figure}
	\includegraphics[width=17cm]{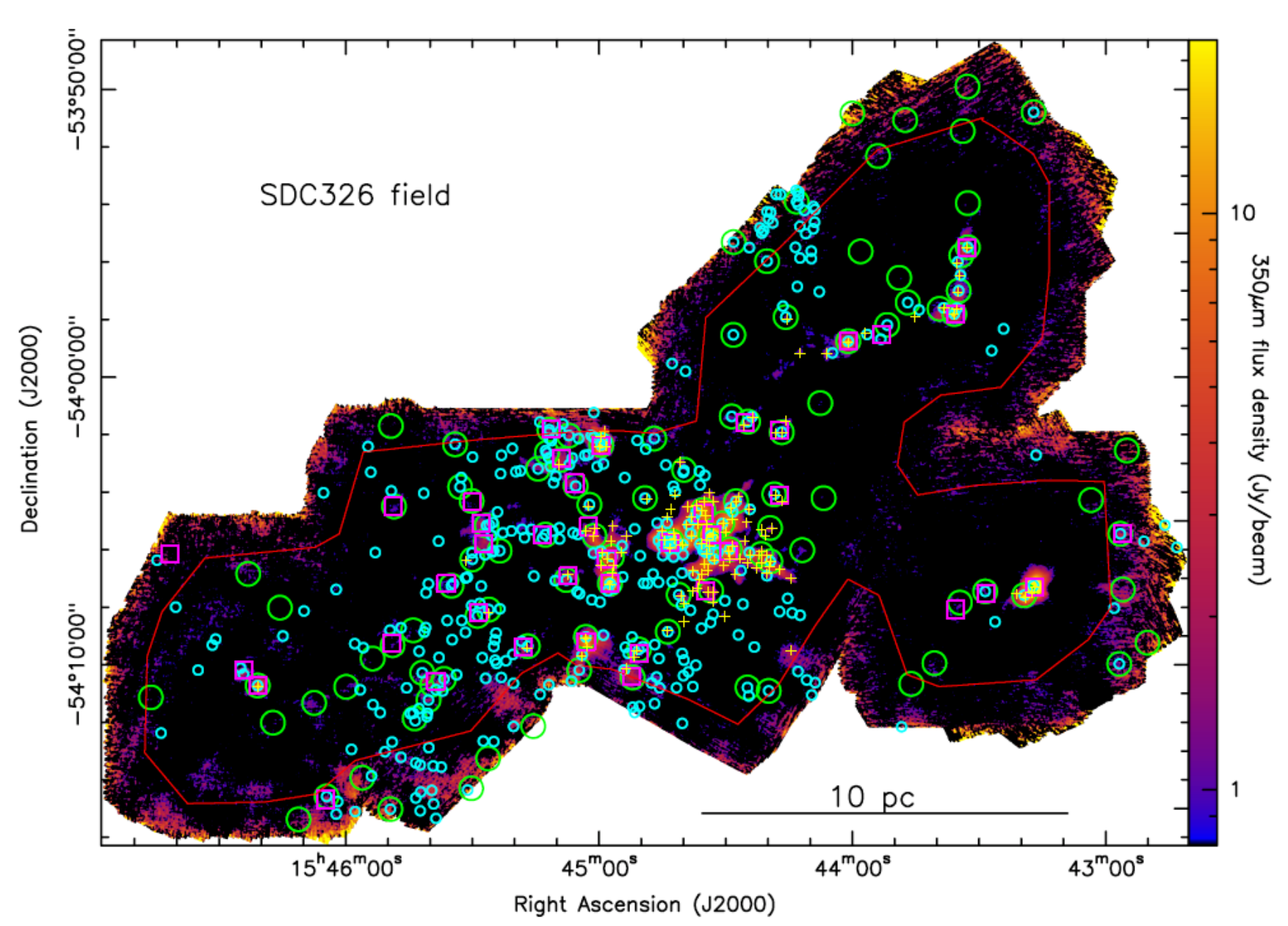}
    \caption{ Background image is the same as in Fig.~\ref{art_sdc326_nosource}. The yellow crosses mark the central positions of the identified ArT\'eMiS sources. The cyan circles mark the central positions of the Hi-GAL 70$\mu$m sources (Molinari et al. 2016). The green circles mark the central positions of the Herschel clumps (Elia et al. 2017). The purple squares mark the central positions of the ATLASGAL sources (Csengeri et al. 2014). The red solid line shows the area over which all source statistics presented in the paper have been calculated (i.e. excluding the noisy edges of the ArT\'eMis image). }
    \label{fig:sdc326_withsource}
\end{figure}

\clearpage

\begin{figure}
	\includegraphics[width=9cm]{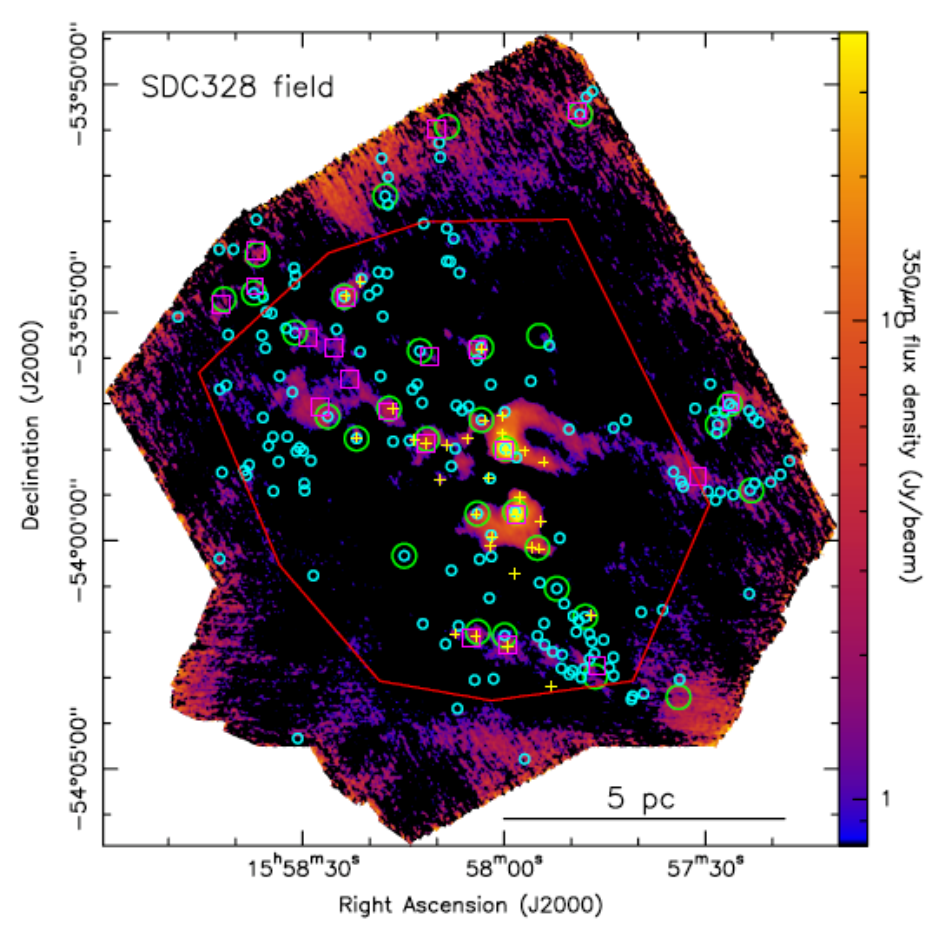}
    \caption{Same as Fig.~\ref{fig:sdc326_withsource} for the SDC328 field}
    \label{fig:example_figure}
\end{figure}

\begin{figure}
	\includegraphics[width=9cm]{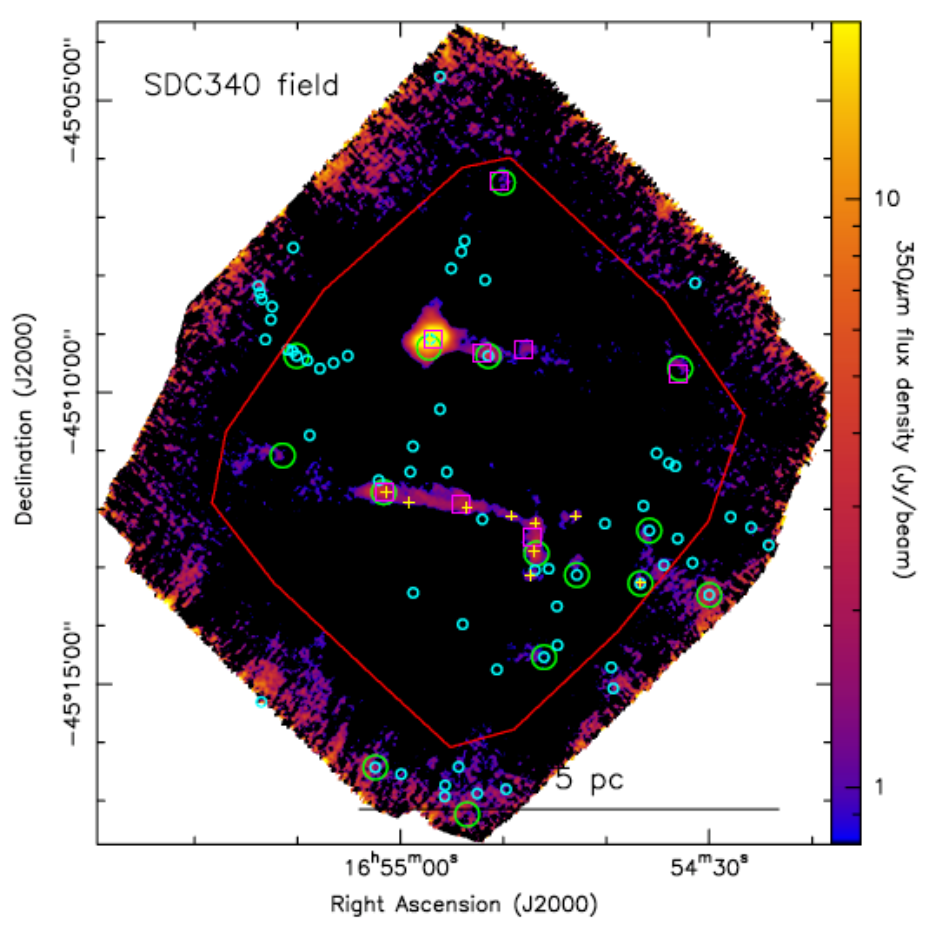}
    \caption{Same as Fig.~\ref{fig:sdc326_withsource} for the SDC340 field}
    \label{fig:example_figure}
\end{figure}

\begin{figure}
	\includegraphics[width=9cm]{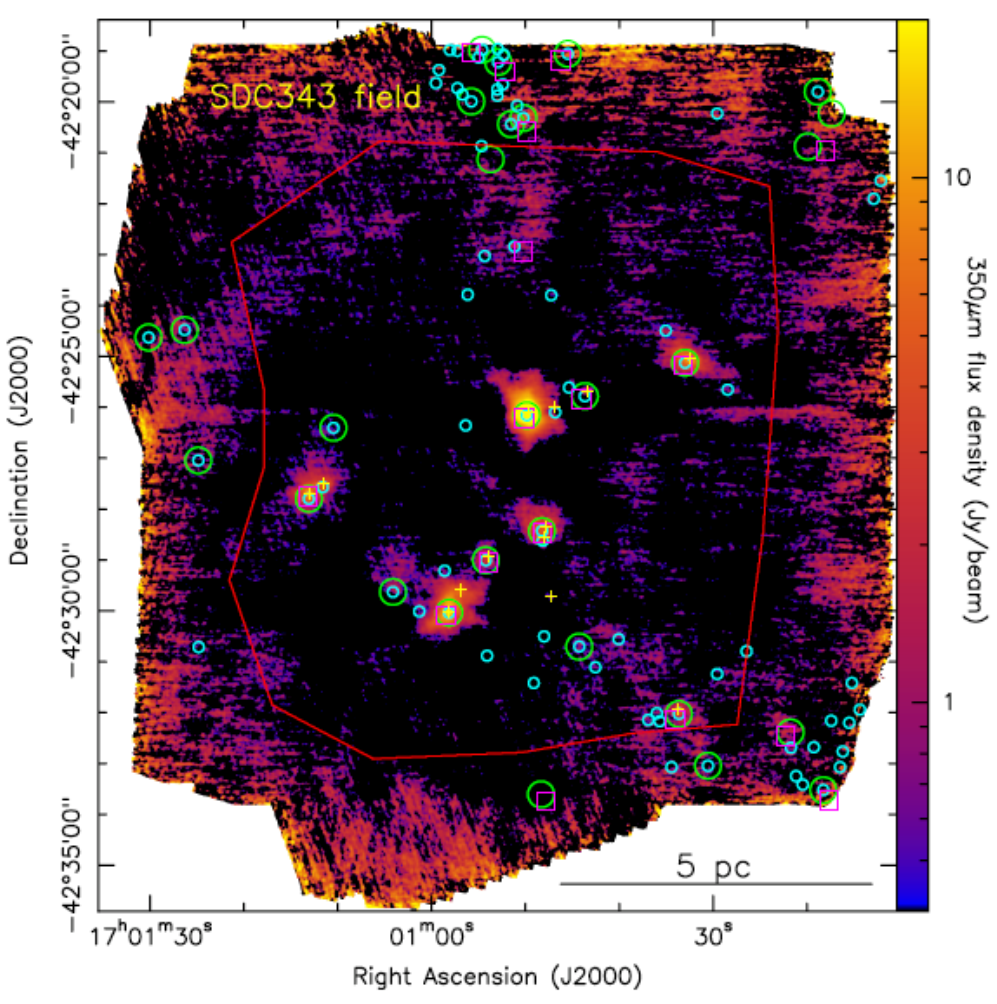}
    \caption{Same as Fig.~\ref{fig:sdc326_withsource} for the SDC343 field}
    \label{fig:example_figure}
\end{figure}

\begin{figure}
	\includegraphics[width=9cm]{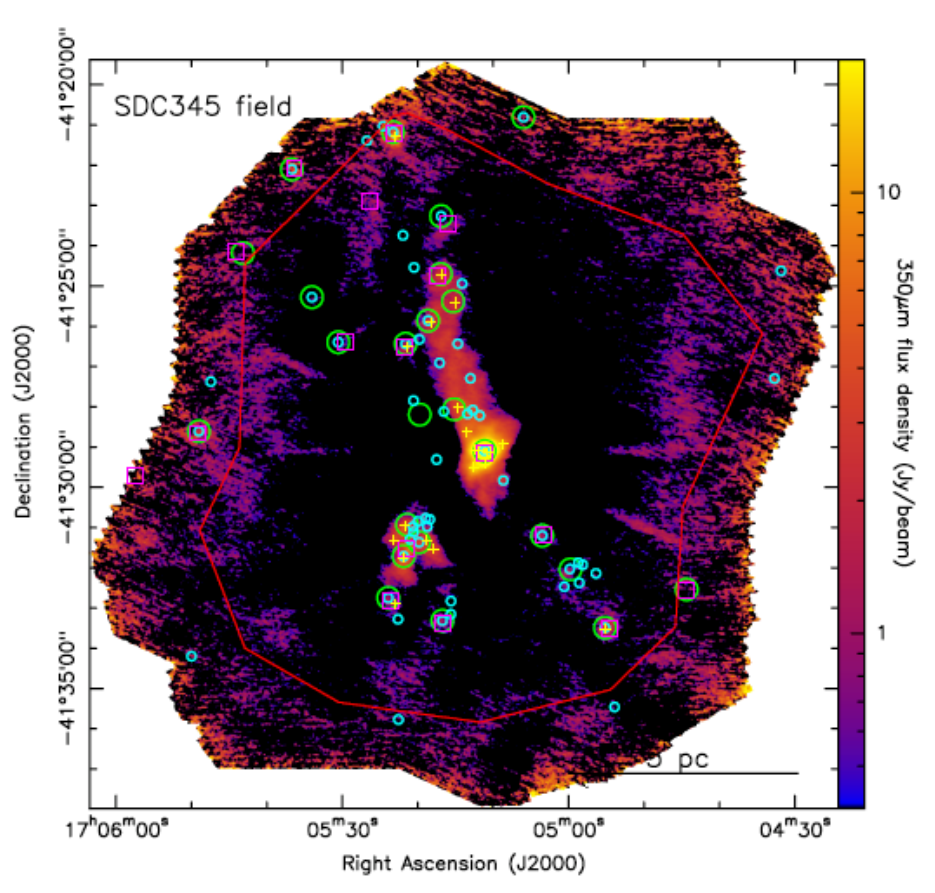}
    \caption{Same as Fig.~\ref{fig:sdc326_withsource} for the SDC345 field}
    \label{fig:example_figure}
\end{figure}

\clearpage

\section{Individual cutout images around each robust ArT\'eMiS source}

In this Appendix we present individual cutout images of each ArT\'eMiS source (see Sec.~3).

\begin{figure}
	\includegraphics[width=8cm]{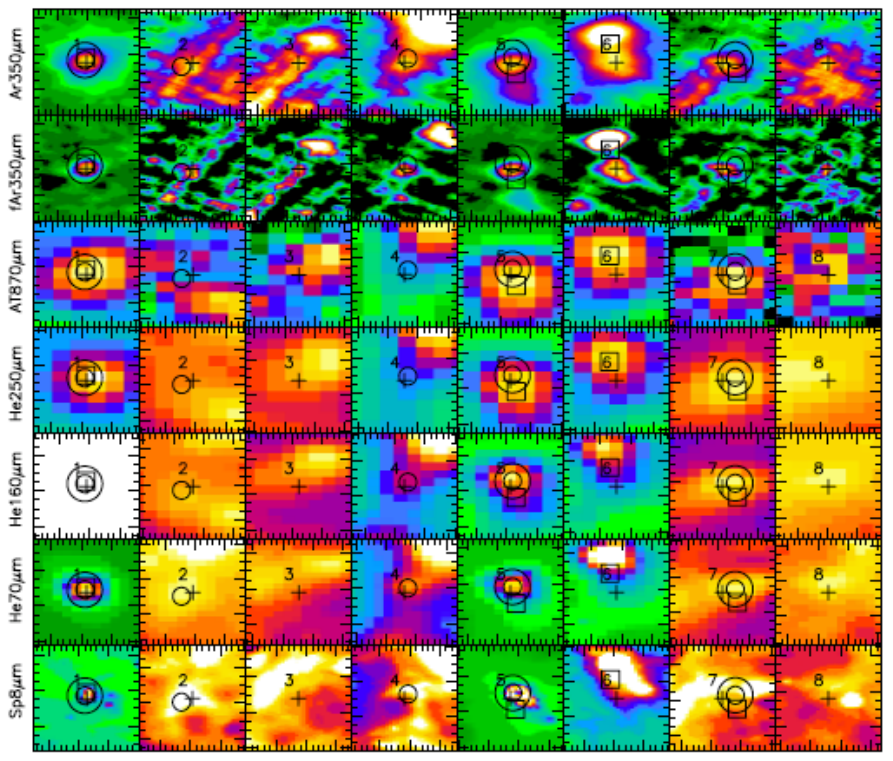}
    \caption{Cutout images of each individual ArT\'eMiS sources identified in the SDC326 field.  Each cutout is a 1arcmin by 1arcmin box centred on the source position. Each column corresponds to a different source, the id number of source is indicated in each panel. Each row corresponds to a different wavelength or image type. The 1st row presents the original ArT\'eMiS image of the source; the 2nd row to the filtered ArT\'eMiS image; the 3rd to ATLASGAL; the 4th to {\it Herschel} 250$\mu$m; the 5th to {\it Herschel} 160$\mu$m;the 6th to{\it Herschel} 70$\mu$m; the 7th to {\it Spitzer} 8$\mu$m. The central black crosses mark the central position of the ArT\'eMiS source. The black squares mark the position of ATLASGAL sources \citep{csengeri2014}. The small black circles mark the positions of {\it Herschel} 70$\mu$m sources \citep{molinari2016}. The large black circles mark the positions of {\it Herschel} clumps \citep{elia2017}   }
    \label{fig:example_figure}
\end{figure}

\begin{figure}
	\includegraphics[width=8cm]{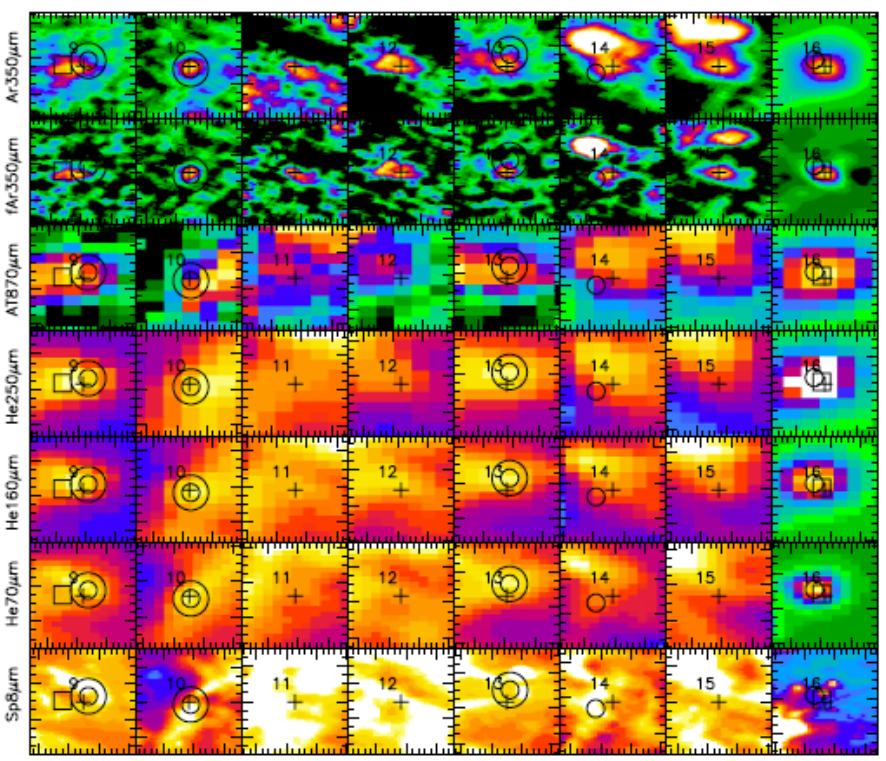}
    \caption{Figure C1 continued.}
    \label{fig:example_figure}
\end{figure}

\begin{figure}
	\includegraphics[width=8cm]{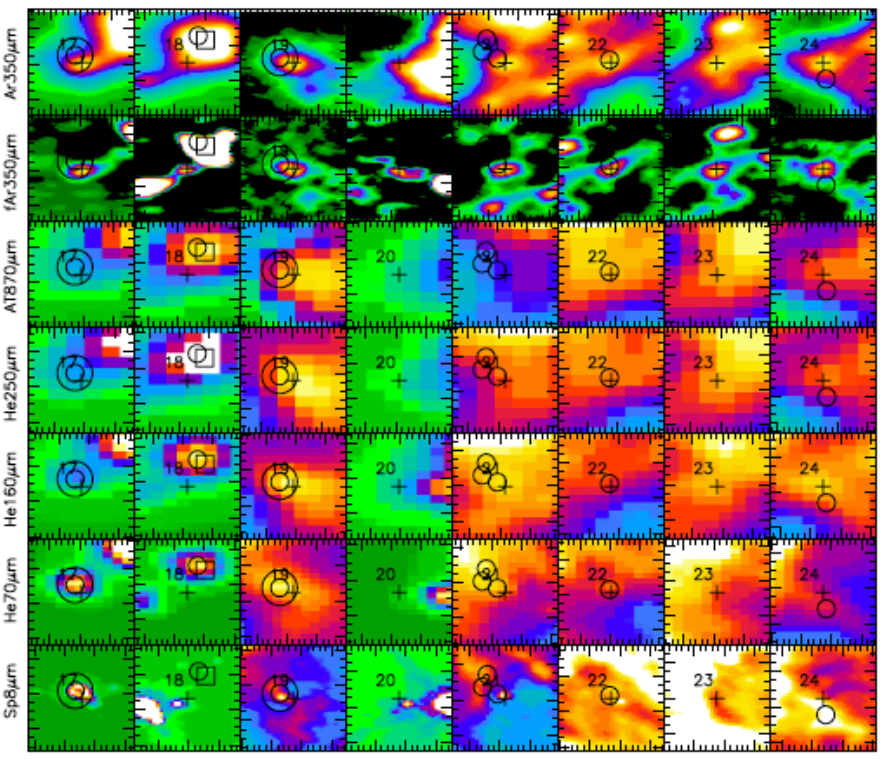}
    \caption{Figure C1 continued.}
    \label{fig:example_figure}
\end{figure}

\begin{figure}
	\includegraphics[width=8cm]{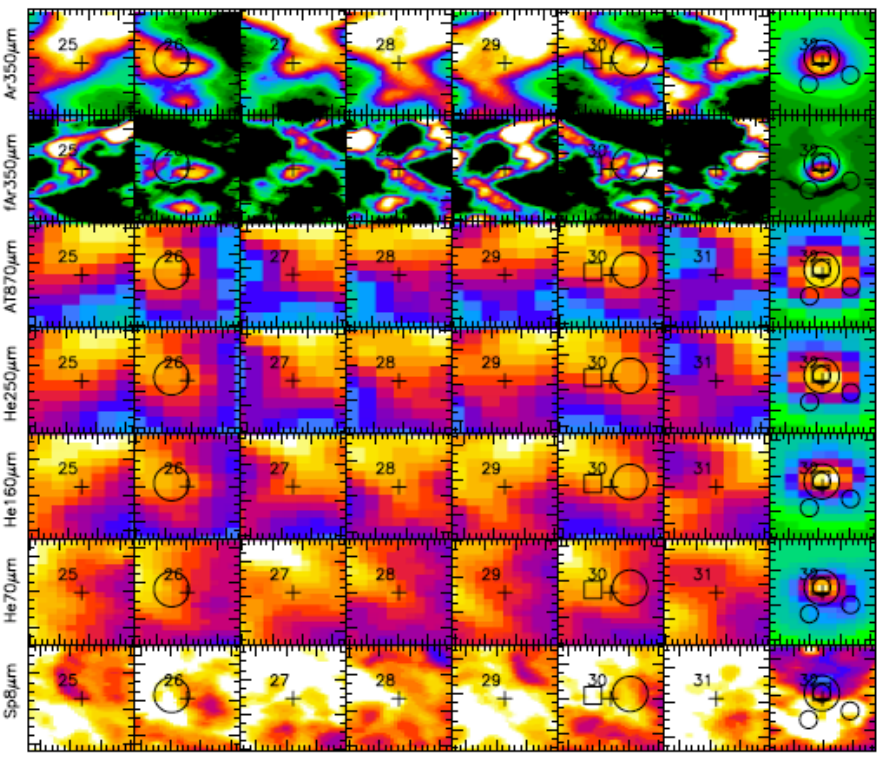}
    \caption{Figure C1 continued.}
    \label{fig:example_figure}
\end{figure}

\begin{figure}
	\includegraphics[width=8cm]{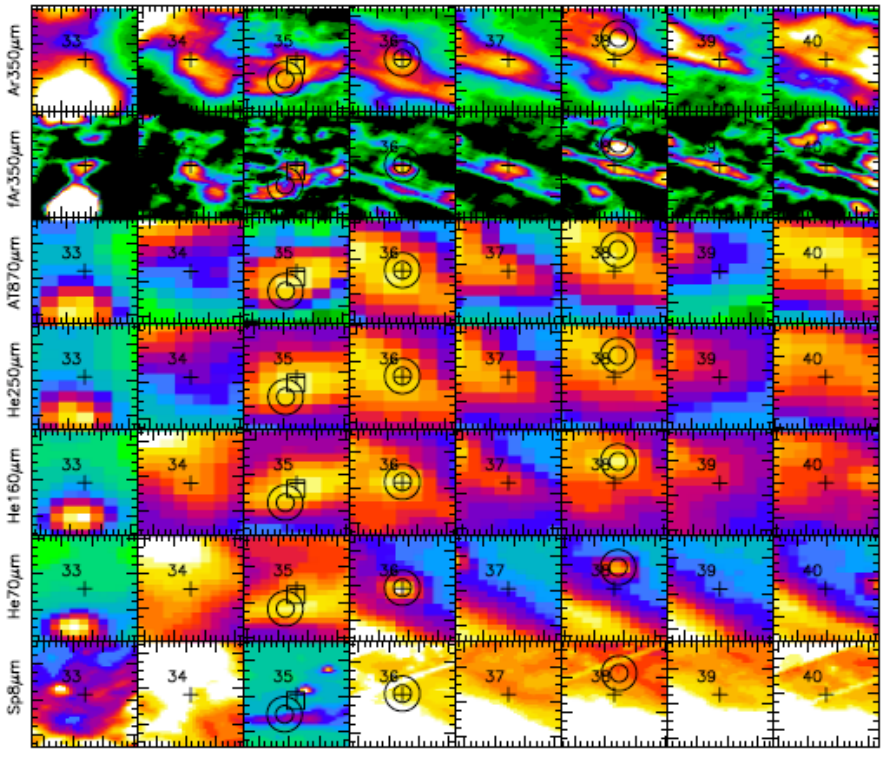}
    \caption{Figure C1 continued.}
    \label{fig:example_figure}
\end{figure}

\begin{figure}
	\includegraphics[width=8cm]{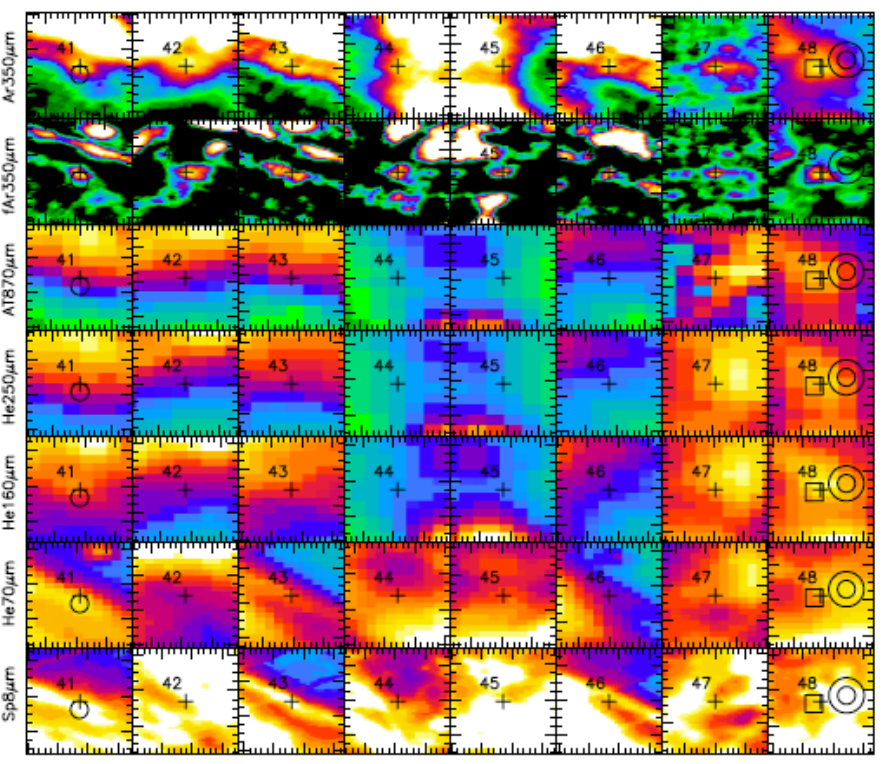}
   \caption{Figure C1 continued.}
    \label{fig:example_figure}
\end{figure}

\begin{figure}
	\includegraphics[width=8cm]{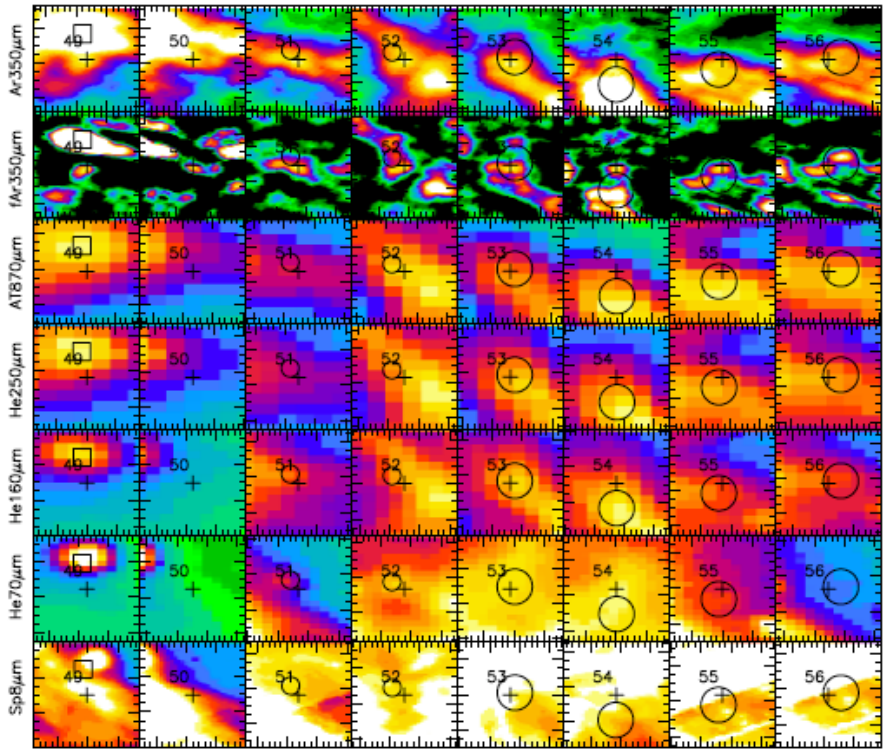}
    \caption{Figure C1 continued.}
    \label{fig:example_figure}
\end{figure}

\begin{figure}
	\includegraphics[width=8cm]{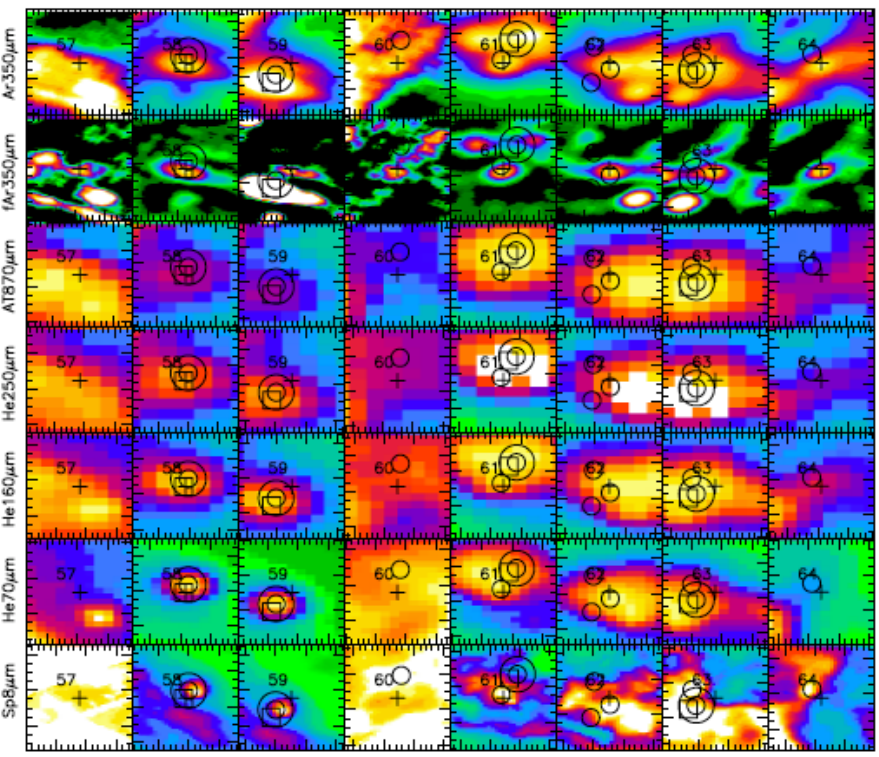}
     \caption{Figure C1 continued.}
    \label{fig:example_figure}
\end{figure}

\begin{figure}
	\includegraphics[width=8cm]{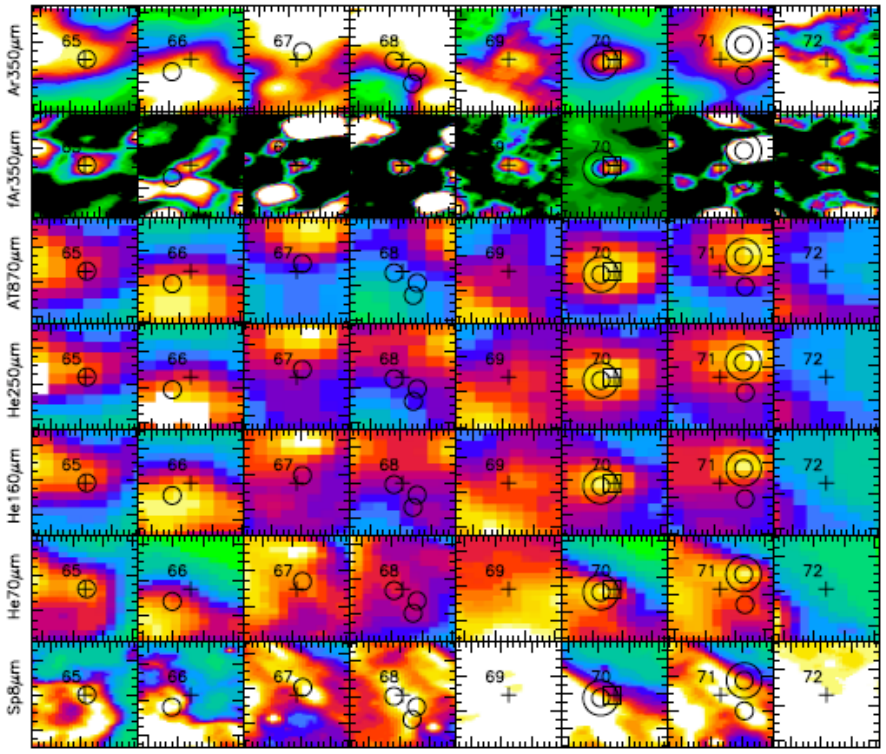}
   \caption{Figure C1 continued.}
    \label{fig:example_figure}
\end{figure}

\begin{figure}
	\includegraphics[width=8cm]{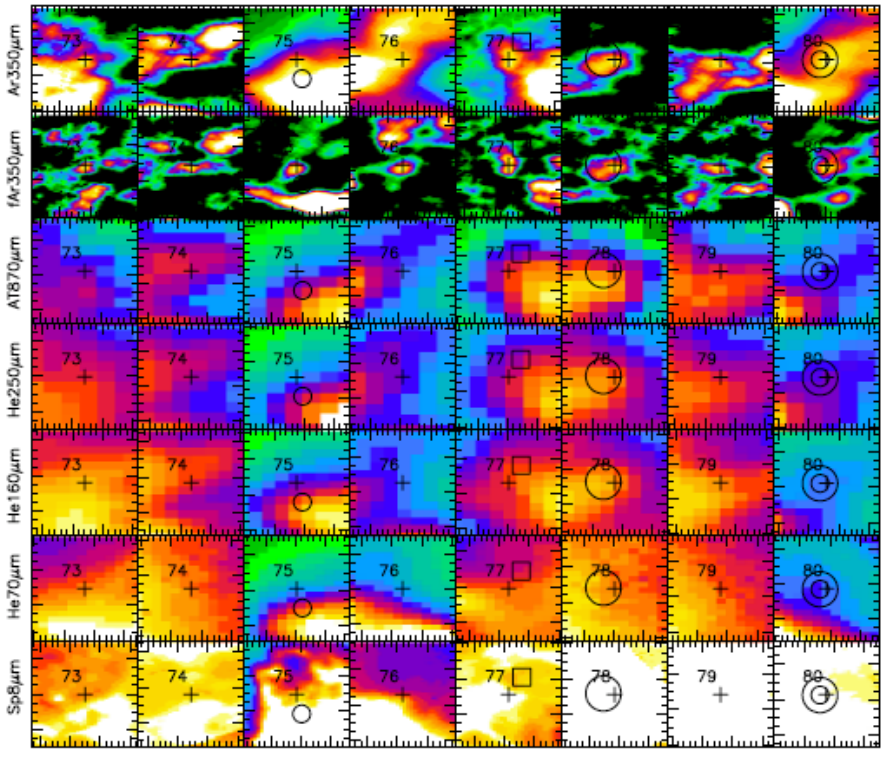}
    \caption{Figure C1 continued.}
    \label{fig:example_figure}
\end{figure}

\begin{figure}
	\includegraphics[width=8cm]{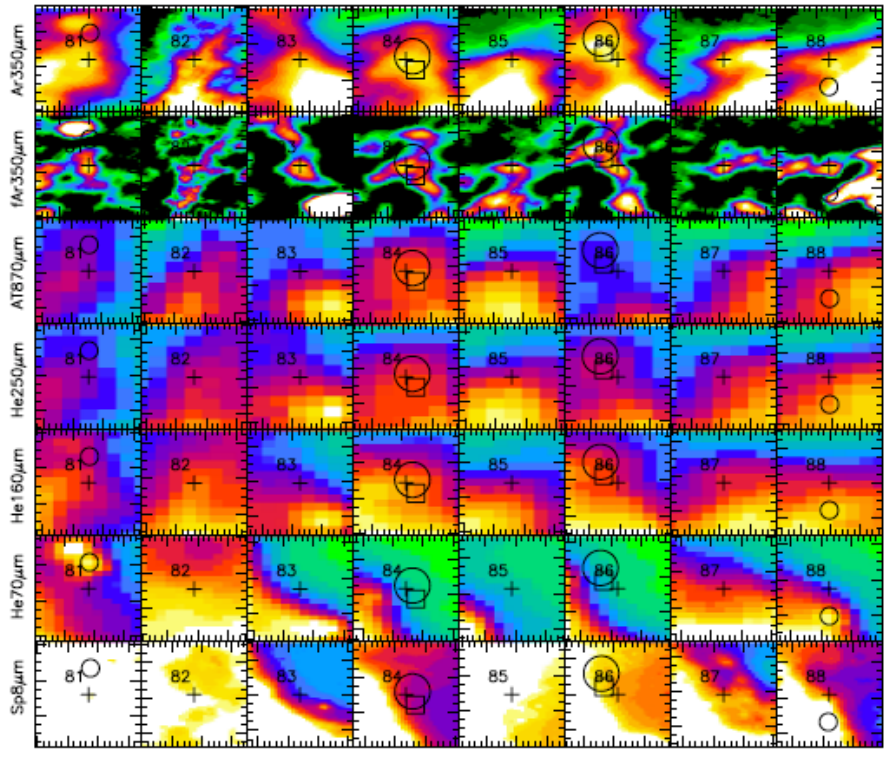}
    \caption{Figure C1 continued.}
    \label{fig:example_figure}
\end{figure}

\begin{figure}
	\includegraphics[width=8cm]{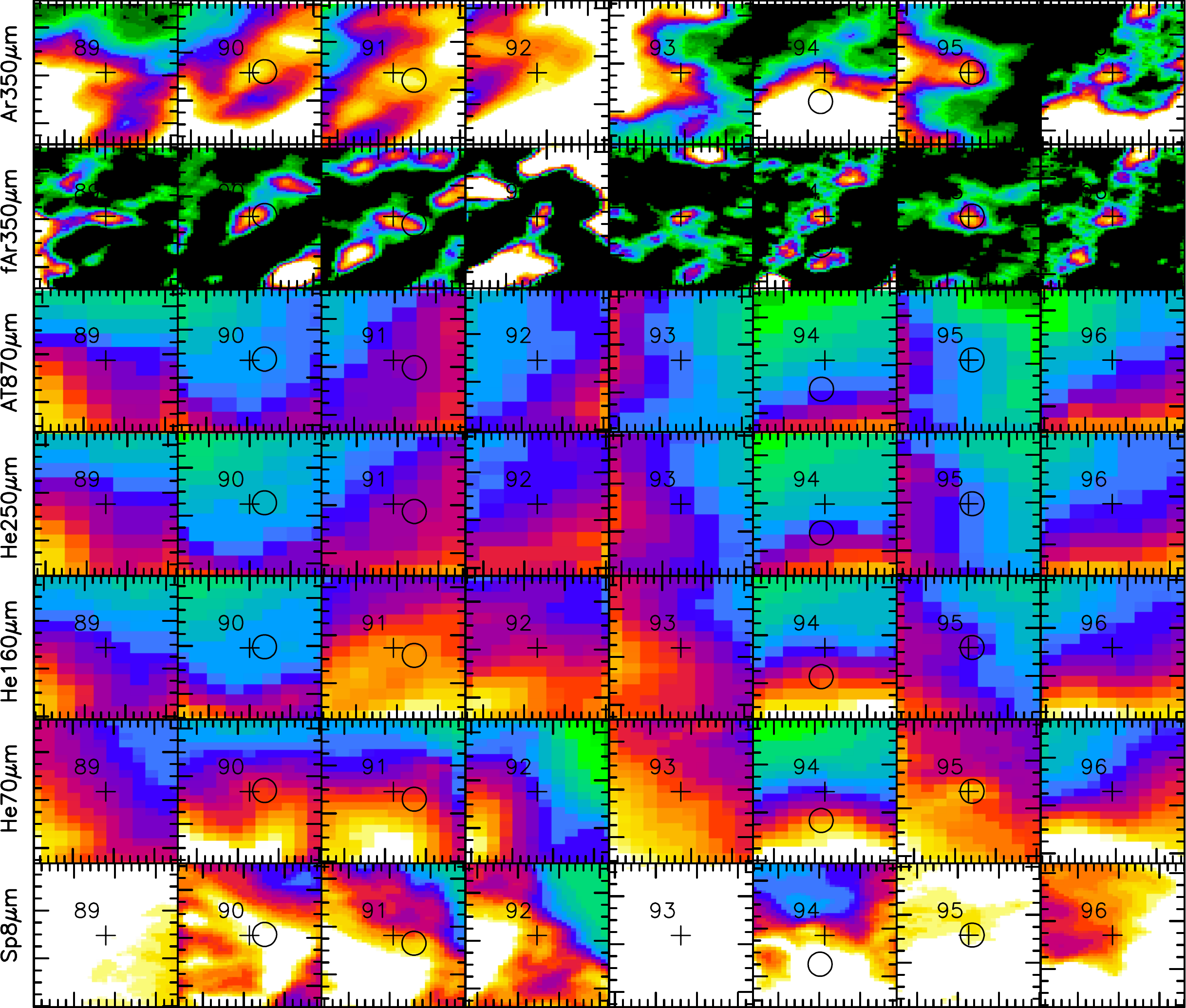}
    \caption{Figure C1 continued.}
    \label{fig:example_figure}
\end{figure}

\begin{figure}
	\includegraphics[width=8.cm]{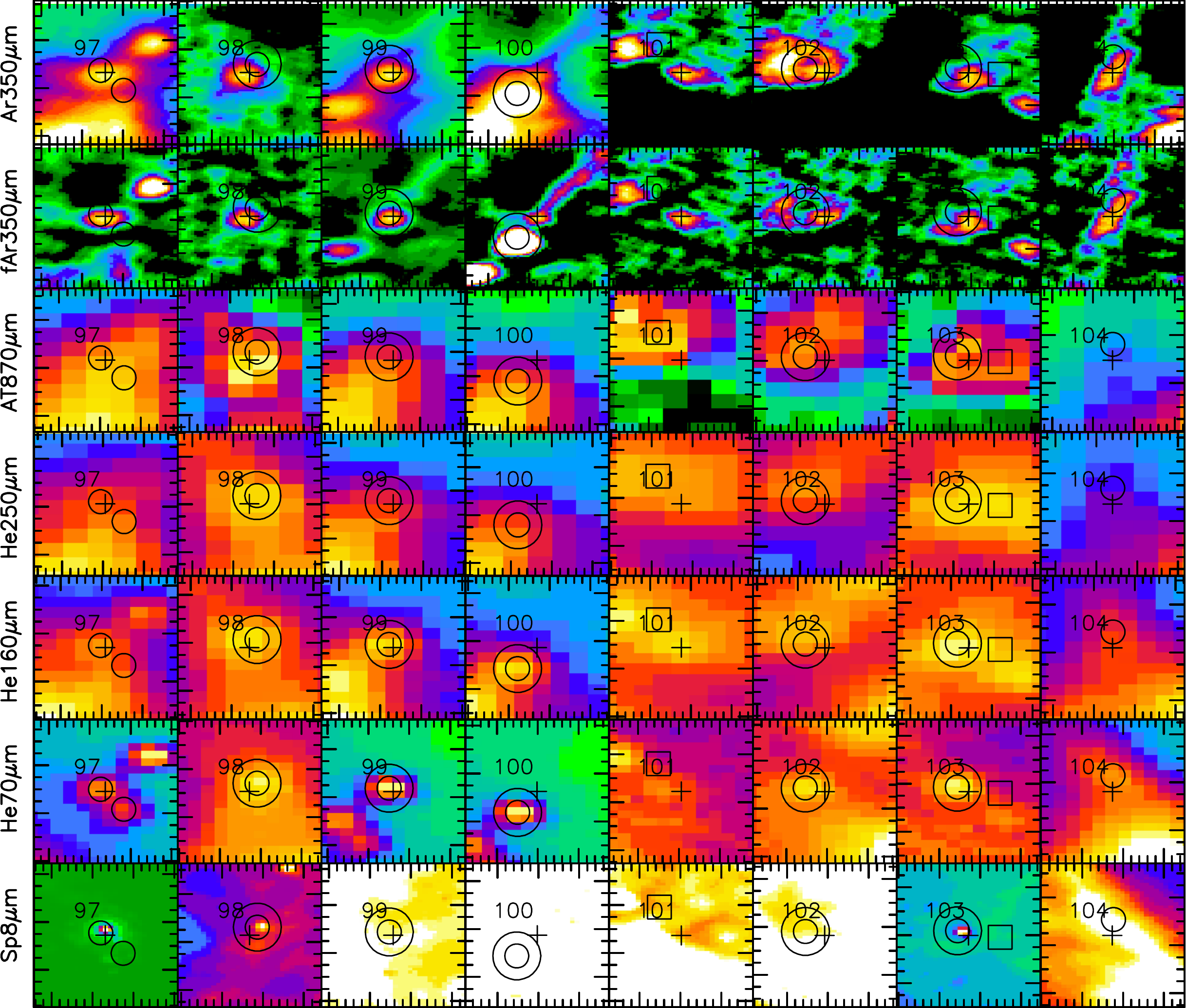}
    \caption{Figure C1 continued.}
    \label{fig:example_figure}
\end{figure}

\begin{figure}
	\includegraphics[width=8.cm]{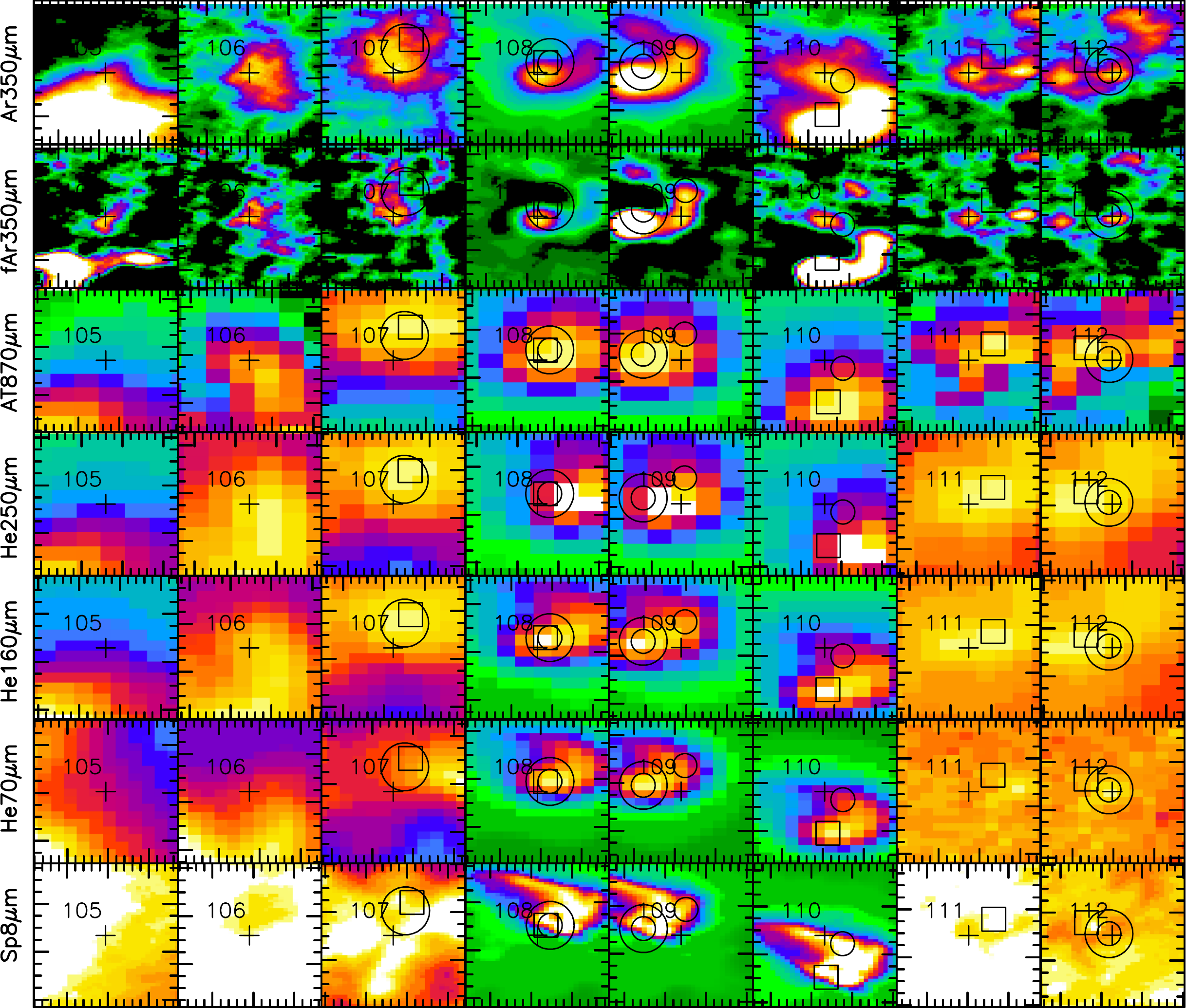}
    \caption{Figure C1 continued.}
    \label{fig:example_figure}
\end{figure}

\begin{figure}
	\includegraphics[width=8.cm]{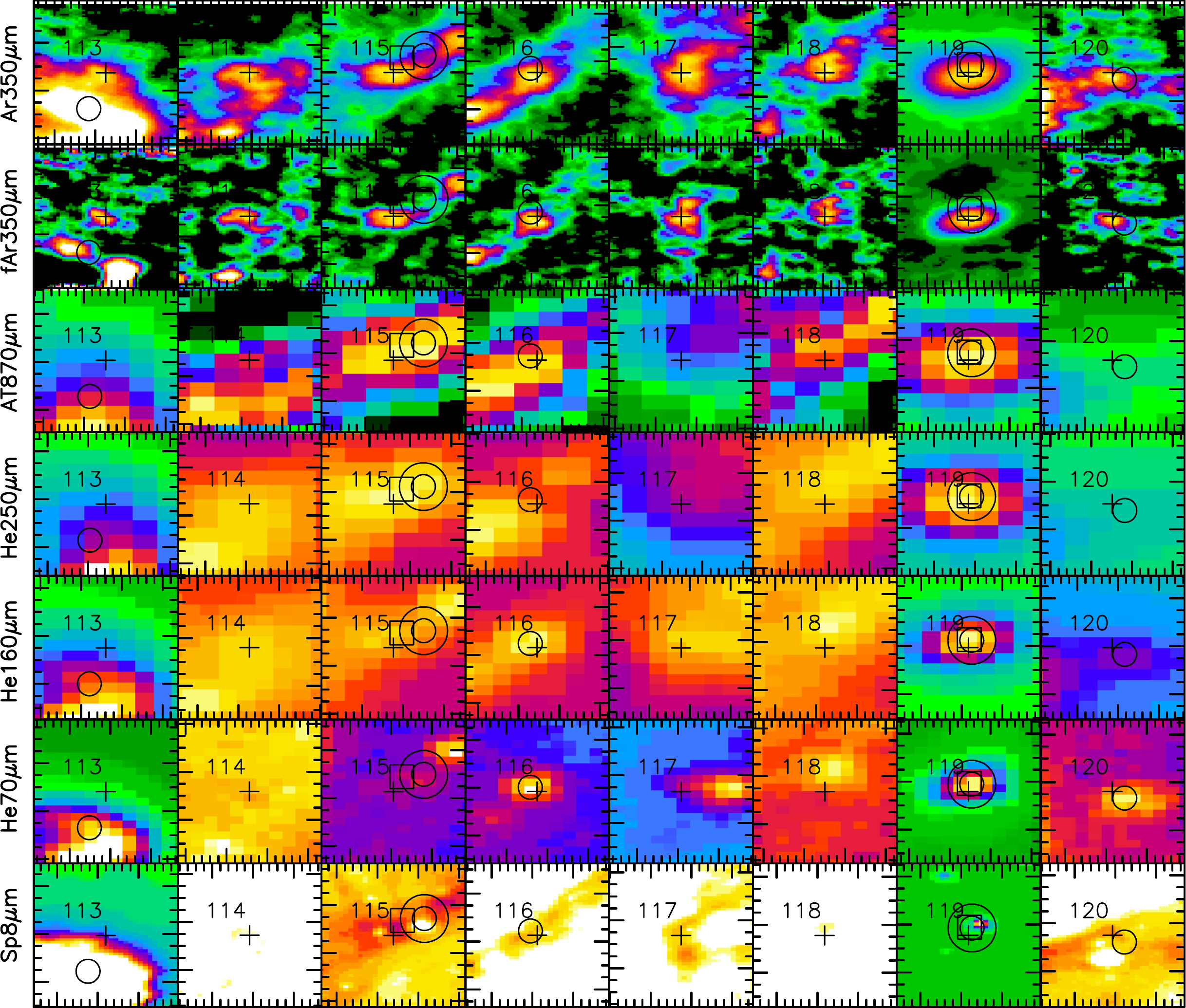}
    \caption{Figure C1 continued.}
    \label{fig:example_figure}
\end{figure}

\begin{figure}
	\includegraphics[width=8.cm]{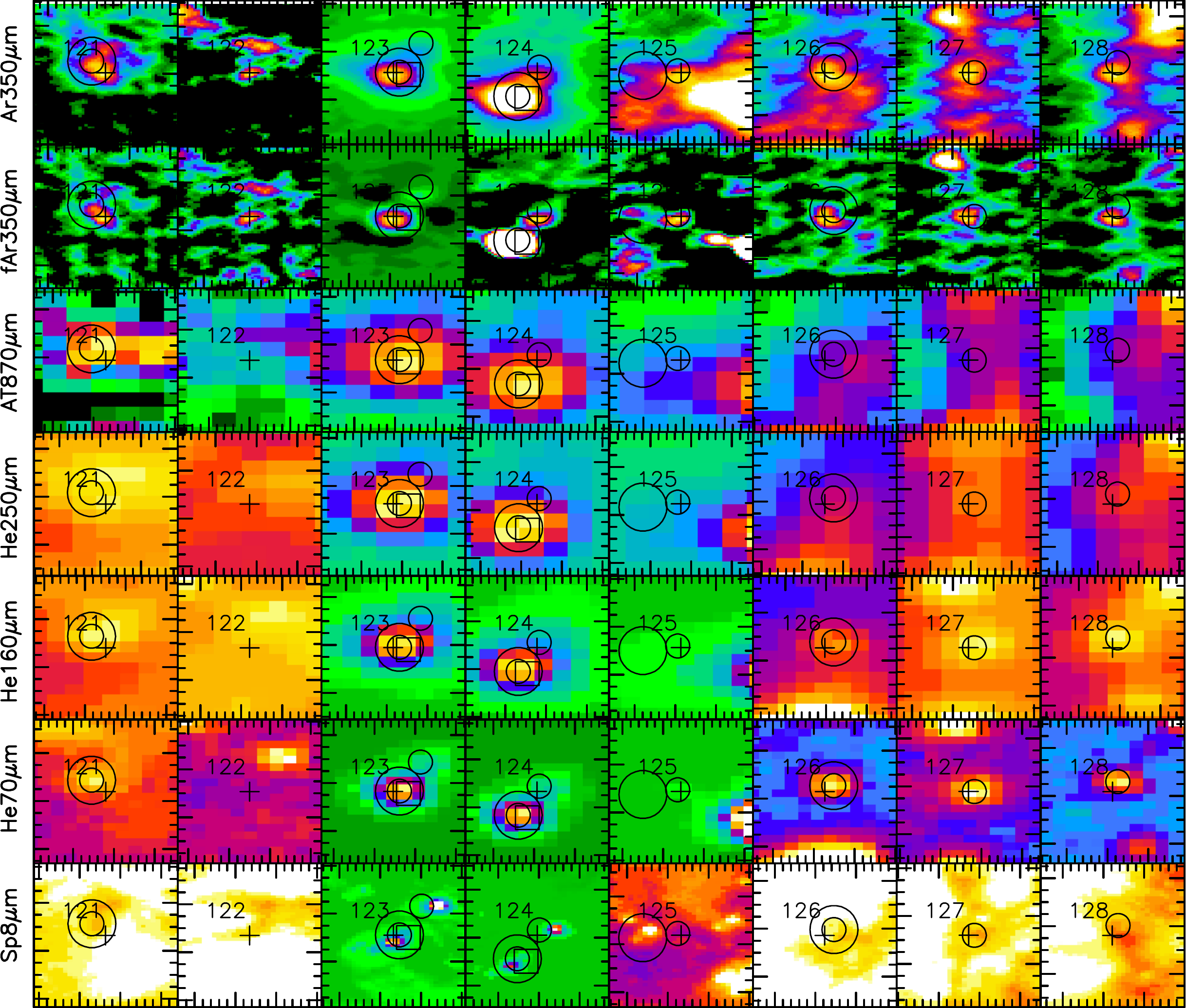}
    \caption{Figure C1 continued.}
    \label{fig:example_figure}
\end{figure}

\begin{figure}
	\includegraphics[width=1.62cm]{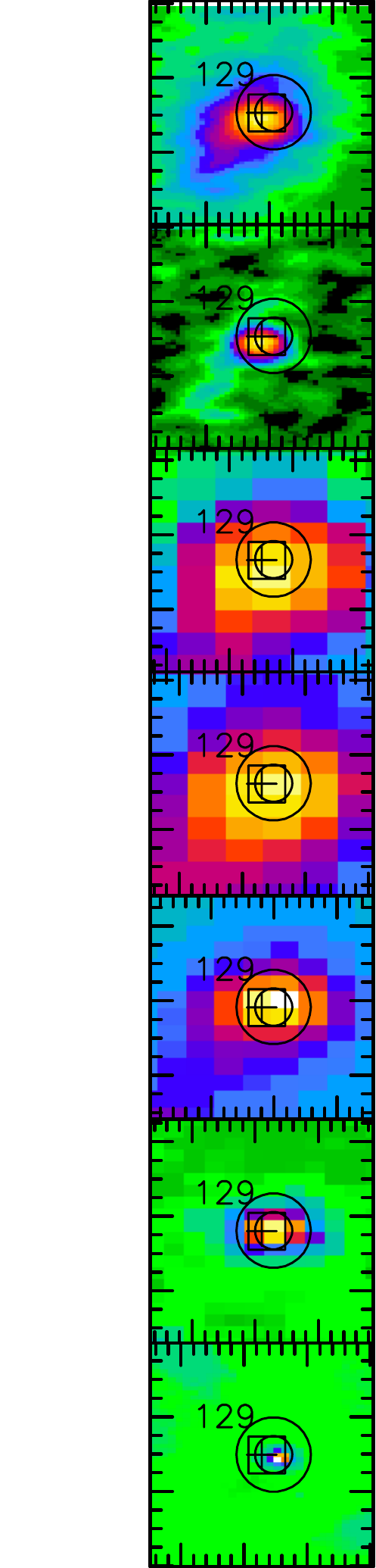}
    \caption{Figure C1 continued.}
    \label{fig:example_figure}
\end{figure}

\begin{figure}
	\includegraphics[width=8cm]{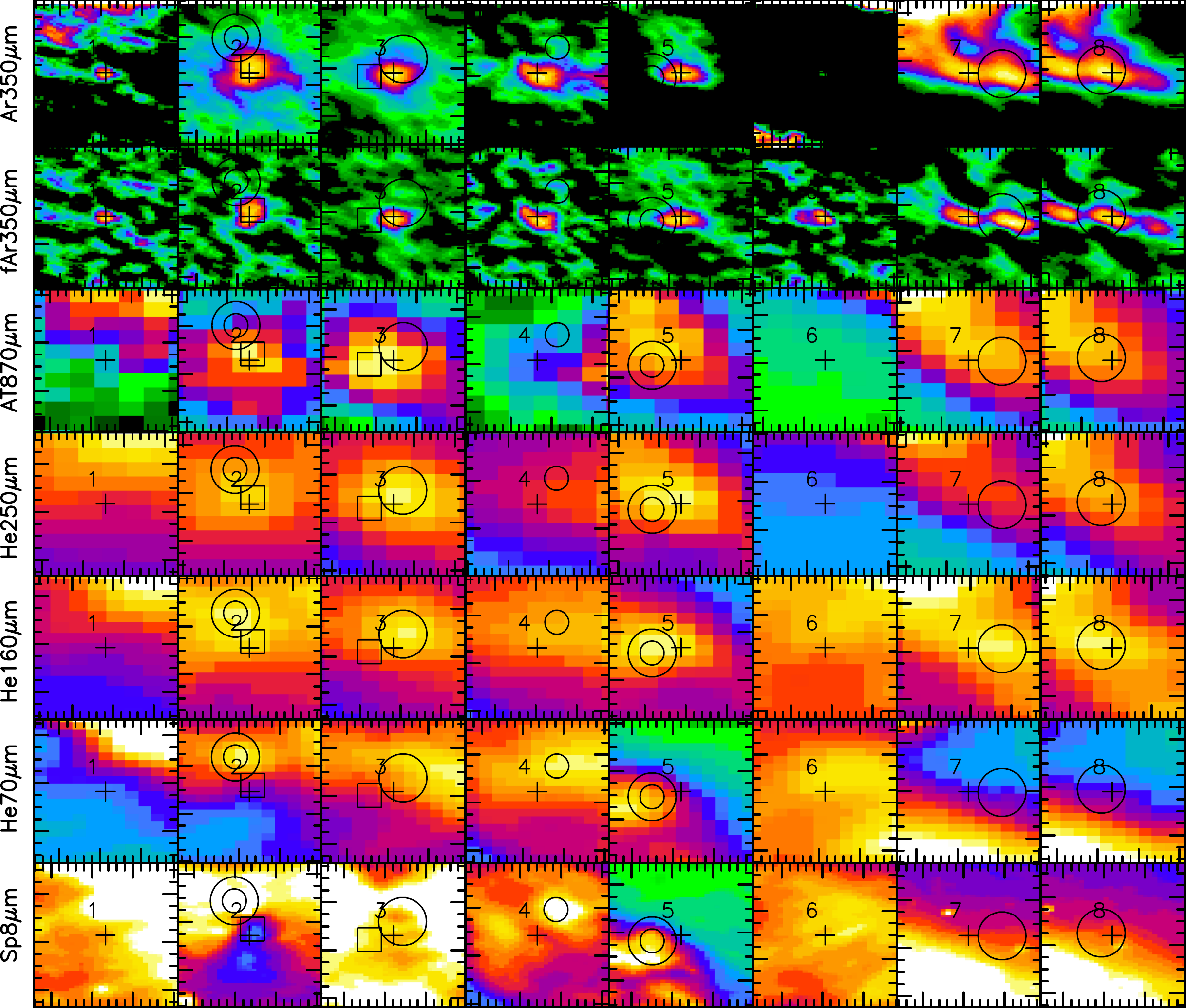}
    \caption{Same as Figure C1 but for the SDC328 field}
    \label{fig:example_figure}
\end{figure}

\begin{figure}
	\includegraphics[width=8cm]{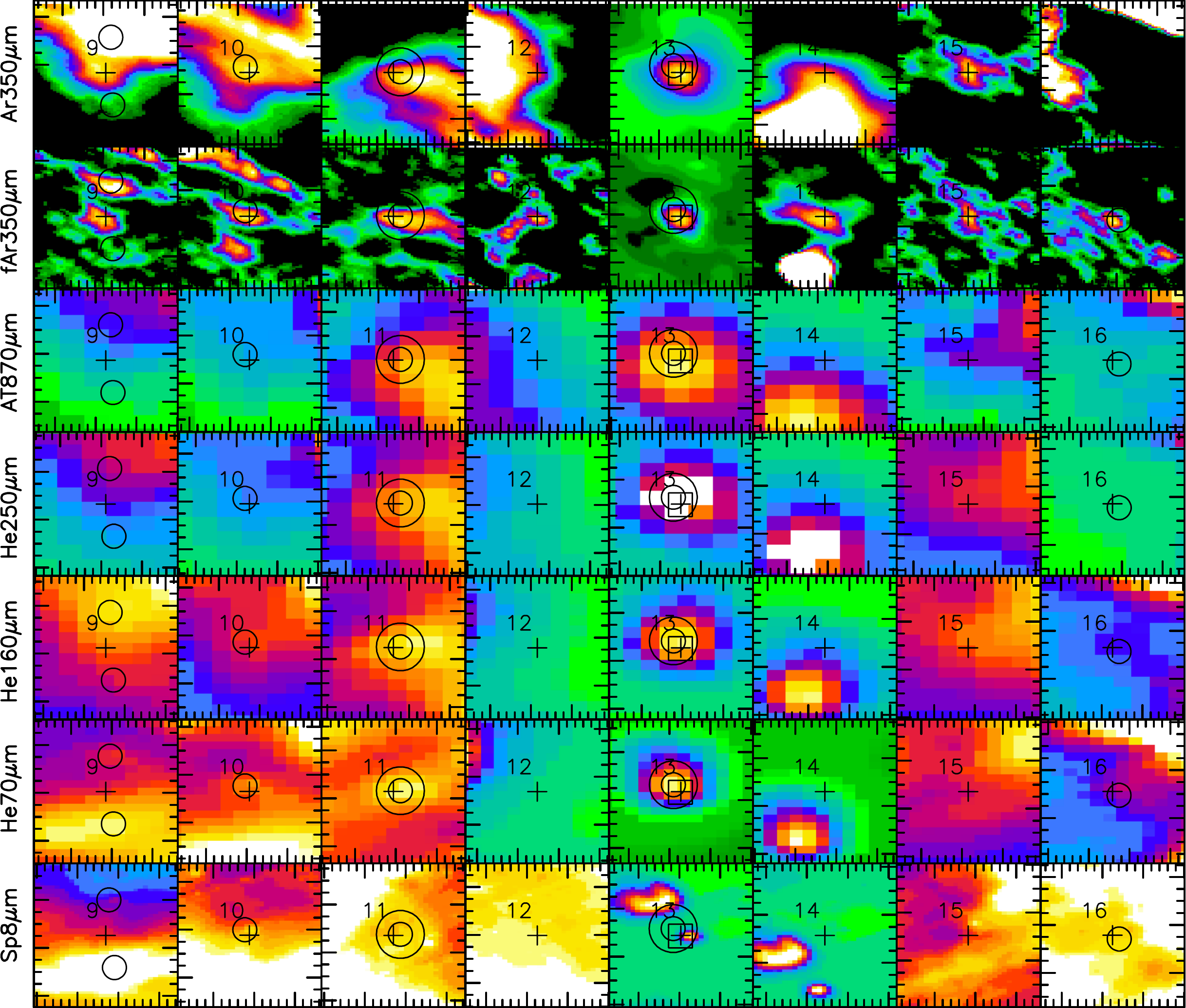}
     \caption{Figure C14 continued.}
    \label{fig:example_figure}
\end{figure}

\begin{figure}
	\includegraphics[width=8cm]{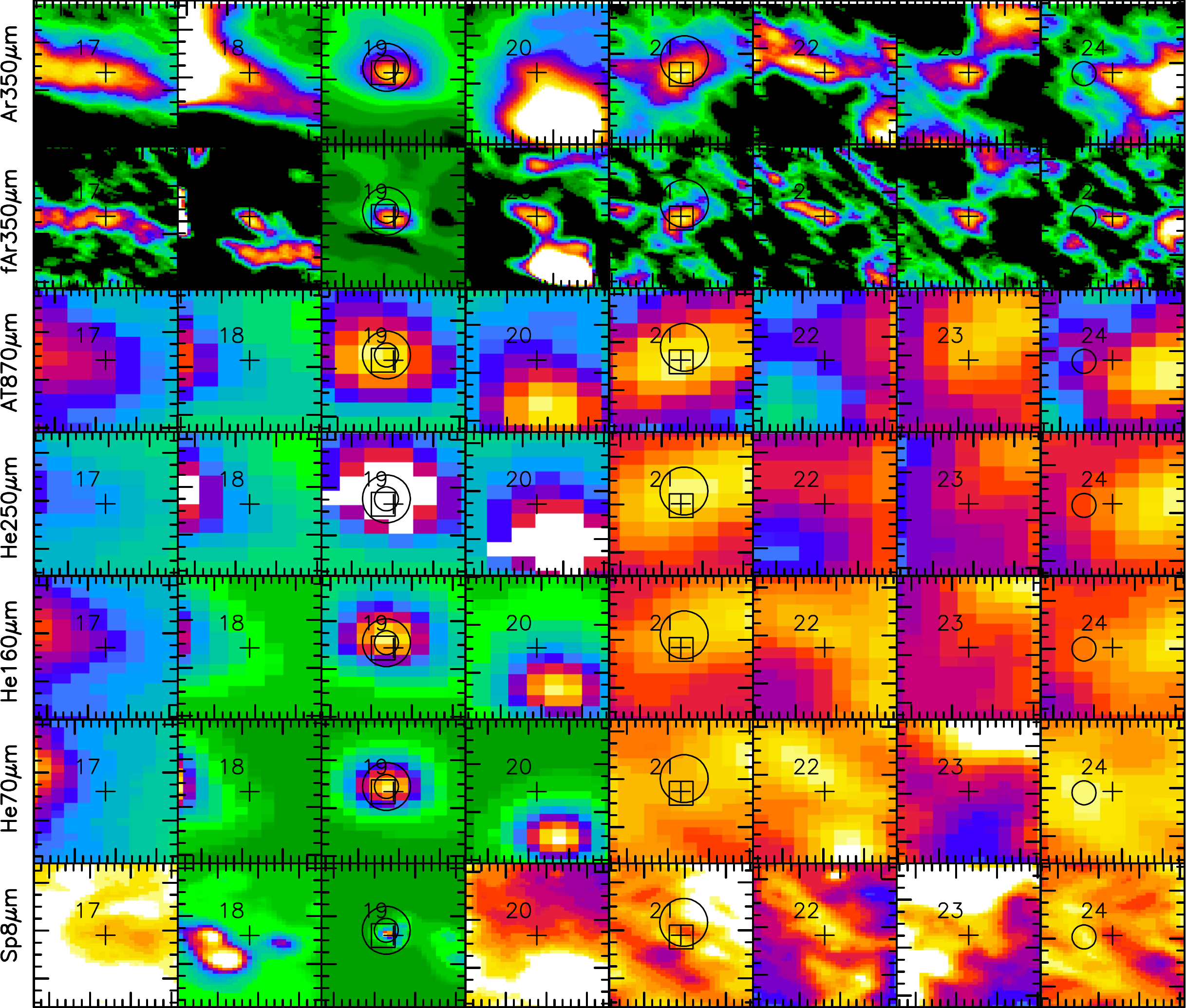}
     \caption{Figure C14 continued.}
    \label{fig:example_figure}
\end{figure}

\begin{figure}
	\includegraphics[width=7.5cm]{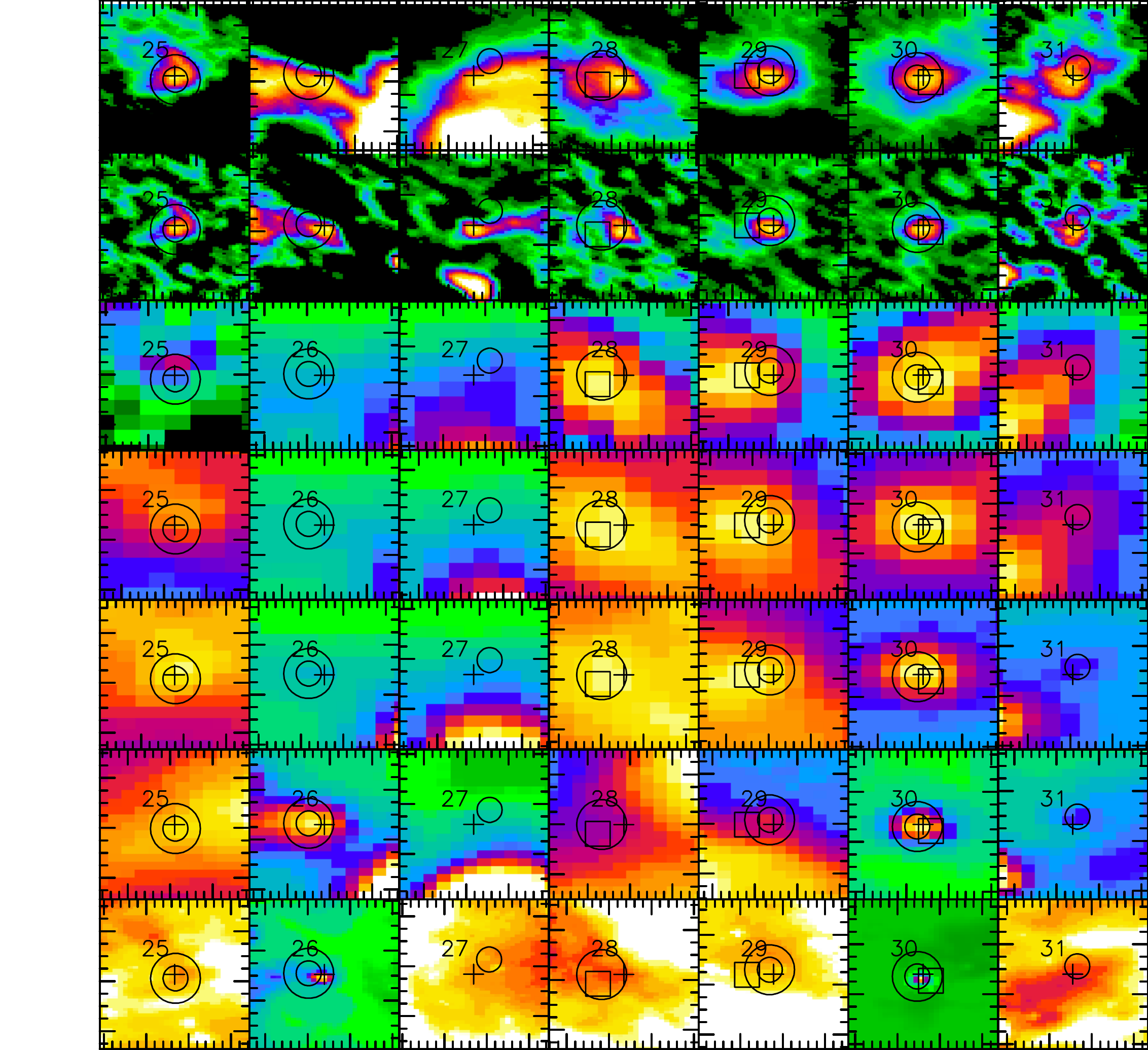}
     \caption{Figure C14 continued.}
    \label{fig:example_figure}
\end{figure}

\begin{figure}
	\includegraphics[width=8cm]{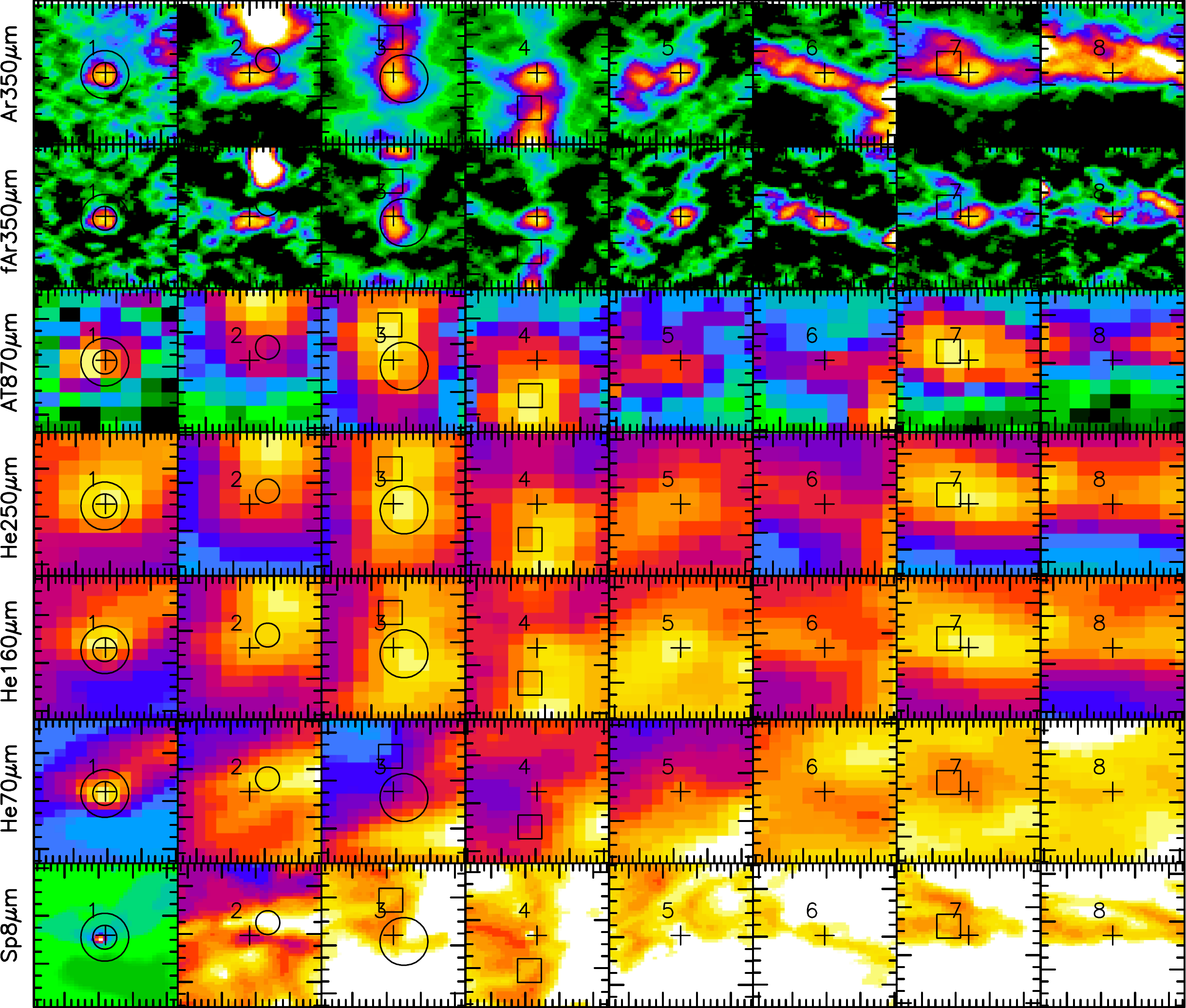}
    \caption{Same as Figure C1 but for the SDC340 field}
    \label{fig:example_figure}
\end{figure}

\begin{figure}
	\includegraphics[width=3.6cm]{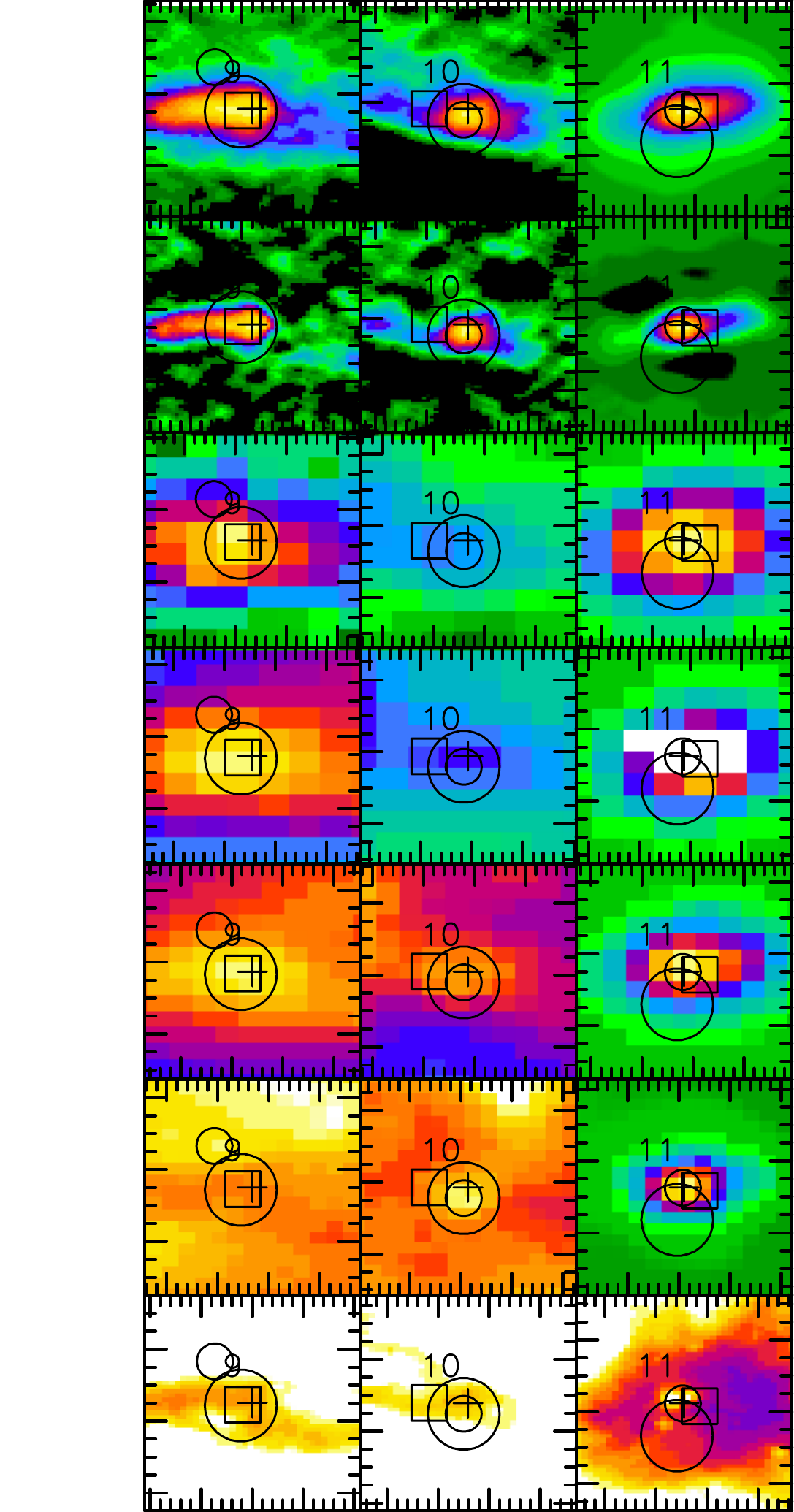}
      \caption{Figure C18 continued.}
    \label{fig:example_figure}
\end{figure}

\begin{figure}
	\includegraphics[width=8cm]{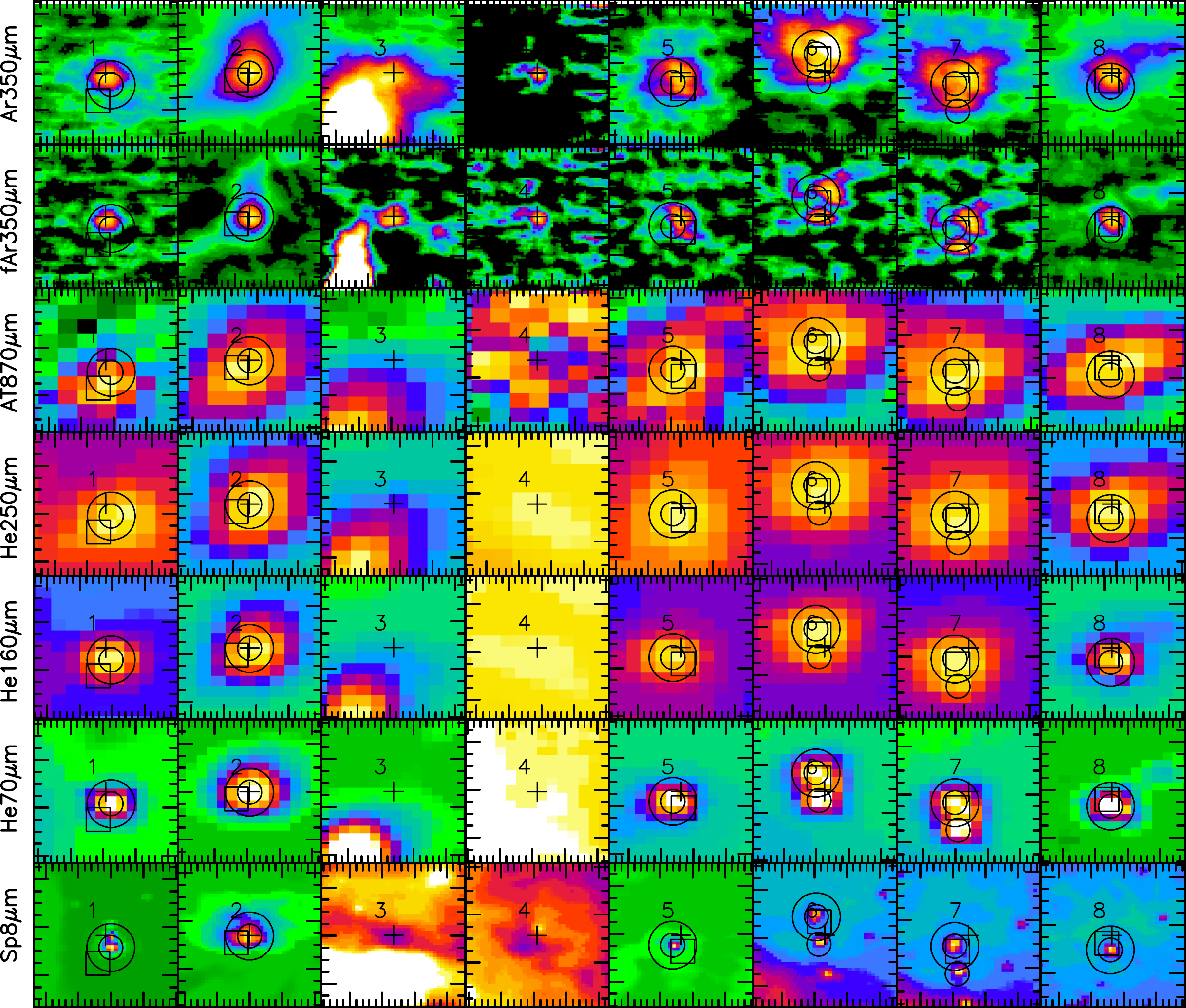}
   \caption{Same as Figure C1 but for the SDC343 field}
    \label{fig:example_figure}
\end{figure}

\begin{figure}
	\includegraphics[width=5.5cm]{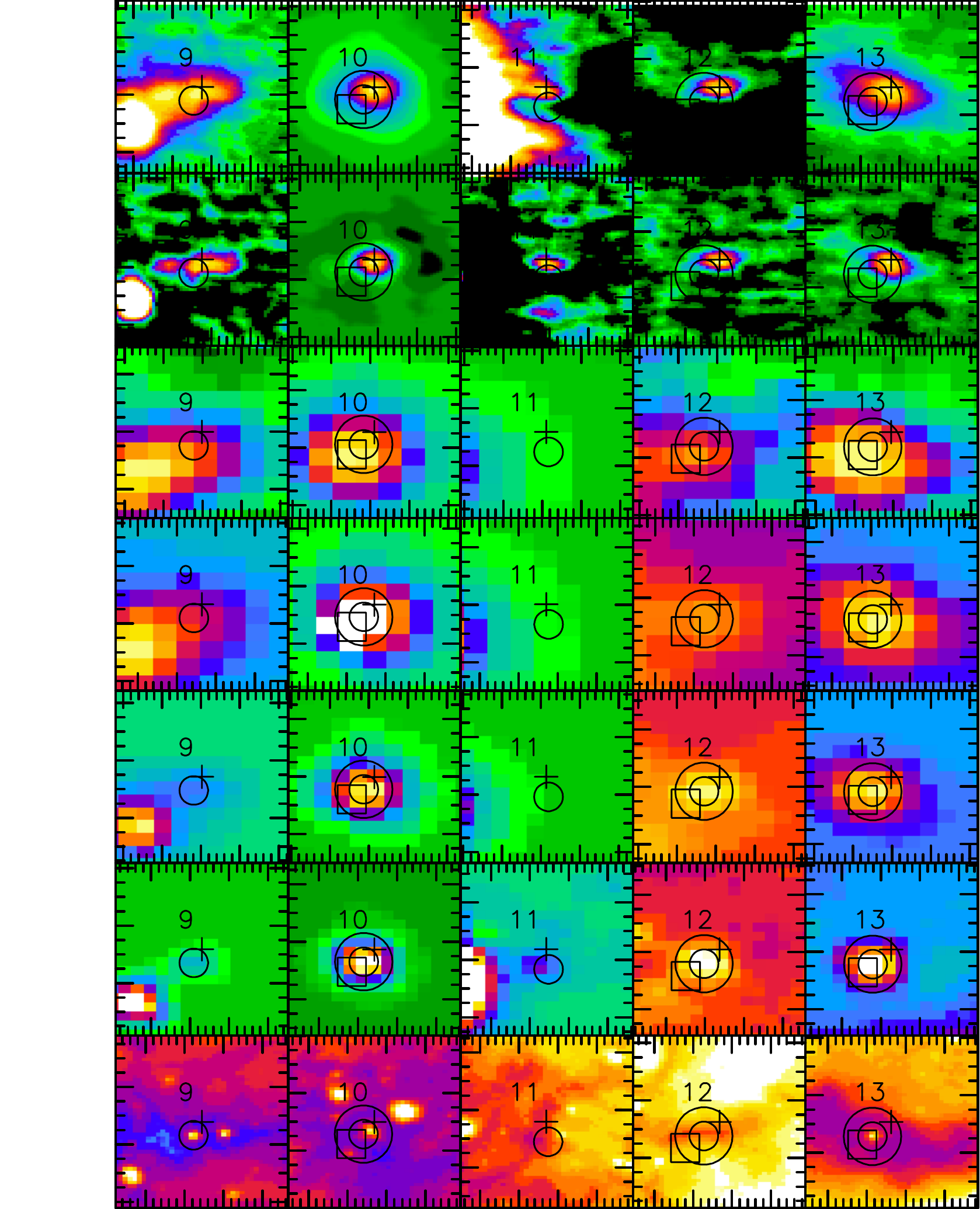}
    \caption{Figure C20 continued.}
    \label{fig:example_figure}
\end{figure}

\begin{figure}
	\includegraphics[width=8cm]{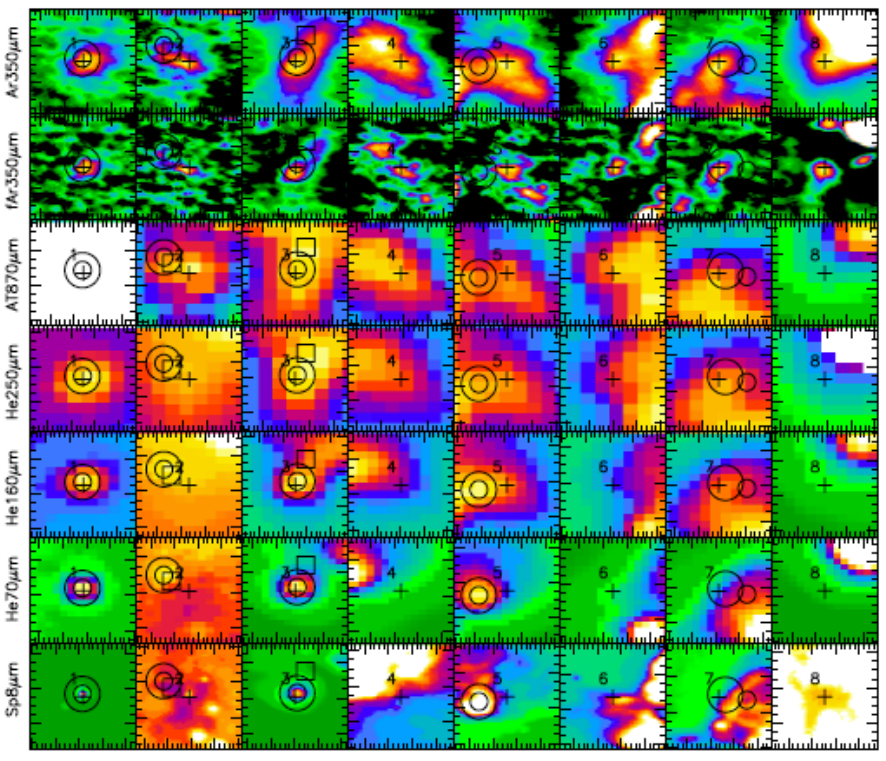}
    \caption{Same as Figure C1 but for the SDC345 field}
    \label{fig:example_figure}
\end{figure}

\begin{figure}
	\includegraphics[width=8.cm]{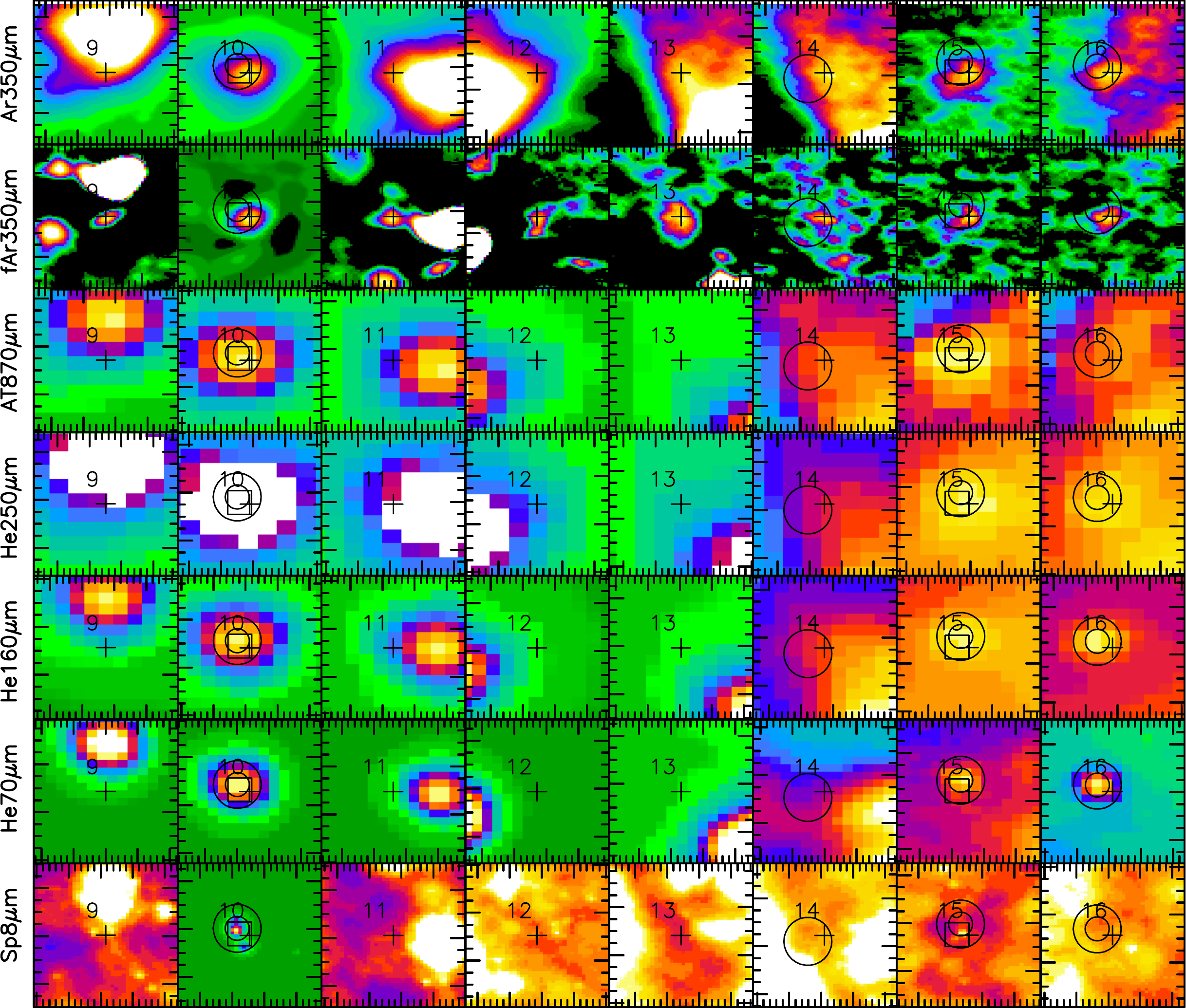}
        \caption{Figure C22 continued.}
    \label{fig:example_figure}
\end{figure}

\begin{figure}
	\includegraphics[width=3.6cm]{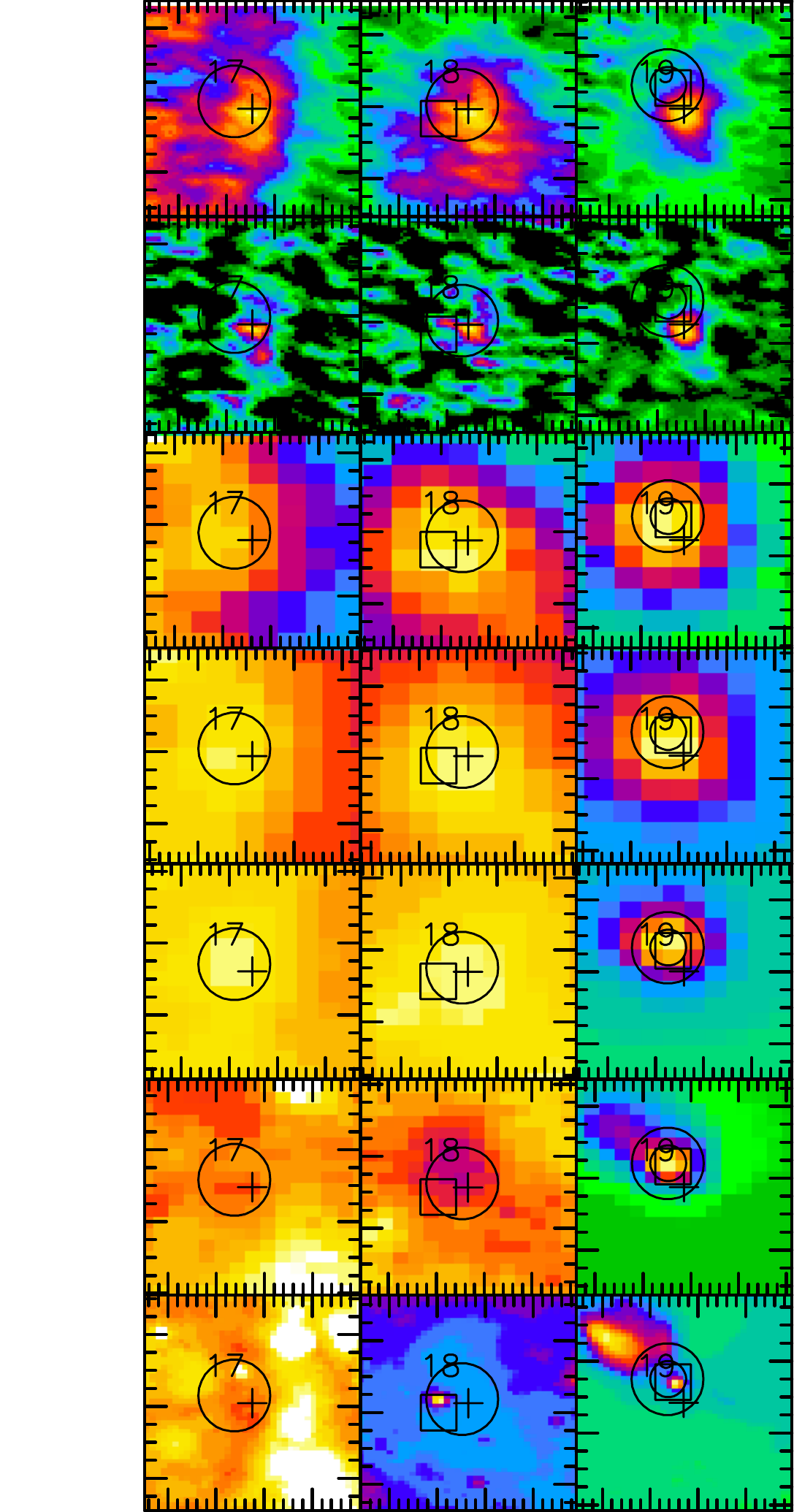}
        \caption{Figure C22 continued.}
    \label{fig:example_figure}
\end{figure}

\clearpage

\section{Alternative dust temperature assumption}

{As mentioned in Sec.~6, {\it all} core temperatures displayed in Fig.~\ref{mass_temp} are derived from Eq.~(5). This relationship has been partly inferred from the observed correlation between the internal temperature and the colour temperature of protostellar sources (see Fig.~\ref{tdustcomp}). The choice of applying Eq.~(5) to both protostellar and starless sources is justified by the absence of correlation between the ratio $\overline{T}_{\rm{int}}/T_{\rm{col}}$  and the source internal luminosity. However, for completeness, we here show the mass vs. temperature diagram where the dust temperatures of starless sources are estimated using $T_{\rm{dust}}=T_{\rm{col}}$ while using $T_{\rm{dust}}=1.2\times1.32\times T_{\rm{col}}$ for protostellar sources (as in Fig.~\ref{mass_temp_0p05pc}). The 1.2 factor is taken from Eq.~(5), while the 1.32 factor corresponds to the rescaling from 0.23\,pc (the original resolution of the temperature data) to 0.1\.pc (see section 6). The resulting mass vs. temperature diagram is shown in Fig.~\ref{masstemp_extreme}.

\begin{figure}
	\includegraphics[width=9.cm]{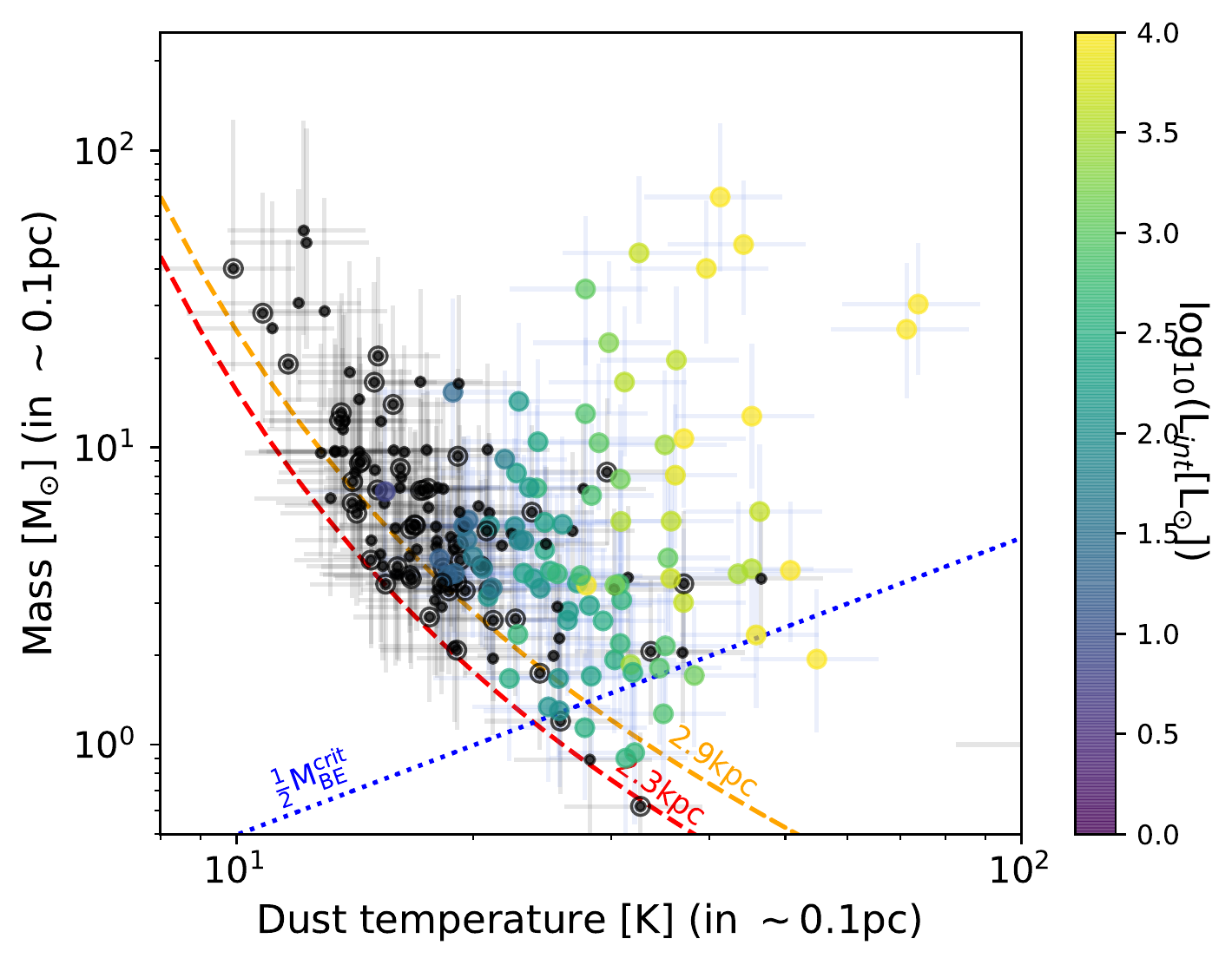}
   \caption{Same as Fig.~\ref{mass_temp_0p05pc} but with starless source's temperatures estimated using $T_{\rm{dust}}=T_{\rm{col}}$ as opposed to  $T_{\rm{dust}}=1.2T_{\rm{col}}$}
    \label{masstemp_extreme}
\end{figure}

\clearpage

\section{Models with $R_{\rm{core}}=1000$~AU}

\begin{figure}
	\includegraphics[width=9.cm]{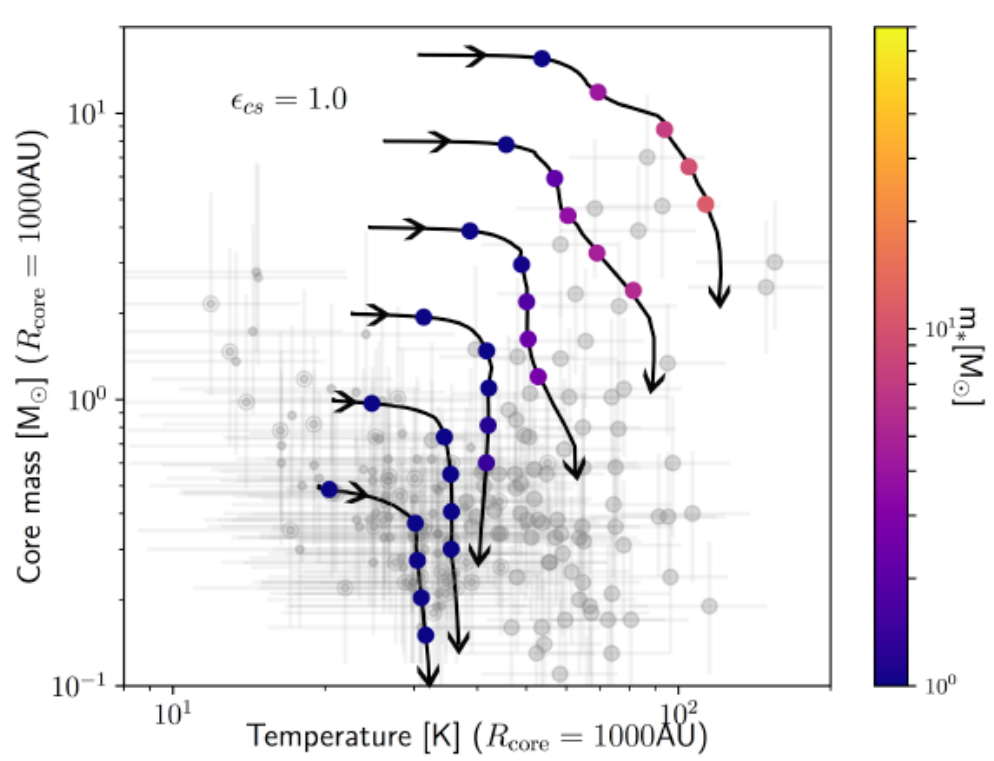}
   \caption{Core-fed models. Each track has been computed for a different initial core mass, from bottom to top ${m}_{\rm{core}}(t=0)= [0.5,1,2,4,8,16]$~M$_{\odot}$. The coloured symbols represent the position of the cores at times $t=[9\times10^3,9\times10^4,1.8\times10^5,2.7\times10^5,3.6\times10^5]$~yr. The colour codes the stellar mass at these times as displayed by the colour bar. The background grey symbols are  the ArT\'eMiS sources whose properties have been rescaled to 0.01pc (see text). Note that sources with $M_{\rm{gas}} < \frac{1}{2}M_{\rm{BE}}^{\rm{crit}}$ have been removed. }
    \label{corefed_1000AU}
\end{figure}

\begin{figure}
	\includegraphics[width=9.cm]{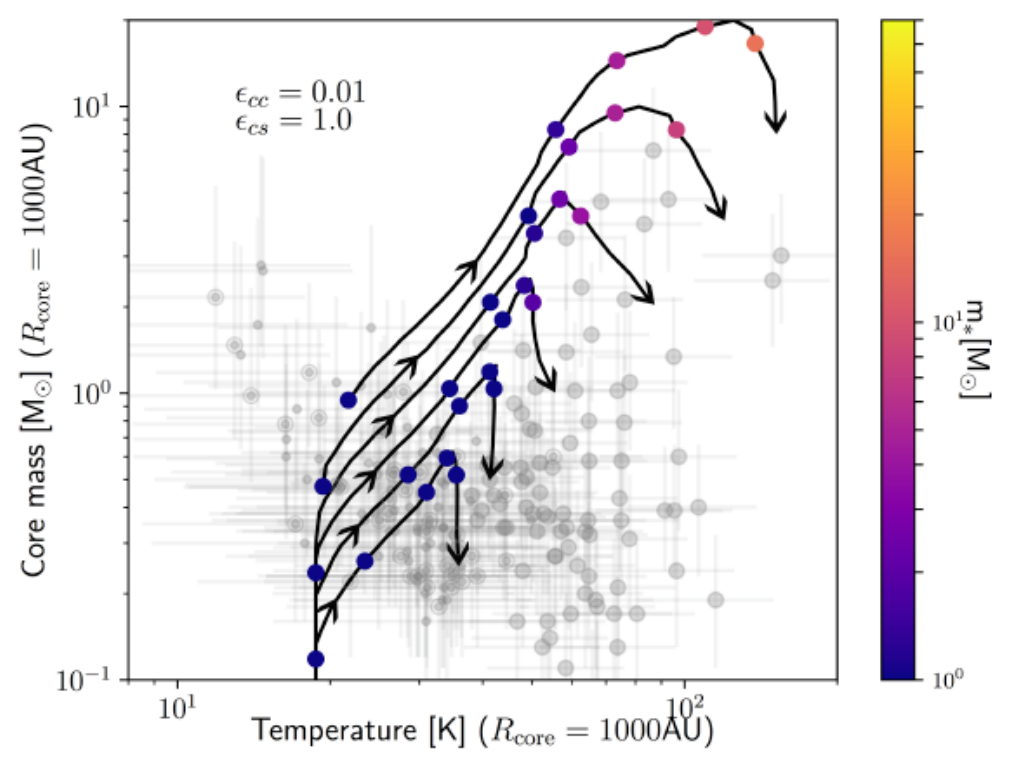}
    \caption{Clump-fed models. Each track has been computed for a different clump mass, from bottom to top $m_{\rm{clump}}=[100, 200, 400, 800, 1600, 3200]$~M$_{\odot}$.The coloured symbols represent the position of the cores at times $t=[3\times10^4,3\times10^5,6\times10^5,9\times10^5,1.2\times10^6]$~yr. The colour codes the stellar mass at these times as displayed by the colour bar. The background grey symbols are those presented in Fig.~\ref{corefed_1000AU}. Note that sources with $M_{\rm{gas}} < \frac{1}{2}M_{\rm{BE}}^{\rm{crit}}$ have been removed.}
    \label{clumpfed_1000AU}
\end{figure}

Fragmentation on scales of a couple of thousands AU scale \citep[e.g.][]{bontemps2010,motte2018b,beuther2018}, or even smaller scale \citep[e.g.][]{palau2013} is routinely observed in massive star-forming regions. 
In an attempt to produce similar model/data comparisons as those presented in Figs.~\ref{corefed} and \ref{clumpfed} but at a core scale of 0.01pc (i.e. $R_{\rm{core}}=1000$~AU) we rescaled the data as follows. For all sources, we assumed a density profile scaling as $\rho\propto r^{-2}$, which in practice implies a decrease of the core masses by a factor of 10 compared to the $R_{\rm{core}}=0.05$~pc case. Regarding the temperatures of protostellar sources, we used Eq.~(1) with the relevant radius, which in practice means an increase of the temperature by a factor 2.1 compared to the  $R_{\rm{core}}=0.05$~pc case. Finally, we leave unchanged the temperatures of starless sources. We here keep the same fractional temperature uncertainties of 20\%, however these are most likely much larger. The resulting observed core temperatures and masses are displayed as grey symbols in Figs.~\ref{corefed_1000AU} and \ref{clumpfed_1000AU}.

Figure \ref{corefed_1000AU} shows a set of core-fed models, with 6 different initial cores masses,  ${m}_{\rm{core}}(t=0)= [0.5,1,2,4,8,16]$~M$_{\odot}$. We kept the timescale $\tau_{\rm{core}}$ the same as in the $R_{\rm{core}}=0.05$~pc, but increased the core to star formation efficiency to $\epsilon_{cs}=1$, the maximum allowed for core-fed models. Unsurprisingly, the conclusions here are similar to those drawn from the $R_{\rm{core}}=0.05$~pc models, which is that they fail to explain the formation of the most massive stars (no massive prestellar cores), but may be compatible with the formation of intermediate-mass stars. The fact that one needs to use $\epsilon_{cs}=1$ to get a reasonable match with the data does show that massive star-forming cores on these sort of scales do need to accrete mass from radii that are larger than the last fragmentation scale. This is somewhat explicit given the low core masses.  

Figure \ref{clumpfed_1000AU} on the other hand, shows clump-fed tracks with an initial core mass ${m}_{\rm{core}}(t=0)=0.1$~M$_{\odot}$, a core formation efficiency $\epsilon_{cc}=0.01$ and a core to star formation efficiency $\epsilon_{cs}=1$. Clump masses are identical to those used for the  $R_{\rm{core}}=0.05$~pc models. Here again, as far as the most massive objects are concerned, we see that the clump-fed models are in better agreement with the observations. And similarly to the core-fed models, the use of  $\epsilon_{cs}=1$ shows that larger scale accretion is required.


\bsp	
\label{lastpage}
\end{document}